\documentclass[usenatbib]{mn2e}
\usepackage{graphicx,natbib,amssymb,amsmath}
\usepackage{txfonts, stfloats}
\usepackage{xcolor, ulem}

\newcommand{\expf}[1]{{{\rm e}^{#1}}}

\newcommand{\id}{{\,\rm d}}

\newcommand{\beq}{\begin{equation}}   %

\newcommand{\eeq}{\end{equation}}   %

\newcommand{\beqa}{\begin{eqnarray}}   %

\newcommand{\eeqa}{\end{eqnarray}}   %

\newcommand{\beal}{\begin{align}}

\newcommand{\enal}{\end{align}}

\newcommand{\bspl}{\begin{split}}

\newcommand{\espl}{\end{split}}

\newcommand{\bsub}{\begin{subequations}}

\newcommand{\esub}{\end{subequations}}

\newcommand{\bmulti}{\begin{multline}}   %

\newcommand{\beqm}{\begin{mathletters}}   %

\newcommand{\eeqm}{\end{mathletters}}   %

\newcommand{\me}{m_{\rm e}}

\newcommand{\Ne}{N_{\rm e}}

\newcommand{\Te}{T_{\rm e}}

\newcommand{\The}{\theta_{\rm e}}

\newcommand{\sigT}{\sigma_{\rm T}}

\newcommand{\vek} [1]{\mbox{\boldmath${#1}$\unboldmath}}

\newcommand{\lit}[1]{{\color{red}[{\sc REF}]}}

\title[The Compton scattering kernel]{Dissecting the Compton scattering kernel I: Isotropic media}
\author{Abir Sarkar and Jens Chluba}

\begin{document}

\author[Sarkar, Chluba \& Lee]{
Abir Sarkar$^1$\thanks{E-mail:abir.sarkar@manchester.ac.uk}, 
Jens Chluba$^1$\thanks{E-mail:jens.chluba@manchester.ac.uk} and 
Elizabeth Lee$^1$\thanks{E-mail:elizabeth.lee-2@postgrad.manchester.ac.uk}
\\
$^1$Jodrell Bank Centre for Astrophysics, School of Physics and Astronomy, The University of Manchester, Manchester M13 9PL, U.K.
}

\date{\vspace{-6.2mm}{Accepted 2019 --. Received May 2019}}

\maketitle

\begin{abstract}
Compton scattering between electrons and photons plays a crucial role in astrophysical plasmas. Many important aspects of this process can be captured by using the so-called Compton scattering kernel. For isotropic media, exact analytic expressions (valid at all electron and photon energies) do exist but are hampered by numerical issues and often are presented in complicated ways.
In this paper, we summarize, simplify and improve existing analytic expressions for the Compton scattering kernel, with an eye on clarity and physical understanding. We provide a detailed overview of important properties of the kernel covering a wide range of energies and highlighting aspects that have not been appreciated as much previously. We discuss analytic expressions for the moments of the kernel, comparing various approximations and demonstrating their precision. We also illustrate the properties of the scattering kernel for thermal electrons at various temperatures and photon energies, introducing new analytic approximations valid to high temperatures. 
The obtained improved formulae for the kernel and its moments should prove useful in many astrophysical computations, one of them being the evolution of spectral distortions of the cosmic microwave background in the early Universe. A novel code, {\tt CSpack}, for efficient computations of the Compton scattering kernel and its properties (in the future also including anisotropies in the initial electron and photon distributions) is being developed in a series of papers and will be available within one month.
\end{abstract}


\section{Introduction}
Compton scattering is one of the most important processes in astrophysical plasmas \citep[e.g.,][]{Blumenthal1970, Rybicki1979}. It controls the energy exchange between photons and electrons, leads to the redistribution of these particles both in energy and direction, and plays a crucial role for the thermalization of their distribution functions.
One important application of Compton scattering is in computations of spectral distortions of the cosmic microwave background (CMB) caused by energy release in the primordial Universe \citep[e.g.,][]{Sunyaev1970mu, Burigana1991, Hu1993, Chluba2011therm}. Another is related to the Sunyaev-Zeldovich (SZ) effect from the scattering of CMB photons by the hot electron plasma inside clusters of galaxies \citep{Zeldovich1969, Carlstrom2002, SZreview2019}, which today allows studying objects out to high redshifts \citep[e.g.,][]{Vanderlinde2010, STA11, Planck2013SZ}. Compton scattering furthermore plays an important role in shaping the spectra of cosmic-ray particles and for electromagnetic cascades inside dilute plasmas \citep[e.g.,][]{Shull1985, Slatyer2009, Valdes2010, Slatyer2015}. It is therefore important to understand this process for a wide range of energies and physical conditions. 
 
 All relevant aspects of the Compton process can be captured by the scattering kernel. It describes how a photon of a given energy is redistributed in a collision with an electron. Generally, the scattering between anisotropic photon and electron distributions has to be considered; however, in many astrophysical situations (e.g., for the thermal SZ effect and for CMB spectral distortions) it is usually sufficient to consider the isotropic case unless high precision is required. This limit is also far easier to handle analytically and numerically, and we shall focus on it in the present work.
 
The Compton kernel in isotropic media has been studied quite extensively in the literature for different astrophysical situations of interest, both numerically, with different levels of simplifications \citep[e.g.,][]{Pomraning1972, Buchler1976, Pozdniakov1979, Guilbert1981, Madej2017}, and analytically in various approaches \citep[e.g.,][]{Aharonian1981, Brinkmann1984, Kershaw1986, Nagirner1994, Sazonov2000, Ensslin2000, Dolgov2001, Pe'er2005}. To our knowledge, the first exact calculation of the kernel for mono-energetic electrons and photons was performed by \citet[][hereafter J68]{1968PhRv..167.1159J}, who carried out all integrals analytically, expressing the kernel in terms of elementary functions. However, the expressions given in J68 appear complicated, contain a few misprints and suffer from numerical instabilities. 
The task of tidying these expression up was later undertaken by several authors \citep[e.g,][]{Aharonian1981, Brinkmann1984, Kershaw1986, Nagirner1994}. We refer to \citet[][hereafter NP94]{Nagirner1994} for an extensive review of previous works, with general expressions for several relevant quantities\footnote{A few minor typos in NP94 will be mentioned in Sect.~\ref{sec:moment}.} (e.g., intensity redistribution function, the moments and the thermally-averaged moments). 
The numerical issues of J68 were considered more recently in a few papers \citep[e.g.,][]{Pe'er2005,2009A&A...506..589B} yielding numerically more stable expressions. However, the formulae provided there and in NP94 still remain cumbersome and extracting physical insight from them is difficult.
The literature is furthermore full of simpler approximations for limiting cases \citep[e.g., J68,][]{Blumenthal1970, Sazonov2000, Ensslin2000} and a comprehensive comparative discussion of all regimes seems beneficial.
 
In this paper, we present a compact expression for the kernel that is numerically stable and easy to interpret. We start from the expressions given by \citet[][hereafter B09]{2009A&A...506..589B} and then reformulate them (see Appendix~\ref{app:Belmont} for details). 
We explicitly demonstrate that for fixed incident electron and photon momenta the scattering kernel at most has three energy zones (Sect.~\ref{sec:energy_zones}). This is one of the main steps that permit further simplifications over B09, as the expressions given there contain several switches through conditions, suggesting that naively some 16 cases have to be distinguished. 
The energy zones are separated by singular points (e.g., cusps and knees), which naturally arise from restrictions in the scattering angles (see J68). One cusp is located at the initial photon energy, a second may appear red- or blue-ward of this energy, depending on the ratios of the electron and photon momenta. Interestingly, the appearance of a cusp blue-ward of the initial photon energy is less known as it only appears when the photon has a sufficient initial momentum.
After solving the conditions for the energy zones, only one unifying algebraic expression in terms of elementary functions can be used to compute the kernel by simply switching the appropriate variables (Sect.~\ref{sec:final_kernel_ex}). The final expression is general and applicable at all energies, overcoming some of the remaining limitations of B09. We also compute the thermally-averaged kernel and discuss the first five kernel moments using our formulae, confirming their numerical stability and independently deriving explicit expressions for the first three moments.
 
This work is the first in a series of papers that will develop a new code, {\tt CSpack}, for efficient computations of the Compton scattering kernel and its properties, covering all cases presented here. In particular,  {\tt CSpack} will ensure that full numerical stability and efficiency is guaranteed at all energies. In the near future, we plan to extend this package to also account for anisotropies in the initial electron and photon distribution. This will allow us to develop novel numerical schemes to solve radiative transfer problems encountered in the early Universe and other astrophysical plasmas.
 
The paper is organized as follows. In Sect.~\ref{sec:kinetic_equation}, we formally define the scattering kernel in an isotropic medium, starting from the photon collision term. 
In Sect.~\ref{sec:ker_domains}, we provide a simple analytic expression of the kernel and discuss its physical properties at different zones. We also summarize existing approximate expressions for the kernel. 
Section~\ref{sec:illus_iso} contains illustrations for the kernel for different combinations photon energies and electron momentum. In that section, we illustrate the applicability of the approximate expressions for some extreme scenarios. In Sect.~\ref{sec:moment}, we study the first five moments of the scattering kernel. We provide general analytic formulae for the first three moments along with some approximate expressions that we graphically justify later in the section.
In Sect.~\ref{sec:th_moments}, we study the first five thermally-averaged moments, with several approximation schemes for the first three of them, which we compare with their exact numerical behaviour. 
We conclude in Sect.~\ref{sec:conclusion}. 

\vspace{-3mm}
\section{Definition of the Compton kernel}
\label{sec:kinetic_equation}
In this section, we give the definition of the Compton scattering kernel and discuss some of its simple properties.
Compton scattering is a process of the form $\gamma_0(k_0) +e_0(p_0) \leftrightarrow \gamma(k) + e(p)$, where in this context $p_0$, $k_0$, $p$ and $k$ denote four-vectors. We assume that the initial electron and photon distributions are both isotropic, and that the electrons are non-degenerate (i.e., Fermi-blocking can be neglected). The kinetic equation for the photon occupation number, $n(\omega_0)$, at energy\footnote{We shall use the common definitions of physical constants ($c, h,\me$, etc).} $\omega_0=h\nu_0/\me c^2$ is then given by \citep[e.g.,][NP94]{Pomraning1972, Buchler1976}
\begin{align}
\label{eq:kin_eq1}
\frac{1}{c}\frac{\text{d}n(\omega_0)}{\text{d}t}  
&= \frac{1}{2 E_{\gamma_0}}\!\int \!\frac{\text{d}^3p_0}{(2\pi)^32E_0}  \frac{\text{d}^3p}{(2\pi)^32E} \frac{\text{d}^3k}{(2\pi)^32E_{\gamma}}  \nonumber
\\[0.5mm] 
\nonumber
&\quad\qquad\times
(2\pi)^4\delta^{(4)}(p+k-p_0-k_0) \,\lvert \mathcal{M} \rvert ^2 
\\[0.5mm] 
&\quad\qquad\qquad\times\Big[f n(1+n_0) - f_0 n_0(1+n)\Big].
\end{align}
Here, the energies of the particles are determined by\footnote{We denote Lorentz-factors by $\gamma=\sqrt{1+p^2}=1/\sqrt{1-\beta^2}$ with speed $\beta=\varv/c=\gamma p$, where here $p$ is the electron momentum in units of $\me c$.} $E_0=\gamma_0 \,\me c^2$, $E_{\gamma_0}=\omega_0 \,\me c^2$, $E=\gamma \,\me c^2$ and $E_\gamma=\omega \,\me c^2$, respectively. The electron distribution functions are $f_0=f(\gamma_0)$ and $f=f(\gamma)$, while those for the photons are $n_0=n(\omega_0)$ and $n=n(\omega)$. The factors $\propto(1+n)$ account for stimulated scattering effects, which are important close to equilibrium.  Finally, $\lvert \mathcal{M} \rvert ^2$ is the squared matrix element for Compton scattering \citep[e.g., see][]{Jauch1976}.

After some simplification and change of variables (we refer to Appendix~\ref{sec:app:kerdef} for details), Eq.~\eqref{eq:kin_eq1} can be cast into the compact form \citep[e.g., see][for comparison albeit with slightly different conventions]{Sazonov2000}
\begin{align}
\!\!\!\!\frac{\text{d}n_0}{\text{d}\tau} 
&\!= \! \int \Bigg[\frac{\omega^2}{\omega_0^2} P(\omega \rightarrow \omega_0) \,n(1+n_0) - P(\omega_0 \rightarrow \omega) \,n_0(1+n) \Bigg]\,\text{d}\omega
\label{eq:kin_eq2}
\end{align}
with Thomson scattering optical depth, $\tau = \int cN_{\rm e}\sigma_{\rm T}\text{d}t$. The factor $\omega^2/\omega_0^2$ ensures photon number conservation, $\int \omega_0^2 \id \omega_0 \text{d}n_0/\text{d}\tau=0$, as can be readily confirmed.
We also introduced the scattering kernel, $P(\omega_0 \rightarrow \omega)$, which describes the redistribution of photons from the initial energy $\omega_0$ to $\omega$, while the reverse process is given by $P(\omega \rightarrow \omega_0)$. These kernels are obtained from the single-momentum scattering kernel, $P(\omega_0 \rightarrow \omega, p_0)$, as defined in Sect.~\ref{sec:final_kernel_ex}, after integrating over all electron momenta:
\bsub
\label{eq:therm_kernel}
\begin{align}
P(\omega_0 \rightarrow \omega) &= \int_{p_0^{\rm {min}}}^{\infty} p_0^2 f(\gamma_0) P(\omega_0 \rightarrow \omega, p_0)\, \text{d}p_0
\\
P(\omega \rightarrow \omega_0) 
&=
\,\int_{p^{\rm {min}}}^{\infty} p^2 f(\gamma) \,P(\omega \rightarrow \omega_0, p)\, \text{d}p,
\\
\nonumber
&\equiv \frac{\omega_0^2}{\omega^2} 
\int_{p_0^{\rm {min}}}^{\infty} p_0^2 f(\gamma) \,P(\omega_0 \rightarrow \omega, p_0)\, \text{d}p_0.
\end{align}
\esub
Here, $\gamma = \gamma_0+\omega_0-\omega$ and $p_0^{\rm {min}}=p_0^{\rm {min}}(\omega_0, \omega)$ is the minimally required electron momentum in the redistribution process, as will be explained in Sect.~\ref{sec:thermal_average}. We also used the kernel relation, Eq.~\eqref{app:kernel_prop}, to switch variables.
The phase space distribution function of the electrons, $f(\gamma)$, is assumed to only depend on the electron energy but otherwise can have a general form (e.g., thermal or non-thermal), determined by extra parameters, i.e, the temperature of the electrons or the spectral index of the energy spectrum, which we shall suppress in our notation. 

In the present work, we will illustrate the kernel for non-degenerate, thermal electrons. We shall assume a relativistic Maxwell-Boltzmann (rMB) distribution, also known as Maxwell-Juttner  distribution function at a temperature $\Te$, determined by:
\begin{equation}
f(\gamma_0) = \frac{\exp\Big(-\gamma_0/\The\Big)}{\theta_{\rm e} K_2(1/\theta_{\rm e})}
\quad \text{with} \quad \theta_{\rm e} = \frac{kT_{\rm e}}{m_{\rm e}c^2}.
\label{eq:rmbdist}
\end{equation}
Here $K_2(1/\theta_{\rm e})$ is the modified Bessel function of second order, necessary to ensure the normalization $\int_0^{\infty}p_0^2f(\gamma_0)\,\text{d}p_0 = 1$. For thermal electrons, one has $f(\gamma)=f(\gamma_0)\,\exp(\frac{
\gamma_0-\gamma}{\theta_{\rm e}})\equiv f(\gamma_0)\,\exp(\frac{
\omega-\omega_0}{\theta_{\rm e}})$, such that with Eq.~\eqref{eq:therm_kernel} the kernel obeys the {\it detailed balance relation} 
\begin{equation}
P^{\rm th}(\omega \rightarrow \omega_0) 
= \frac{\omega_0^2}{\omega^2}\,\expf{\frac{\omega-\omega_0}{\theta_{\rm e}}}\,
P^{\rm th}(\omega_0 \rightarrow \omega),
\label{eq:det_bal}
\end{equation}
however, generally this relation is not applicable.

We will describe the properties of the kernel for different combinations of $\omega_0$ and $p_0$ below. One of them is related to the kernel moment of order $m$, which is given by
\begin{equation}
\label{eq:moment}
\Sigma_m(\omega_0, p_0)  = \int \Bigg(\frac{\omega-\omega_0}{\omega_0}\Bigg)^m P(\omega_0 \rightarrow \omega, p_0) \, \text{d}\omega.
\end{equation}
In this paper, we shall consider the zeroth, first and second moments of the kernel analytically, while a numerical study for the third and fourth moment is carried out for further illustration. The first three moments can be identified with the total cross-section, the net energy exchange with the scattered photon and the dispersion of the scattered photon distribution, respectively. Along with the exact expressions, we also provide several simpler versions for the moments in Doppler- and recoil-dominated scattering processes. The moments can also be averaged over the electron distribution function, as we discuss assuming thermal electrons.

\section{Anatomy of the Compton kernel}
\label{sec:ker_domains}

\subsection{Critical energies and kernel domains}
\label{sec:energy_zones}
To understand the anatomy of the kernel better, we first consider the energy of the scattered photon, $\omega=h\nu/\me c^2$, which is determined by \citep[e.g.,][]{Jauch1976}
\begin{equation}
\omega = \frac{1-\beta_0\mu_0}{1-\beta_0\mu}\frac{\omega_0}{1+\frac{\omega_0}{\gamma_0}\frac{1-\mu_{\rm sc}}{1-\beta_0\mu}}.
\label{omscat}
\end{equation}
Here $\mu_0$ and $\mu$ respectively are the direction cosines of the angles between the incident and scattered photon with the incident electron, while $\mu_{\rm{sc}}$ denotes the direction cosine between the two photons. Energy conservation ensures that $p=\sqrt{(\gamma_0+\omega_0-\omega)^2-1}$ and $\gamma=\gamma_0+\omega_0-\omega$ for the scattered electron.

\begin{figure}
\includegraphics[width=\linewidth ]{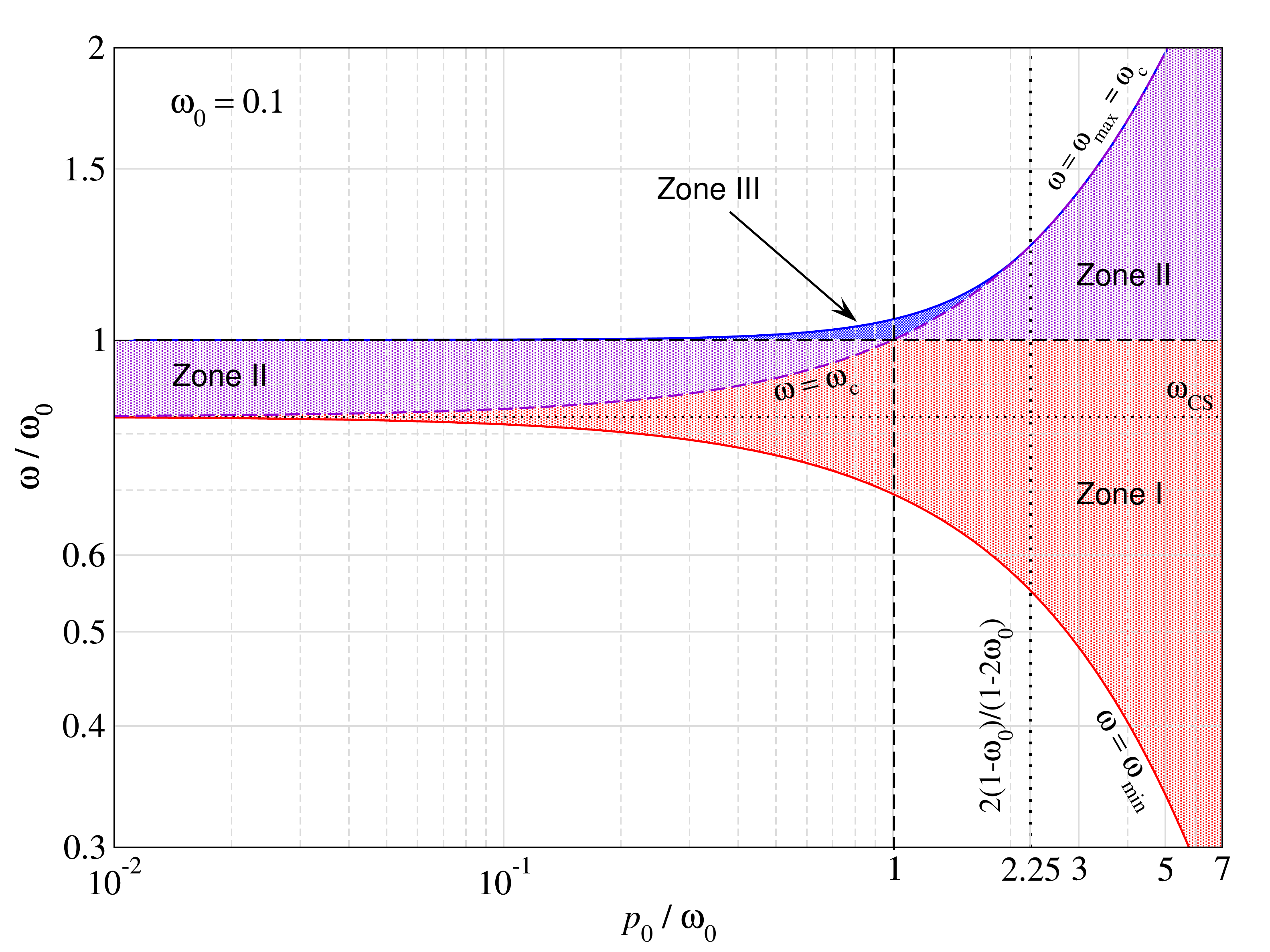}
\\[0mm]
\includegraphics[width=\linewidth ]{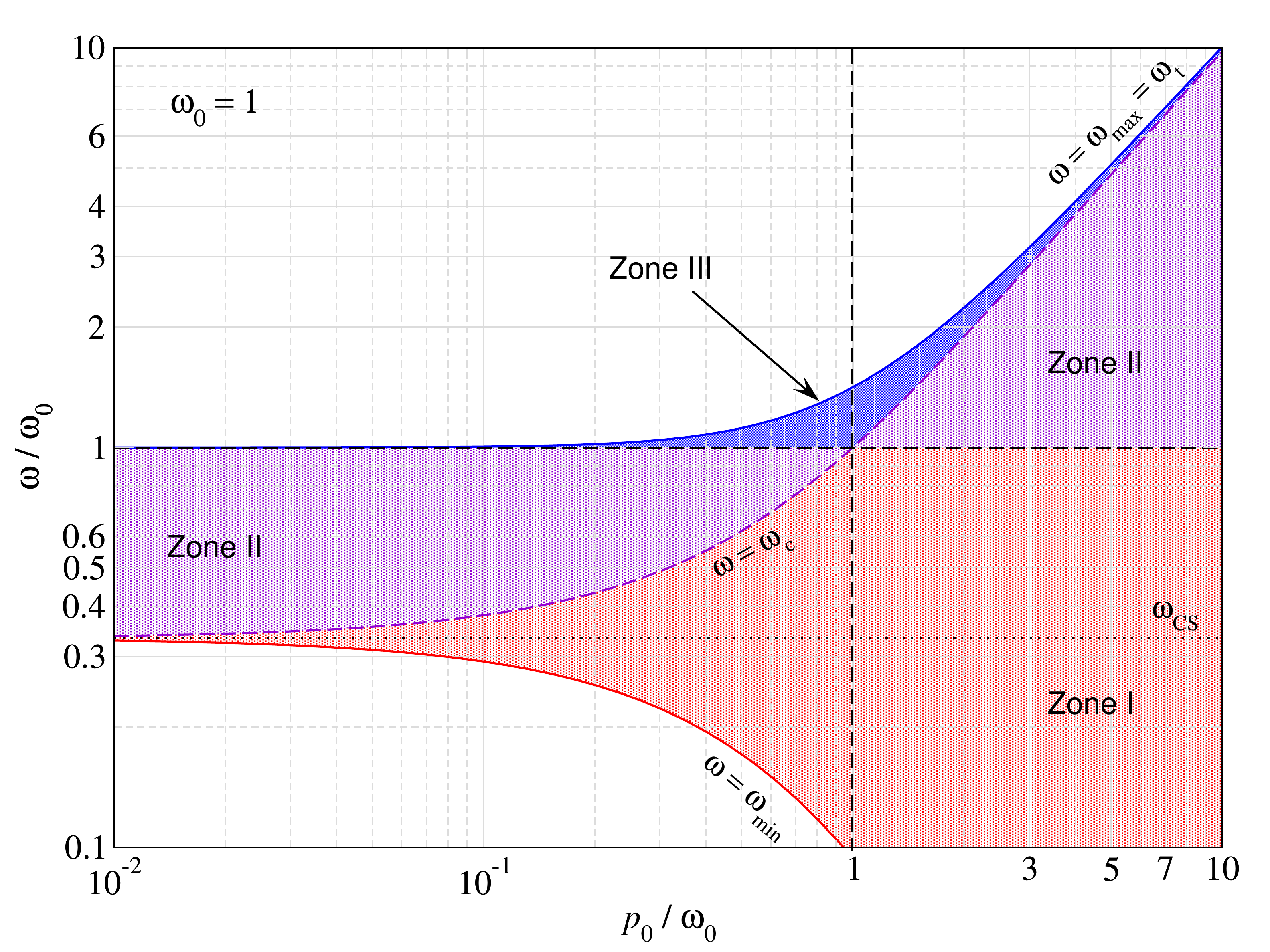}
\caption{Domains of the Compton kernel for representative examples. For $p_0=\omega_0$, only two zones are present (zone I and III), while for $\omega_0>1/2$ three zones are relevant at all values of $p_0$. The vertical dotted line in the upper panel denotes the electron momentum where zone-III disappears.}
\label{fig:domains}
\end{figure}
Considering all possible scattering constellations, the scattered photon energy reaches its minimal value when the incident photon and electron travel in the same direction ($\mu_0=1$) and the photon is then back-scattered ($\mu=\mu_{\rm sc}=-1$):
\begin{align}
\label{ommin}
\omega_{\rm{min}}
&=  \frac{(\gamma_0-p_0)\,\omega_0}{\gamma_0+p_0+2\,\omega_0}.
\end{align}
This expression simply reflects that kinematically it is impossible to transfer all the energy of the incident photon to the scattered electron, which in the extreme case only carries a total energy of $\gamma_{\rm max}=\gamma_0+2\omega_0(\omega_0+p_0)/(\gamma_0+p_0+2\omega_0)<\gamma_0+\omega_0$. We note that the minimal photon energy is also lower than the Compton back-scattering peak for resting electrons, i.e., $\omega_{\rm CS}=\omega_0/(1+2\omega_0)$.
For $\omega_0\ll p_0$, one furthermore has $\omega^{\rm D}_{\rm{min}}\approx \omega_0(1-\beta_0)/(1+\beta_0)$, which turns out to be valid even when $\omega_0\gtrsim 1$ as long as $\omega_0\ll p_0$.

In contrast, it is kinematically possible to transfer {\it all} the initial kinetic energy of the incident electron to the scattered photon. In this limiting case, one maximally obtains $\omega_{\rm max}\equiv\omega_{\rm t}=\gamma_0+\omega_0-1$; however, this regime is only accessible if $\omega_0>(1+p_0-\gamma_0)/2$ or equivalently $p_0<2\omega_0(1-\omega_0)/(1-2\omega_0)$ are satisfied, since otherwise restrictions on the scattering angles apply (see also, B09). If $\omega_0>1/2$, this implies that $\omega_{\rm max}\equiv\omega_{\rm t}$ for all $p_0>0$. One thus obtains the maximal energy of the scattered photon as
\begin{subequations}
\begin{align}
\label{ommax}
\omega_{\rm{max}} &=
\begin{cases}
\gamma_0+\omega_0-1 &\text{for}\; \omega_0 > \frac{1}{2}(1+p_0-\gamma_0)
\\[1mm]
\omega_{\rm c} &\text{for}\; \omega_0 \leq \frac{1}{2}(1+p_0-\gamma_0),
\end{cases}
\\[0.5mm]
\label{omcrit}
\omega_{\rm c}
&=\frac{(\gamma_0+p_0)\,\omega_0}{\gamma_0-p_0+2\,\omega_0}.
\end{align}
\end{subequations}
For convenience, we also introduced the critical frequency, $\omega_{\rm c}$, which is obtained for scattering constellations in which the incident electron and photon experience a head-on collision ($\mu_0=-1$) with full back-scattering ($\mu=\mu_{\rm sc}=1$).\footnote{This constellation covers a small phase space and hence should create a feature in the scattering kernel.}
For $\omega_0=\frac{1}{2}(1+p_0-\gamma_0)<1/2$, one finds $\omega_{\rm t}=\omega_0+\gamma_0-1 \equiv  \omega_{\rm c}$. It is also easy to show that $\omega_{\rm{max}}$ has a smooth first derivative around $\omega_0=\frac{1}{2}(1+p_0-\gamma_0)$.

Put together, our discussion reveals a maximum of four critical energies. The maximal and minimal frequencies lie on the boundaries of the scattering domain, where the kernel vanishes. Thus, the only singular points (i.e., cusps and corners) are present at $\omega=\omega_{\rm c}$ and $\omega=\omega_0$. This implies that the Compton kernel generally has three distinct energy zones:
\bsub
\label{zonedef}
\begin{align}
&\text{Zone I:} & \omega_{\rm min}&\leq \omega<\omega_{\rm I},
\\
&\text{Zone II:} &\omega_{\rm I}&\leq \omega < \omega_{\rm II},
\\
&\text{Zone III:} &\omega_{\rm II}&\leq \omega \leq \omega_{\rm max},
\end{align}
\esub
One can furthermore expect $p_0=\omega_0$ to represent a special case for which $\omega_{\rm c}\equiv \omega_0$ such that only two main zones remain. These findings suggest that the final expressions for the kernel can indeed be greatly simplified, as we show next.

In Fig.~\ref{fig:domains}, we illustrate the variation of the critical frequencies with $p_0$ for two representative cases, $\omega_0=0.1$ and $\omega_0=1$. For $p_0\ll \omega_0$, zone II dominates and $\omega_{\rm min}\approx \omega_{\rm CS}=\omega_0/(1+2\omega_0)$, while $\omega_{\rm max}\approx \omega_0$. In this regime, the recoil-dominated scattering kernel provides a good approximation (Sect.~\ref{sec:rec_K}). When increasing $p_0$, also the regimes $\omega <\omega_{\rm CS}<\omega_{\rm c}$ (zone I) and $\omega>\omega_0$ (zone III) open up. Around $p_0\simeq \omega_0$ these two domains reach their largest size, fully dominating the shape of the kernel for $p_0= \omega_0$.
At $p_0 \gg \omega_0$, zone III becomes sub-dominant and the Doppler-dominated (Sect.~\ref{sec:Doppler_K}) and ultra-relativistic (Sect.~\ref{sec:ur-kernel}) scattering kernel approximations become applicable. We show, however, that neither of them capture the general asymptotic behaviour (see Fig.~\ref{fig:kernels_comp_UR}).
Also assuming $\omega_0\ll 1$, we have $\omega_{\rm max}\rightarrow \omega_{\rm c}$ when increasing $p_0$. In this case, the transition of $\omega_{\rm max}$ from $\omega_{\rm t}$ to $\omega_{\rm c}$ occurs smoothly at $p_0\simeq 2\omega_0$.
In the limit $\omega_0\rightarrow 1/2$, the transition moves towards $p_0\rightarrow \infty$, while for $\omega_0>1/2$ one always has $\omega_{\rm max}=\omega_{\rm t}>\omega_{\rm c}$ for all $p_0$, although asymptotically $\omega_{\rm max}\simeq \omega_{\rm t}\simeq \omega_{\rm c}\simeq p_0$ for large $p_0$. In this regime, the subdominant zone III is always present (lower panel of Fig.~\ref{fig:domains}), while for $\omega_0< 1/2$, zone III closes at $p_0>2 \omega_0 (1 - \omega_0)/(1 - 2 \omega_0)$ with $\omega_{\rm max}=\omega_{\rm c}<\omega_{\rm t}$ (upper panel of Fig.~\ref{fig:domains}).

\vspace{-2mm}
\subsection{Simplified exact kernel expression}
\label{sec:final_kernel_ex}
Starting from B09, we obtained a further simplified algebraic expression for the kernel (Appendix~\ref{app:Belmont}). It is based on the auxiliary function (suppressing explicit dependence on $\omega_0, \omega$ and $p_0$)
\begin{subequations}
\label{eq:kernel_all}
\begin{align}
\label{eq:kernel_main}
\mathcal{G}(\omega^*_0, \omega^*, \kappa) &=
\kappa\left\{
2+(\omega^* -\omega^*_0)^2 \frac{(1+\omega \omega_0)}{\omega^2 \omega_0^2}
\right.\nonumber\\
&
\,+ 2
\!\left[\frac{1}{\omega^*} \mathcal{S}\left(\frac{\kappa^2 \lambda_+}{{\omega^*}^2}\right)
-\frac{1}{\omega^*_0}\mathcal{S}\left(\frac{\kappa^2 \lambda_-}{{\omega^*_0}^2}\right)\right]
\\ \nonumber
&\, +(1+\omega \omega_0) \!\left[
\frac{1}{\omega^* \lambda_+}\mathcal{F}\!\left(\frac{\kappa^2 \lambda_+}{{\omega^*}^2}\right)
- \frac{1}{\omega^*_0 \lambda_-}\mathcal{F}\!\left(\frac{\kappa^2 \lambda_-}{{\omega^*_0}^2}\right)\right]\Bigg\}
\end{align}
with the simple definitions
\begin{align}
p &= \sqrt{p_0^2 + 2\gamma_0(\omega_0-\omega) + (\omega_0-\omega)^2}
\\
\lambda_{+}
& = p_0^2 +2\gamma_0\omega_0 + \omega_0^2  
\\
\lambda_{-}
& = p_0^2 -2\gamma_0\omega + \omega^2 
\\[0.5mm]
\mathcal{S}(x)
&=\frac{\sinh^{-1}\!\!\sqrt{x}}{\sqrt{x}}\equiv\frac{\sin^{-1}\!\!\sqrt{-x}}{\sqrt{-x}},
\\
\mathcal{F}(x) &= \mathcal{S}(x) - \sqrt{1+x},
\end{align}
where $p$ is the momentum of the scattered electron. We recommend using the expressions as given here to avoid cancellation issues. It is also worth noting that $\lambda_{+}>0$ for all energies, while $\lambda_{-}$ vanishes if $\omega=\gamma_0\pm1$, however, without causing a real pole (see also B09).
Finally, we will need the mean photon energies $\bar{\omega}$ and $\bar{\omega}_0$, 
\begin{align}
\bar{\omega}&=\sqrt{\frac{\omega\omega_0(\gamma+p)}{\gamma_0+p_0}},
\quad \bar{\omega}_0=\sqrt{\frac{\omega\omega_0(\gamma_0+p_0)}{\gamma+p}},
\end{align}
which satisfy the useful identity $\bar{\omega}\bar{\omega}_0=\omega\omega_0$, and the additional function $\Lambda$ as given by
\begin{align}
&\Lambda(\omega_0, \omega, p_0, p)
=\frac{p_0-p+\omega_0+\omega}{2}.
\end{align}
To avoid numerical issues, we furthermore use
\begin{align}
\mathcal{S}_{\rm low}(x) &\approx 1- \frac{x}{6} + \frac{3x^2}{40} - \frac{5x^3}{112}, 
\\[-0.0mm]
\mathcal{F}_{\rm low}(x) &\approx  -\frac{2x}{3} + \frac{x^2}{5} - \frac{3x^3}{28},
\end{align}
\end{subequations}
whenever $x\ll 1$. The expressions~\eqref{eq:kernel_all} 
  allow computing the general scattering kernel, in all energy zones for any combination of $\omega_0$ and $p_0$. The kernel for a desired zone is found by switching the arguments of $\mathcal{G}(\omega^*_0, \omega^*, \kappa)$, as we explain now.

\vspace{-3mm}
\subsubsection{$\omega_{\rm c}\leq\omega_0$}
Assuming $\omega_{\rm c}\leq\omega_0$ (or $p_0\leq\omega_0$), with Eq.~\eqref{eq:kernel_all} the kernel in the three energy zones can be written as
\begin{subequations}
\label{eq:P_ww0<1}
\begin{align}
P_{\rm I}(\omega_0 \rightarrow \omega, p_0)
&=\frac{3}{8\gamma_0 p_0\omega_0^2}\, \mathcal{G}\left(\bar{\omega}_0, \bar{\omega}, \kappa_1\right),
\\[-0.5mm]
\label{eq:recoil_case}
P_{\rm II}(\omega_0 \rightarrow \omega, p_0)
&=\frac{3}{8\gamma_0 p_0\omega_0^2}\, \mathcal{G}\left(\omega, \omega_0, p_0\right),
\\[-0.5mm]
P_{\rm III}(\omega_0 \rightarrow \omega, p_0)
&=\frac{3}{8\gamma_0 p_0\omega_0^2}\, \mathcal{G}\left(\omega_0, \omega, p\right),
\end{align}
\end{subequations}
where $\kappa_1=\Lambda(\omega_0, \omega, p_0, p)$.
For $p_0=\omega_0$, zone II vanishes and one is left only with $P_{\rm I}$ and $P_{\rm III}$. This covers all the cases for $\omega_{\rm c}\leq\omega_0$.

\vspace{-3mm}
\subsubsection{$\omega_{\rm c}>\omega_0$}
Assuming $\omega_{\rm c}>\omega_0$ (or $p_0>\omega_0$), again with Eq.~\eqref{eq:kernel_all} the kernel in the three energy zones is given by
\begin{subequations}
\label{eq:P_ww0>1}
\begin{align}
P_{\rm I}(\omega_0 \rightarrow \omega, p_0)
&=\frac{3}{8\gamma_0 p_0\omega_0^2}\, \mathcal{G}\left(\bar{\omega}_0, \bar{\omega}, \kappa_1\right),
\\[-0.5mm]
P_{\rm II}(\omega_0 \rightarrow \omega, p_0)
&=\frac{3}{8\gamma_0 p_0\omega_0^2}\, \mathcal{G}\left(\bar{\omega}, \bar{\omega}_0, \kappa_2\right),
\\[-0.5mm]
P_{\rm III}(\omega_0 \rightarrow \omega, p_0)
&=\frac{3}{8\gamma_0 p_0\omega_0^2}\, \mathcal{G}\left(\omega_0, \omega, p\right),
\end{align}
\end{subequations}
where $\kappa_2=\Lambda(\omega, \omega_0, p, p_0)$.
For $\omega_0<1/2$, zone III vanishes at $p_0>2 \omega_0 (1 - \omega_0)/(1 - 2 \omega_0)$ and only zones I and II, based on $\mathcal{G}$ by switching variables, remain. This covers all cases for $\omega_{\rm c}>\omega_0$.
We remark that zones I and II can be joined by redefining $\Lambda$ as $\tilde{\Lambda}=\frac{1}{2}(\omega_0+\omega-|p_0-p|)$, which then means that $\kappa_1$ and $\kappa_2$ are both obtained from the same expression.

\vspace{-3mm}
\subsubsection{Symmetries of the kernel}
As already proven in Appendix~\ref{app:symmetries}, the kernel obeys the following symmetry relation
\begin{align}
P(\omega_0\rightarrow \omega, p_0)
&= \frac{\gamma p\omega^2}{\gamma_0 p_0\omega_0^2}\,P(\omega \rightarrow \omega_0, p).
\end{align}
Looking at the above expressions for the kernel, one can immediately confirm this relation if it is possible to show that a switching of variables leaves the function $\mathcal{G}$ unaltered. 

As already explicitly stated in Eq.~\eqref{eq:lamplus} and \eqref{eq:lamminus}, both $\lambda_+$ and $\lambda_-$ remain invariant when simultaneously switching $\omega_0\leftrightarrow \omega$ and $p_0\leftrightarrow p$.
Similarly, it is easy to show that $\bar{\omega}$ and $\bar{\omega}_0$ are not affected when simultaneously interchanging $\omega_0\leftrightarrow \omega$ and $p_0\leftrightarrow p$. These two statements already solve the issue with the pre-factors of the functions $\mathcal{S}$ and $\mathcal{F}$. The arguments are also immediately consistent, since the roles of the energy zones are too reversed. This independently proves the statement, which was crucial when obtaining the detailed balance relation, Eq.~\eqref{eq:det_bal}.

\vspace{-3mm}
\subsection{Approximate kernel expressions}
\label{sec:kernel_approximations_moments}

\subsubsection{Recoil-dominated scattering kernel ($p_0\ll 1$ and $p_0\ll\omega_0$)}
\label{sec:rec_K}
When both $p_0\ll1$ and $p_0\ll \omega_0$, the kernel mainly contains zone II and is strongly recoil-dominated. 
While the approximation can be rather easily derived from the definition of the kernel, here we directly start from Eq.~\eqref{eq:recoil_case}.
To obtain the relevant expression, we compute $\mathcal{G}(\omega, \omega_0, p_0)$ in the limit $p_0\ll 1$. Then $\mathcal{F}(x)\approx -2 x/3$ for $x\ll 1$, such that the terms $\propto\mathcal{F}$ can be neglected, while one can replace $\mathcal{S}(x)\approx 1$. For $\omega_{\rm CS}=\omega_0/(1+2\omega_0)\leq \omega\leq \omega_0$, this yields
\begin{align}
\label{eq:P_rec}
\mathcal{G}_{\rm recoil}(\omega_0\rightarrow \omega)
&=
p_0\left\{
2+(\omega -\omega_0)^2\frac{(1+\omega \omega_0)}{\omega^2\omega_0^2}
+ 2
\!\left[\frac{1}{\omega_0} -\frac{1}{\omega}\right]\right\},
\nonumber \\
P_{\rm recoil}(\omega_0\rightarrow \omega)
&=
\frac{3}{8 \omega_0^2}\,\left\{1+\frac{\Delta_1^2}{1+\Delta_1}+\left[1-\frac{\Delta_1}{\omega_0}\right]^2\right\},
\end{align}
with $\Delta_1=(\nu_0-\nu)/\nu$. As expected, this expression is closely related to the well-known differential Compton cross-section for resting electrons \citep{Jauch1976}. For $\omega_0\ll 1$, it furthermore reduces to the case discussed in \citet{Pozdniakov1979}.

Below we list the first three moments of the kernel, calculated from Eq.~\eqref{eq:moment} by integrating in the range $\omega_0/(1+2\omega_0)\leq \omega\leq \omega_0$: 
\begin{subequations}
\label{eq:moment_rec}
\begin{align}
\label{eq:moment_rec_a}
\Sigma_0^{\rm Rec} 
&\!=\! \frac{3 \big(1 -\xi  + 15 \xi^2 + \xi^3\big)}{8 \xi^2 (1- \xi)^2} + \frac{3\big(3 + 6\xi - \xi^2\big) \ln \xi}{4 (1- \xi)^3},
\\[1mm]
\Sigma_1^{\rm Rec} 
\label{eq:moment_rec_b}
&\!=\! \frac{2 - 5 \xi - 3 \xi^2 - 71 \xi^3 +5 \xi^4}{8 \xi^3(1 - \xi )^2} - \frac{3 (7 + 6\xi - \xi^2)  \ln\xi}{4 (1 - \xi )^3}, 
\\[1mm]
\label{eq:moment_rec_c}
\Sigma_2^{\rm Rec} 
&\!=\! \frac{ 3 - 11 \xi +12 \xi^2 +28 \xi^3 +   177 \xi^4 - 17 \xi^5}{16 \xi^4 (1 - \xi)^2}  
\nonumber \\ 
&\qquad\qquad\qquad\qquad +  \frac{3 (11+ 6\xi- \xi^2) \ln \xi }{4 (1 - \xi)^3}.
\end{align}
\end{subequations}
with $\xi=1+2\omega_0=\omega_0/\omega_{\rm CS}$. The zeroth moment simply gives the well-known Klein-Nishina cross section, while the others describe the energy shift and dispersion. $\Sigma_1^{\rm Rec}$ can also be obtained from \citet{Evans1955, Barbosa1982}. For $\omega_0\ll 1$, one finds
\begin{subequations}
\label{eq:moment_rec_low}
\begin{align}
\label{eq:moment_rec_low_a}
\Sigma_0^{\rm Rec} 
&\approx
1 -2 \omega_0 + \frac{26}{5}\omega_0^2 - \frac{133}{10}\omega_0^3 + \frac{1144}{35}\omega_0^4 +\mathcal{O}(\omega_0^5)
\\[0mm]
\label{eq:moment_rec_low_b}
\Sigma_1^{\rm Rec} 
&\approx -\omega_0+\frac{21}{5}\omega_0^2-\frac{147}{10} \omega_0^3+ \frac{1616}{35}\omega_0^4+\mathcal{O}(\omega_0^5)
\\[0mm]
\label{eq:moment_rec_low_c}
\Sigma_2^{\rm Rec} 
&\approx \frac{7}{5}\omega_0^2-\frac{44}{5} \omega_0^3+ \frac{1364}{35}\omega_0^4+\mathcal{O}(\omega_0^5).
\end{align}
\end{subequations}
The general behavior of the moments will be discussed in Sect.~\ref{sec:moment}, where we will also compare the approximations with the exact expressions and study their range of validity. 

\vspace{-3mm}
\subsubsection{Doppler-dominated scattering kernel ($\omega_0\ll p_0$)}
\label{sec:Doppler_K}
Assuming $\omega_0\ll p_0$ and $\omega_0<1/2$, the contributions from zone III vanish at $p_0\gg 2 \omega_0$, such that a simpler, two-zone expression can be found. The corresponding derivation is rather cumbersome since many identities have to be used. However, the approximation can also be obtained by starting from the Compton collision term. This derivation was carried out previously \citep[e.g.,][]{Rephaeli1995, Ensslin2000, Colafrancesco2003}, and yields
\begin{align}
\label{eq:P_Doppler}
P_{\rm D}(\omega_0, \omega)
& =\frac{3}{8 \omega_0}\Bigg\{
\frac{(1+t)}{p_0^5}\left[\frac{3+2p_0^2}{2p_0}\left(\left|\ln t \right|-\ln t_{\rm m}\right)\!+\!\frac{3+3p_0^2+p_0^4}{\gamma_0}\right]
\nonumber\\
&\qquad\quad-\frac{|1-t|}{4p_0^6 t}\left[1+(10+8p_0^2+4p_0^4) t + t^2\right]
\Bigg\}
\end{align}
with $t=\omega/\omega_0$ and $t_{\rm m}=(\gamma_0+p_0)/(\gamma_0-p_0)=(1+\beta_0)/(1-\beta_0)$. The kernel is valid for $1/t_{\rm m}<t<t_{\rm m}$. We will see below that this approximation works very well as long as $\gamma_0\omega_0\ll 1$. For higher electron energies, Klein-Nishina corrections and recoil start to become important but have been neglected in the derivation of Eq.~\eqref{eq:P_Doppler}. However, even then we find that at $\omega\leq \omega_0$ the Doppler kernel works.

One can also compute the moments for this kernel finding:
\begin{equation}
\Sigma_0^{\rm Dop} = 1\,,\,\,\,\,\Sigma_1^{\rm Dop} = \frac{4}{3}p_0^2\,,\,\,\,\,\Sigma_2^{\rm Dop} = \frac{2}{3}p_0^2 + \frac{14}{5}p_0^4,
\label{eq:moment_dop}
\end{equation}
after integrating for $(1-\beta_0)/(1+\beta_0)\leq \omega/\omega_0 \leq (1+\beta_0)/(1-\beta_0)$.
This is in agreement with previous results \citep[e.g.,][]{Blumenthal1970, Ensslin2000}, =but is found to have rather limited applicability (see Sect.~\ref{sec:moment}). 

\vspace{-3mm}
\subsubsection{Scattering kernel for ultra-relativistic electrons ($1\ll p_0$)}
\label{sec:ur-kernel}
In the ultra-relativistic limit $1\ll p_0$ (and $\omega_0$ sufficiently low), one can find \citep[see,][]{1968PhRv..167.1159J, Blumenthal1970}
\begin{align}
\label{eq:P_Urel}
P_{\rm ur}(\omega_0, \omega)
& =\frac{3}{4\gamma_0 p_0 \omega_0}
\Bigg\{
2 q\ln q + \left(1+2q+\frac{\Gamma^2 q^2}{2(1+\Gamma q)}\right)(1-q)
\Bigg\},
\nonumber \\
\Gamma &= 4 \omega_0\gamma_0,\qquad q=\frac{\omega/\Gamma}{\gamma_0-\omega}
\end{align}
for the scattering kernel. Although often presented as general ultra-relativistic case, for larger $\omega_0$ it is no longer valid. In particular once $\omega_0\geq 1/2$, the approximation breaks down, as we illustrate below. We also find that at $\omega/\omega_0\leq 1$ one can always use Eq.~\eqref{eq:P_Doppler} to correctly capture the behavior of the kernel for $1\ll p_0$, while $P_{\rm ur}(\omega_0, \omega)$ incorrectly tends to a constant.

Expressions for the moments in this limit can in principle be derived but they are cumbersome and contain special functions similar to those in the exact expressions given below (i.e., Polylogarithm). Hence, we shall only present asymptotic expansions directly derived from the exact kernel (see Sect.~\ref{sec:moment}).

\vspace{-3mm}
\subsubsection{Thermally-averaged scattering kernel}
\label{sec:therm-kernel-ana}
In this work, we only consider the thermally-averaged scattering kernel numerically, integrating Eq.~\eqref{eq:therm_kernel} over a rMB distribution function, Eq.~\eqref{eq:rmbdist} (see Sect.~\ref{sec:thermal_average}). For analytic approximations valid at mildly-relativistic energies we refer the reader to \citet{Sazonov2000}. Additional approximate expressions for the thermally-averaged kernel can be found in \citet{1970JQSRT..10.1277B, Arutyunyan1980, 1996ApJ...470..249P, 2012A&A...545A.120S} for various limiting cases. 
In our discussion we go well beyond the regimes at which these expressions are accurate. We also explicitly demonstrate the contribution of varying electron momenta to different parts of the thermally-averaged kernel at high energies (Fig.~\ref{fig:kernels_therm_cons}). For additional numerical studies of the thermally-averaged kernel please refer to \citet{Pomraning1972, Guilbert1981, Madej2017}.

\begin{figure}
\includegraphics[width=\linewidth ]{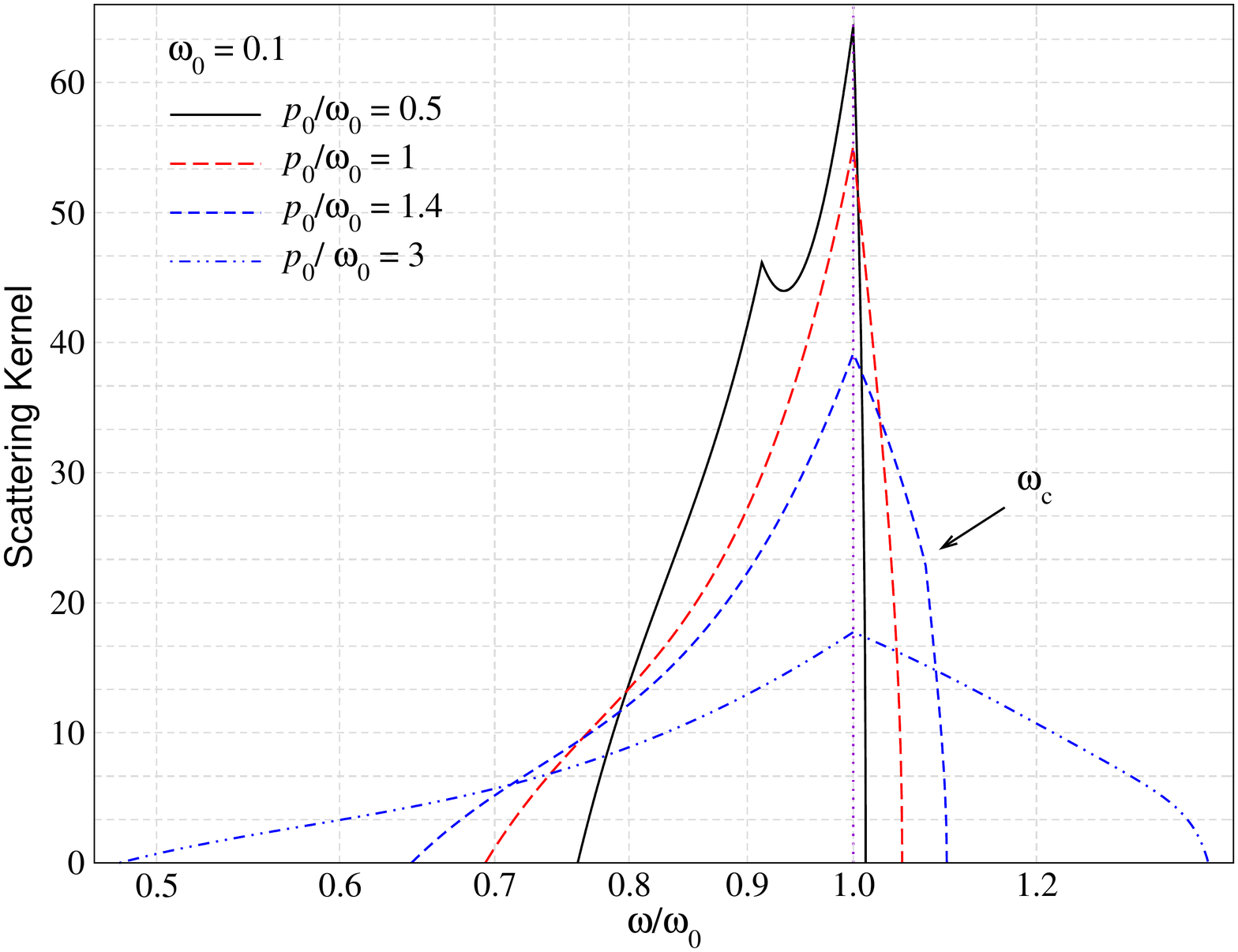}
\\[-2mm]
\includegraphics[width=\linewidth ]{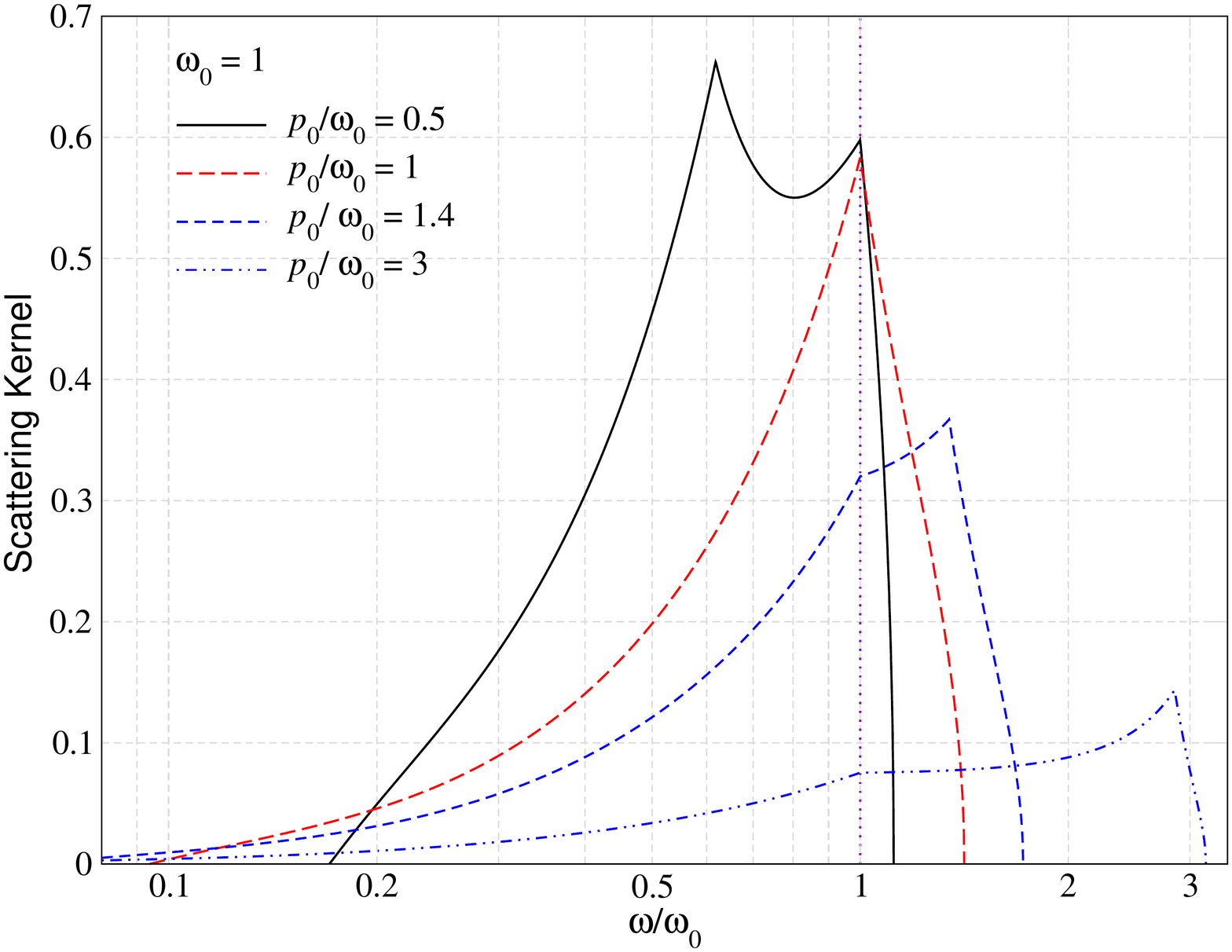}
\caption{Illustrations of the Compton scattering kernel, $P(\omega_0 \rightarrow \omega, p_0)$, for different values of $\omega_0$ and $p_0/\omega_0$. The upper and lower panels contain the kernels for $\omega_0 = 0.1$ and $\omega_0 = 1$, respectively. The values of $p_0/\omega_0$ are provided at the top-left corner of each plot. The violet dotted line in each plot marks the cusp at $\omega_0$, the other cusp being at $\omega_{\rm c}$. The most interesting phenomenon is the movement of the cusp at $\omega_{\rm c}$ from the low-energy side of $\omega_0$ to the high-energy side, while coinciding with $\omega_0$ when $\omega_0 = p_0$. Also, the third zone disappears for $p_0/\omega_0 = 3$ when $\omega_0 = 0.1$ but never for $\omega_0=1$, as explained in the text in more detail. In the upper panel, we have marked the cusp at $\omega_{\rm c}$ with an arrow for $p_0/\omega_0=1.4$ for convenience.}
\label{fig:kernels}
\end{figure}

\section{Illustrations of the kernel}
\label{sec:illus_iso}
In this section, we illustrate the shape of Compton kernel for different values of $\omega_0$ and $p_0$. We use combinations for which $\omega_0$ and $p_0$ are of the same order of magnitude (non-extreme scenarios) and where one is much higher than the other (extreme scenarios). For the latter, we also compare the approximate Eqs.~\eqref{eq:P_rec}, \eqref{eq:P_Doppler} and \eqref{eq:P_Urel}, with our expression to assess their range of validity. 
In Sect.~\ref{sec:thermal_average} we briefly consider the thermally-averaged kernel for varying temperature. All figures can be reproduced with {\tt CSpack}.

\subsection{Non-extreme scenarios}
\label{kernel-nex}
We start with non-extreme scattering scenarios, when $\omega_0$ and $p_0$ are comparable to each other. In each panel of Fig.~\ref{fig:kernels}, the kernels are plotted for $p_0/\omega_0 = \{0.5,1,1.4,3\}$. For better comparison, the values of $\omega_0$ are chosen to be the same as that of Fig.~\ref{fig:domains}.

First, we shall point out some common features of the kernel. According to Fig.~\ref{fig:domains}, there should at most be three distinct zones in the kernel, separated by cusps located at $\omega_0$ and $\omega_{\rm c}$. When $\omega_0 > p_0$, one has $\omega_{\rm c} < \omega_0$, and thus, the cusp is in the down-scattering domain. This cusp will be referred to as the recoil peak hereafter. The two cusps coincide when $\omega_0 = p_0$, eliminating the second zone from the kernel. When $\omega_0 < p_0$, generally three distinct zones are present since $\omega_{\rm c} > \omega_0$. Therefore, now the cusp at $\omega_{\rm c}$ lies in the up-scattering domain of the kernel, which we refer to as the Doppler peak. For all of the combinations, the third zone is usually the smallest and hence has not received much attention in the literature. All the aforementioned features are visible in Fig.~\ref{fig:domains}. 

Additional trends can immediately be appreciated from Fig.~\ref{fig:kernels}. Increasing the particle energies leads to an overall broadening of the kernel (second moment increases). Increasing the energy of the photon usually leads to stronger down-scattering and hence more loss of energy (net transfer of energy to the electron), while increasing the momentum of the electron exhibits the opposite trend. These features will also be discussed for the kernel moments (Sect.~\ref{sec:moment}) and are a natural consequence of Compton scattering \citep[e.g.,][]{Blumenthal1970, Pomraning1972, Nagirner1994, Sazonov2000, Sazonov2001}.

Considering the cases with $p_0/\omega_0=0.5$ one can see that the probability of a photon being up-scattered by an electron is rather low, leading to a kernel with a slim zone III (black solid lines in Fig.~\ref{fig:kernels}). On the other hand, the motion of the electron does broaden the low-energy tail of the kernel significantly beyond the usual recoil peak for resting electrons, which for $\omega_0=0.1$ and $\omega_0=1$ would be located at $\omega/\omega_0\simeq 0.83$ and $\omega/\omega_0\simeq 0.33$, respectively. In contrast, energetic electrons ($p_0/\omega_0>1$) can strongly up-scatter photons, broadening the high-energy tail of the kernel beyond the Doppler peak. This peak becomes more pronounced when the photon is also energetic (lower panel Fig.~\ref{fig:kernels}).

\begin{figure}
\includegraphics[width=\linewidth ]{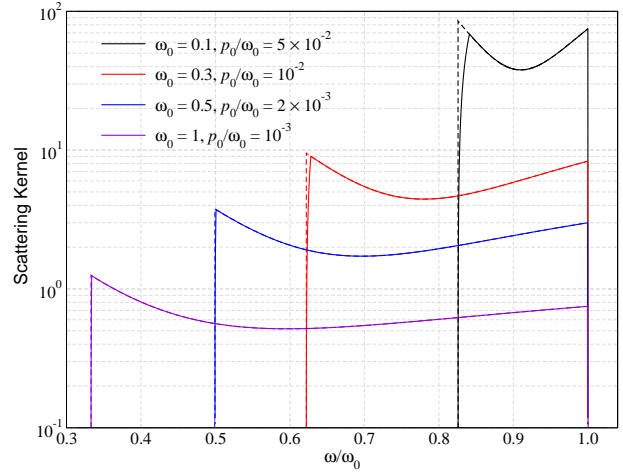}
\caption{Illustration of the Compton scattering kernel (solid lines), $P(\omega_0 \rightarrow \omega, p_0)$ in the recoil-dominated regime for different combinations of $\omega_0$ and $p_0$. The dashed lines show the simple analytic approximation, Eq.~\eqref{eq:P_rec}, which clearly works well for $p_0\ll 1$ once $\omega_0\gg p_0$.}
\label{fig:kernels_recoil_dom}
\end{figure}

While having the aforementioned similarities, comparison of the upper panel with the lower in Fig.~\ref{fig:kernels} reveals that the kernels for the lower energy photons show visible differences, especially when the electron momentum is increased. As discussed in Sect.~\ref{sec:ker_domains}, the Doppler peak merges with the high-energy boundary ($\omega_{\rm c}\equiv \omega_{\rm max}$) of the kernel if $\omega_0 \leq (1+p_0-\gamma_0)/2$. Thus, we expect the third zone to disappear at a certain $p_0$, which for $\omega_0=0.1$ occurs at $p_0 \simeq 2.25\omega_0$. We can observe this effect in the upper panel by comparing the cases for $p_0/\omega_0 = 1.4$ and $p_0/\omega_0 = 3$: for $p_0/\omega_0 = 1.4$, the cusp at $\omega=\omega_{\rm c} < \omega_{\rm {max}}$ is present (indicated by an arrow), implying that some photon are boosted beyond the Doppler peak. Increasing the electron momentum pushes the peak to the maximal energy, leading to only two zones in the kernel.

\begin{figure}
\hspace{-0mm}\includegraphics[width=1.01 \linewidth ]{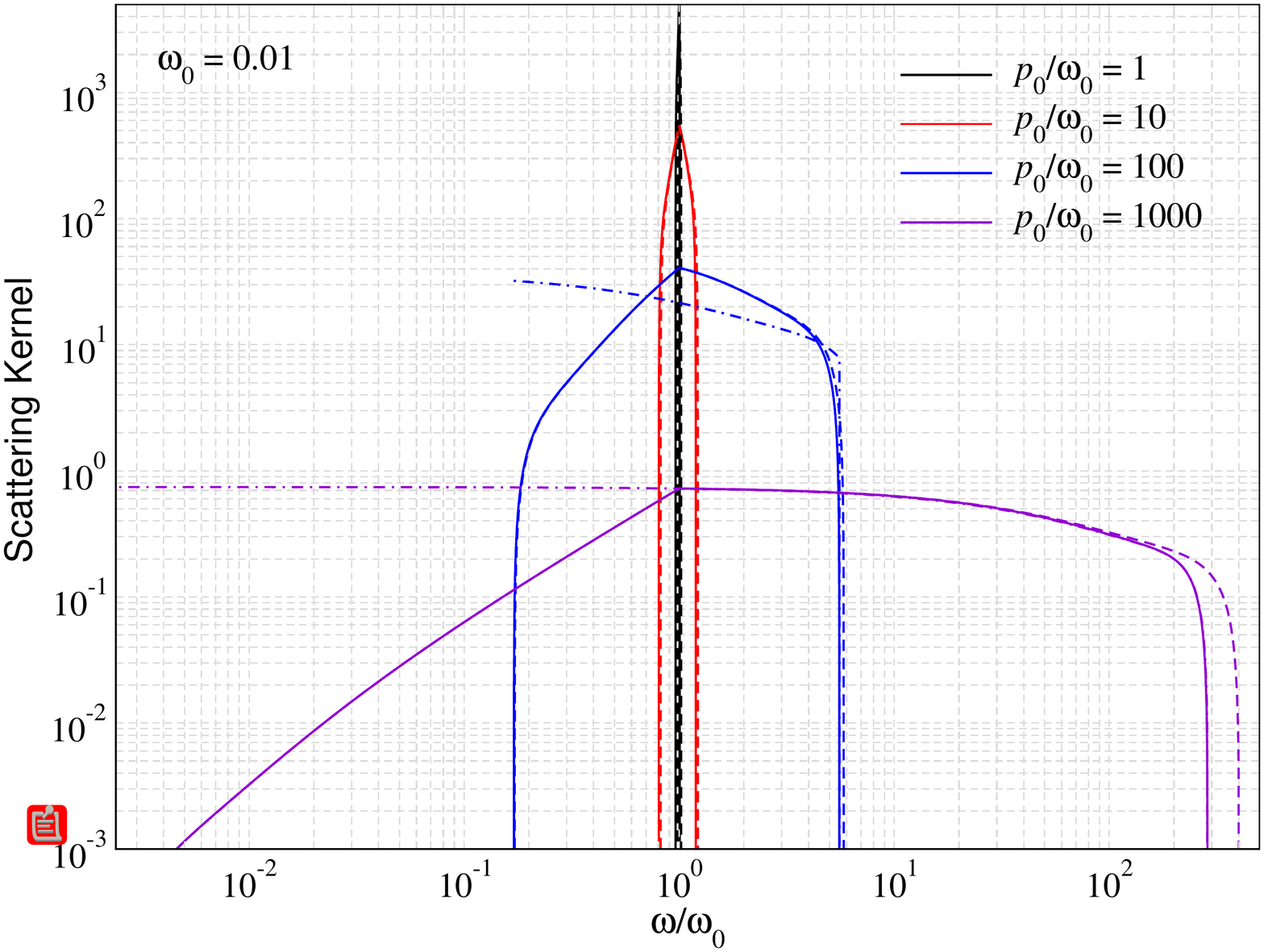}
\\
\hspace{-1mm}\includegraphics[width=1.01 \linewidth ]{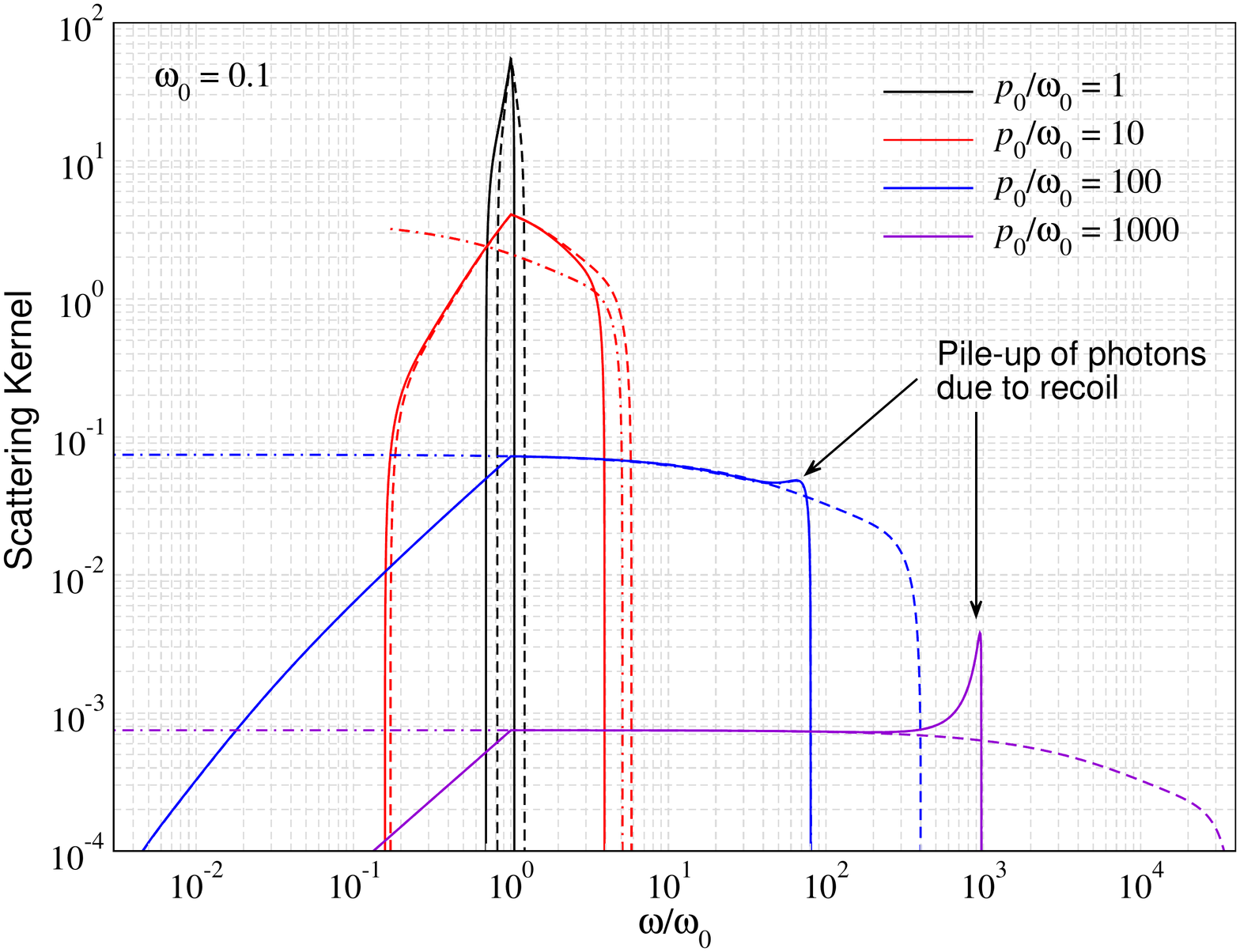}
\caption{Illustration of the Compton scattering kernel (solid lines), $P(\omega_0 \rightarrow \omega, p_0)$ in the Doppler-dominated regime for $\omega_0=\{0.01, 0.1\}$ (note the difference in scales) and increasing values of $p_0$. The dashed lines show the simple analytic approximation, Eq.~\eqref{eq:P_Doppler}, which does not work well at $\omega>\omega_0$ since recoil effects become kinematically important. In contrast, the ultra-relativistic approximation (dashed-dotted lines, Eq.~\eqref{eq:P_Urel}; not shown for $p_0/\omega_0=1$) does capture the high-frequency behavior of the kernel, while being inaccurate at frequencies $\omega<\omega_0$.}
\label{fig:kernels_UR}
\end{figure}

On the other hand, if $\omega_0 > 1/2$, the third zone never disappears even if it may become extremely slim. As seen from the lower panel of Fig.~\ref{fig:kernels}, the kernel for $p_0/\omega_0 = 3$ does show a cusp at $\omega=\omega_{\rm c}>\omega_0$ and still contains the third zone. However, the broadening beyond the Doppler peak is relatively weak. Also, as a photon with higher energy can experience stronger energy exchange with an electron, kernels are wider for $\omega_0 = 1$, as also is expected from Fig.~\ref{fig:domains}. One additional feature to note is that for $\omega_0=0.1$, the height of the cusp at $\omega_{\rm c}$ is lower than the cusp at $\omega_0$, while for $\omega_0=1$, it is the opposite. This implies that photons with higher energy are more likely to end up close to the Doppler- or recoil-peaks.

\begin{figure}
\includegraphics[width=1.0 \linewidth ]{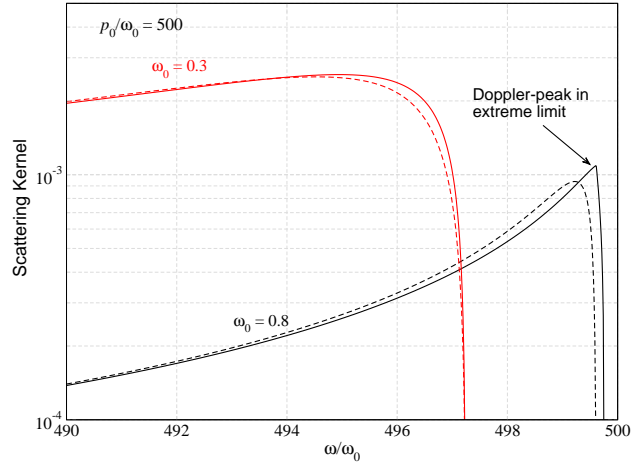}
\caption{Comparison of the exact kernel (solid line), $P(\omega_0 \rightarrow \omega, p_0)$, with the approximation Eq.~\eqref{eq:P_Urel} derived for ultra-relativistic electron (dashed). For $\omega_0=0.8$, a cusp close to the Doppler-peak remains visible, which is not described by the simple approximation.}
\vspace{-3mm}
\label{fig:kernels_comp_UR}
\end{figure}

\vspace{-4mm}
\subsection{Extreme scenarios}
\label{kernel-ex}
The kernels for extreme cases, i.e, the recoil-dominated ($\omega_0 \gg p_0$) and the Doppler-dominated ($\omega_0 \ll p_0$) regime, are presented in Fig.~\ref{fig:kernels_recoil_dom} and \ref{fig:kernels_UR}, respectively. According to Fig.~\ref{fig:domains}, in the recoil-dominated regime, $\omega_{\rm c}$ asymptotically approaches $\omega_{\rm{min}}$, while $\omega_{\rm{max}}$ tends to $\omega_0$. This leads to a situation when all the photons are down-scattered such that the kernel is dominated by zone II with the other two zones practically disappearing if also $p_0\ll 1$ (see Fig.~\ref{fig:kernels_recoil_dom}). Increasing the photon energy decreases the probability of the photon being able to retain its initial energy. As also seen in Fig.~\ref{fig:kernels_recoil_dom}, when the photon energy increases, the recoil peak moves to lower energy, following $\omega\simeq \omega_{\rm CS}=\omega_0/(1+2\omega_0)$. The overall amplitude of the kernel in this regime scales as $P(\omega_0\rightarrow \omega, p_0)\propto 1/\omega_0^2$ (see Eq.~\ref{eq:P_rec}), explaining the large reduction with increasing $\omega_0$. Equation~\eqref{eq:P_rec} provides an excellent approximation for the kernel in the recoil-dominated regime.

Figure~\ref{fig:kernels_UR} illustrates the kernel and various approximation at the transition to the opposite extreme, i.e, $\omega_0 \ll p_0$. For $p_0=\omega_0$, significant recoil contributions are present and none of the simple approximations work well in this case. 
Increasing $p_0$, Doppler-broadening becomes stronger and on average photons are up-scattered blue-ward of the initial photon energy. At $\omega<\omega_0$, the approximation, Eq.~\eqref{eq:P_Doppler}, converges to the exact expression, while at high frequencies it fails to capture the shape of the kernel in particular for very large values of $p_0$. The main reason is that in the Doppler-dominated approximation, $\omega^{\rm D}_{\rm max}/\omega_0\approx (\gamma_0+p_0)/(\gamma_0-p_0)$ increases as $\simeq 4 p_0^2$ for large momenta. Even if $\omega_0\ll p_0$, this strongly overestimates the correct scaling of $\omega_{\rm max}$, which only is $\omega_{\rm max}/\omega_0\approx p_0$ for $\omega_0\leq (1+p_0-\gamma_0)/2$. In Fig.~\ref{fig:kernels_UR}, this effect, which is related to recoil, can be seen as a pile-up of photons close to $\omega\simeq \omega_{\rm max}=\omega_{\rm c}$. In this regime, the approximation, Eq.~\eqref{eq:P_Urel} performs very well, while always failing at $\omega<\omega_0$.

The behavior described above highlights an important difference between the recoil- and Doppler-dominated regimes. When increasing $\omega_0$ for fixed $p_0$ ($\ll 1$) one can always find a value for $\omega_0$ beyond which the recoil-dominated approximation, Eq.~\eqref{eq:P_rec}, works. The opposite is not true for the Doppler-dominated regime, precisely because recoil becomes important again once some large value for $p_0$ is exceeded. This value can be estimated by asking when $\gamma_0-p_0\lesssim2\omega_0$, yielding $p_0\gtrsim (4\omega_0)^{-1}$. For lower $\omega_0$, the Doppler-dominated approximation thus holds longer, as can also be deduced by comparing the upper and lower panels of Fig.~\ref{fig:kernels_UR}.

For similar reasons one expects the ultra-relativistic approximation in Eq.~\eqref{eq:P_Urel} to fail when $\omega_0> 1/2$. In this case, zone III never disappears, which implies that the kernel always has a cusp at $\omega=\omega_{\rm c}\lesssim \omega_{\rm max}\simeq \gamma_0+\omega_0-1$. This is illustrated in Fig.~\ref{fig:kernels_comp_UR}, which shows that the ultra-relativistic approximation becomes inaccurate. Even when increasing $p_0$, the approximation departs from the exact solution. 
However, the differences are relatively small and the dominant features are still captured (i.e., total energy exchange). 

\begin{figure}
\includegraphics[width=1.01 \linewidth ]{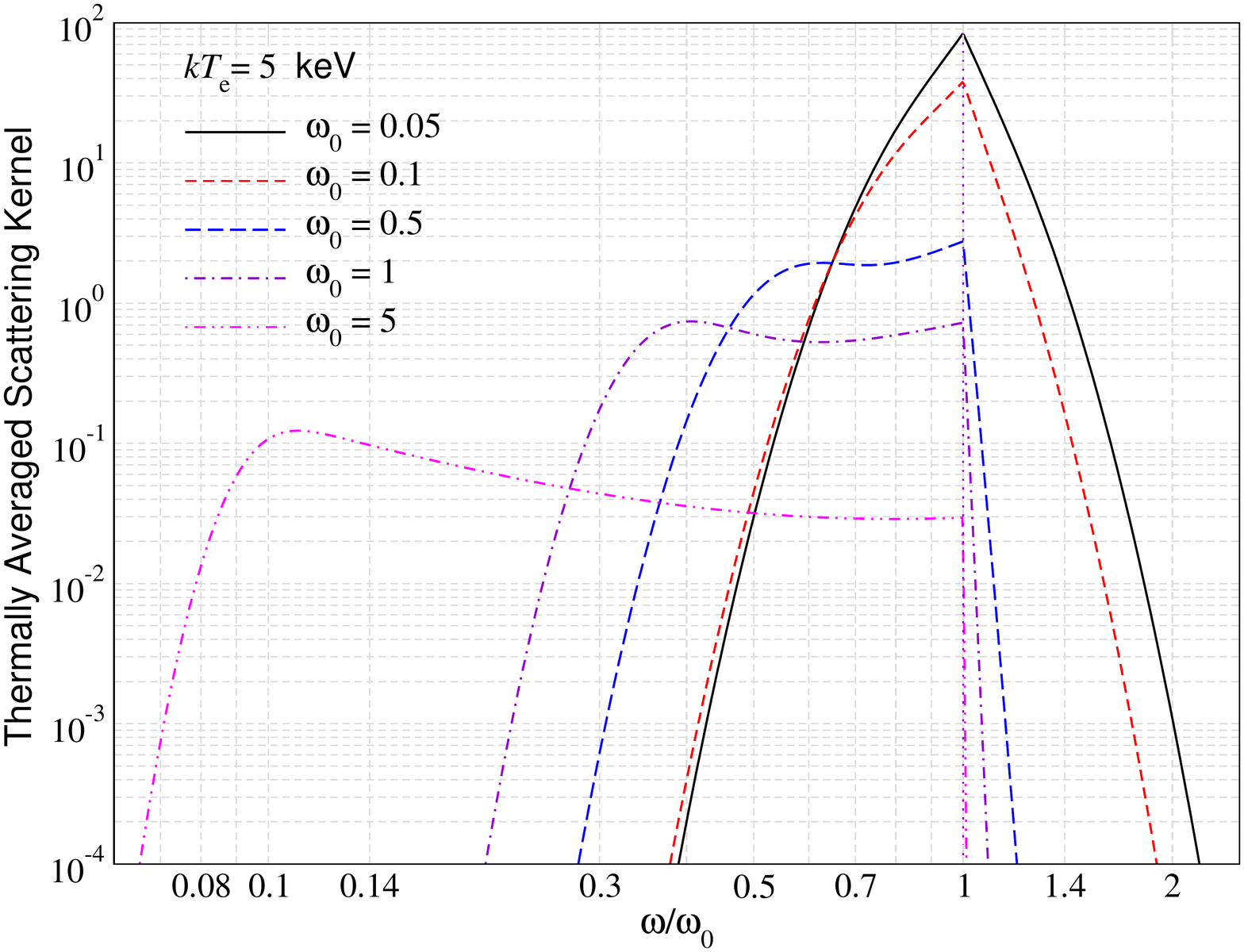}
\\[-2.0mm]
\includegraphics[width=1.01 \linewidth ]{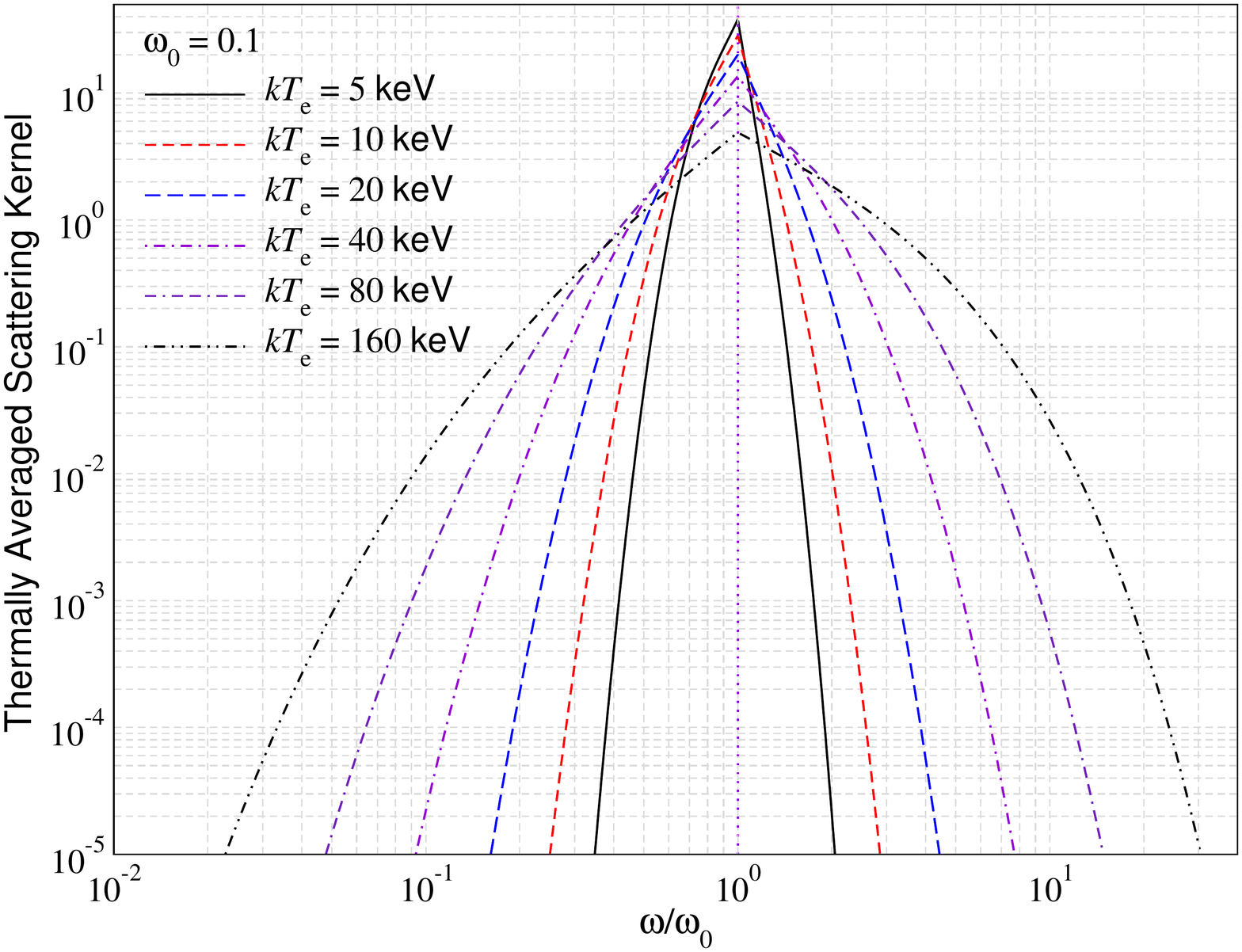}
\caption{Thermally-averaged scattering kernel for a set of electron temperatures, $T_{\rm e}$, and $\omega_0$. 
In the upper panel, we show kernels for $kT_{\rm e} = 5$~keV and $\omega_0=\{0.05,0.1,0.5,1,5\}$. Recoil losses dominate, and on average photons are down-scattered.
In the lower panel we compare kernels for $kT_{\rm e} = \{5,10,20,40,80,160\}$\,keV and $\omega_0=0.1$. The width of the kernel increases with $T_{\rm e}$, owing to more dispersion in the electron momenta. The violet  dotted vertical lines in both panels mark $\omega = \omega_0$.}
\vspace{-2mm}
\label{fig:kernels_therm}
\end{figure}

\begin{figure}
\includegraphics[width=0.98 \linewidth ]{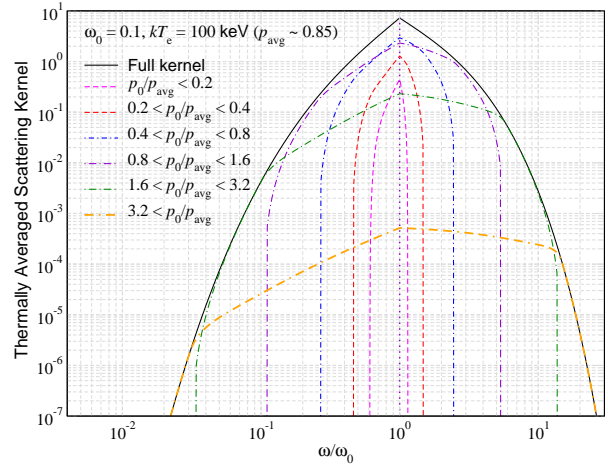}
\caption{Illustration of the partial contributions of different electron momenta to the thermally-averaged kernel. While electrons with momenta lower than $p_{\rm avg}$ mostly constructs the areas around the central peak, faster electrons efficiently scatter the photons into the distant wings of the kernel.}
\label{fig:kernels_therm_cons}
\end{figure}

\vspace{-4mm}
\subsection{Thermally-averaged scattering kernel}
\label{sec:thermal_average}
To compute the thermally-averaged kernel, $P(\omega_0 \rightarrow \omega)$, as given by Eq.~\eqref{eq:therm_kernel}, we carry out the required integral numerically. 
A few words regarding the integration limits are in order. For a certain scattering event, the kernel spreads from $\omega_{\rm {min}}$ to $\omega_{\rm {max}}$. For a given $\omega_0$, the electron thus needs to have a minimal momentum, $p_0^{\rm min}$ to scatter the photon to a certain energy $\omega$. The threshold momentum depends on whether the photon is up-scattered or down-scattered and can be found by solving $\omega = \omega_{\rm {min}}$ and $\omega = \omega_{\rm {max}}$ for $p_0$. For down-scattered photons ($\omega\leq\omega_0$) it is given by
\bsub
\label{eq:p0min_limits}
\begin{align}
\label{eq:p0min_ommin}
p_0^{\rm {min}} 
=
\begin{cases}
\frac{\omega_0-\omega}{2}\sqrt{\frac{1 + \omega \omega_0}{\omega\omega_0}} -\frac{\omega_0+\omega}{2}
&\text{for}\; \omega \leq \frac{\omega_0}{1+2\omega_0}
\\[1mm] 
0
&\text{for}\; \omega > \frac{\omega_0}{1+2\omega_0}
\end{cases}
\end{align}
Similarly, for up-scattered photons ($\omega>\omega_0$) one obtains
\begin{align}
\label{eq:p0min_ommax}
p_0^{\rm {min}}
=
\begin{cases}
\sqrt{(\omega-\omega_0+1)^2-1}
&\text{for}\; \omega \leq \frac{\omega_0}{1-2\omega_0}\wedge \omega_0<1/2
\\[1mm] 
\frac{\omega-\omega_0}{2}\sqrt{\frac{1 + \omega \omega_0}{\omega\omega_0}} +\frac{\omega_0+\omega}{2}
&\text{for}\; \omega > \frac{\omega_0}{1-2\omega_0}\wedge \omega_0<1/2
\\[1mm] 
\sqrt{(\omega-\omega_0+1)^2-1}
&\text{for}\; \omega_0>1/2
\end{cases}
\end{align}
\esub
If an electron has lower momentum than this, it is not energetic enough to scatter the photon to the desired energy. However, there is no upper limits on the momentum and thus $p_0^{\rm {max}} = \infty$.\footnote{For numerical purpose we utilize the fact that the rMB distribution function falls off exponentially with $\gamma_0$. We thus determine $p_0^{\rm {max}}$ using the condition $\expf{-(\gamma-1)/\The} = \epsilon\ll 1$ or $p_0^{\rm {max}}=\sqrt{(\The\ln\epsilon)^2-2\The\ln\epsilon}$, which ensures that higher momenta do not contribute significantly.}

We illustrate the thermally-averaged kernel for different values of $\omega_0$ and electron temperature, $T_{\rm e}$, in Figure~\ref{fig:kernels_therm}. In the upper panel, we show kernels for $kT_{\rm e} = 5$ keV and $\omega_0=\{0.05,0.1,0.5,1,5\}$. This figure illustrates the transition of the kernel from the largely recoil-dominated, with only little smearing due to the motions of the electrons, to the Doppler-dominated regime. For all shown cases, photons are on average down-scattered, implying that the electrons are heated in the interaction. Increasing the temperature or further lowering the energy of the incident photon to $\omega_0\lesssim 4\The$ would cause net up-scattering, as we explain in Sect.~\ref{sec:th_moments}. Additional illustrations of the kernel in this regime can be found in \citet{Pomraning1972} and \citet{Madej2017}.

In the lower panel of Fig.~\ref{fig:kernels_therm} we compare kernels for temperatures $kT_{\rm e} = \{5,10,20,40,80, 160\}$ keV and $\omega_0=0.1$. Increasing $T_{\rm e}$ makes the kernel wider because of significant contributions from electrons with large momenta (see below). For $kT_{\rm e}\leq 10\,{\rm keV}$ and $\omega_0=0.1$, photons are on average more strongly down-scattered (recoil-dominates on average), while for higher temperatures more up-scattering takes place. In the latter case, losses by recoil are compensated by those due to Doppler-boosting, as we will also find reflected in the moments of the kernel (see Sect.~\ref{sec:moment}).
Additional illustrations for the kernel at lower temperature and comparison with approximations can be found in \citet{Sazonov2000}.

In Fig.~\ref{fig:kernels_therm_cons}, we explicitly illustrate the partial contribution of electrons with varying momenta to the final kernel. We considered a medium with electrons at $kT_{\rm e} = 100$~keV, which implies an average momentum $p_{\rm avg}=\langle p_0\rangle \simeq 0.85$ (see Appendix~\ref{app:MMrMB} for general expressions). We computed the partial contributions to the total kernel over different bins in $p_0$, showing these contributions individually.  Electrons with low momenta ($p_0 < p_{\rm avg}$) do not scatter photons strongly, leading to little broadening of the initial distribution. Therefore, electrons with lower momenta mostly contributes to the central peak of the kernel with no contribution to the distant wings, as seen in Fig.~\ref{fig:kernels_therm_cons}. On the other hand, energetic electrons ($p_0 > p_{\rm avg}$) can efficiently scatter photons to much higher and lower energies, thereby leading to strong broadening of the initial distribution. Thus, their contribution to the central peak is very small, while they are the ones constructing the high and low energy wings of the kernel. Overall, the kernel is extremely featureless, even if the individual contributions do show visible kinks, especially for large momenta (see Fig.~\ref{fig:kernels_therm_cons}). This is due to the strong smearing and large dispersion of electron momenta in the rMB distribution.

\vspace{-1mm}
\subsubsection{Extreme temperatures}
\label{sec:extreme_Te}
The expression for the thermally-averaged scattering kernel (see Eq.~\ref{eq:therm_kernel}) is in principle applicable at all photon energies and electron temperatures. However, when $k T_{\rm e} \gtrsim 511$~keV ($\equiv m_{\rm e}c^2$), the rate of pair-production can no longer be neglected, and one expects a pair-plasma to form \citep[e.g.,][]{Svensson1984, Zdziarski1989}. In this case, both electrons and positrons will scatter photons and degeneracy effects (Fermi-blocking) will furthermore quickly become important. Assuming quick thermalization, the latter effect can be added to the computation by modifying Eq.~\eqref{eq:therm_kernel} to 
\begin{align}
P(\omega_0 \rightarrow \omega) 
&= \int_{p_0^{\rm {min}}}^{\infty} \frac{p_0^2 \,f(\gamma_0)}{\Ne}\,\left[1+f(\gamma)\right]\, P(\omega_0 \rightarrow \omega, p_0)\, \text{d}p_0 
\end{align}
where $\gamma = \gamma_0+\omega_0-\omega$ and $f(\gamma)$ is now generalized to a Fermi-distribution, $f(\gamma)=[\expf{(\gamma-\mu)/\The}+1]^{-1}$. The chemical potential, $\mu$, determines the normalization of the distribution and thereby fixes the electron degeneracy. However, here we shall restrict our discussion to lower temperatures, where these effects are not relevant.

\vspace{-1mm}
\section{Moments of the Kernel}
\label{sec:moment}
The kernel moments of order $m$ are defined in Eq.~\eqref{eq:moment}. A general expression for $\Sigma_m$ is difficult to derive. Below we give analytic expressions for $\Sigma_0$, $\Sigma_1$ and $\Sigma_2$, which we derived independently as functions of $\omega_0$ and $p_0$.
We were able to confirm these results numerically and also using the expressions given in Sect.~3 of NP94. For this we had to convert from moments of $\omega^m$ to moments of $(\omega-\omega_0)^m/\omega^m_0$, yielding terms that are closer to the physical quantities of interest here. The expressions given by NP94 could then be greatly simplified and after fixing some typos\footnote{The expressions for $\Psi_{14}$ in Eq.~(3.3.5) and $\psi_{00}$ in Eq.~(3.3.10) of NP94 should respectively be replaced by $\Psi_{14} = 2(2\psi_{-11}-\psi_{-10}-\psi_{-12})/\xi$ and $\psi_{00} = \frac{3}{8\xi}\left[g(\xi)-\frac{2}{\xi}+\left(\frac{1}{2}+\frac{2}{\xi}+\frac{1}{\xi^2}\right)l_{\xi}-\frac{1}{2}R_{\xi} - \frac{3}{2}\right]$, as also noted in B09.}, yielded our results.

We comment, that the moment expressions can in principle be found from Eq.~\eqref{eq:moment} by summing the contributions from the different energy zones. However, it turned out to be easier to directly compute them using the standard scattering cross-section approach. For each moment, we also illustrate the validity of various simpler approximations mentioned in Sect.~\ref{sec:kernel_approximations_moments}. All results were confirmed by numerical integration. We also briefly discuss the third and fourth moments, but restrict ourselves to numerical computations.

\vspace{-3mm}
\subsection{The Zeroth Moment}
The zeroth moment is nothing but the total Compton scattering cross-section, which reads
\begin{align}
\label{eq:mom0_exact}
\Sigma_0(\omega_0, p_0) &=  
\frac{3}{8 \gamma_0 \omega_0}\left\{\frac{4\gamma_0 +9 \omega_0+ 2\gamma_0 \omega_0^2}{4 p_0 \omega_0^2}\,\ln\left(\frac{\alpha_+}{\alpha_-}\right)
\right.
\\[1mm]
\nonumber
&
\left.
\qquad\qquad\;
-\frac{1}{2}\left\{
1+\frac{1}{\lambda}-\left[1-\frac{2}{\omega_0^2}\right]\ln \lambda\right\}
+\frac{\mathcal{F}(\omega_0,p_0)}{p_0 \omega_0}
\right\}.
\end{align}
Here\footnote{Below, we do not explicitly mention taking the real part of the Polylogarithm in the approximate expressions of the moments, but every time only the real part is kept.} $\mathcal{F}(\omega_0,p_0) = \operatorname{Re}{\left[{\rm Li}_2(1-\alpha_+)-{\rm Li}_2(1-\alpha_-)\right]}$ with ${\rm Li}_n(z)$ being the Polylogarithm of $z$ of order $n$, $\alpha_{\pm}=1+2(\gamma_0\pm p_0) \omega_0$ and $\lambda=\alpha_{+}\alpha_{-}=1+4 (\gamma_0+\omega_0) \omega_0$. 

It is straightforward to show that for resting electrons, Eq.~\eqref{eq:mom0_exact} reduces to the Klein-Nishina cross section, explicitly given in Eq.~\eqref{eq:moment_rec_a}. In the non-relativistic limit ($p_0, \omega_0 \ll 1$), one can furthermore find
\begin{align}
&\Sigma_0^{\rm nr}(\omega_0, p_0)
= 1 - 2\omega_0 + \frac{26}{5} \omega_0^2 - \frac{133}{10} \omega_0^3 + \frac{1144}{35}\omega_0^4-\frac{544}{7} \omega_0^5
\nonumber\\
&\qquad
-\left(\frac{5}{3}- \frac{52}{5}\omega_0+\frac{931}{20}\omega_0^2\right) \omega_0 p_0^2
+\frac{7}{12} \omega_0 p_0^4+\mathcal{O}(\omega^2_0 p_0^4).
\label{eq:mom0_nr}
\end{align}
Comparing this with Eq.~\eqref{eq:moment_rec_low_a}, we can confirm the leading order Klein-Nishina terms. There is no leading order correction in terms of $p_0$ alone, as follows from Eq.~\eqref{eq:moment_dop}. The dependence on $p_0$ only enters at $\mathcal{O}(\omega_0 p_0^2)$. The lowest order terms can also be identified with those in Eq.~(16) of \citet{Sazonov2000} after replacing $\left<p_0^2\right>= 3\The+\mathcal{O}(\The^2)$ and $\left<p_0^4\right>= \mathcal{O}(\The^2)$ for the thermal averages.

\begin{figure}
\includegraphics[width=1.0 \linewidth ]{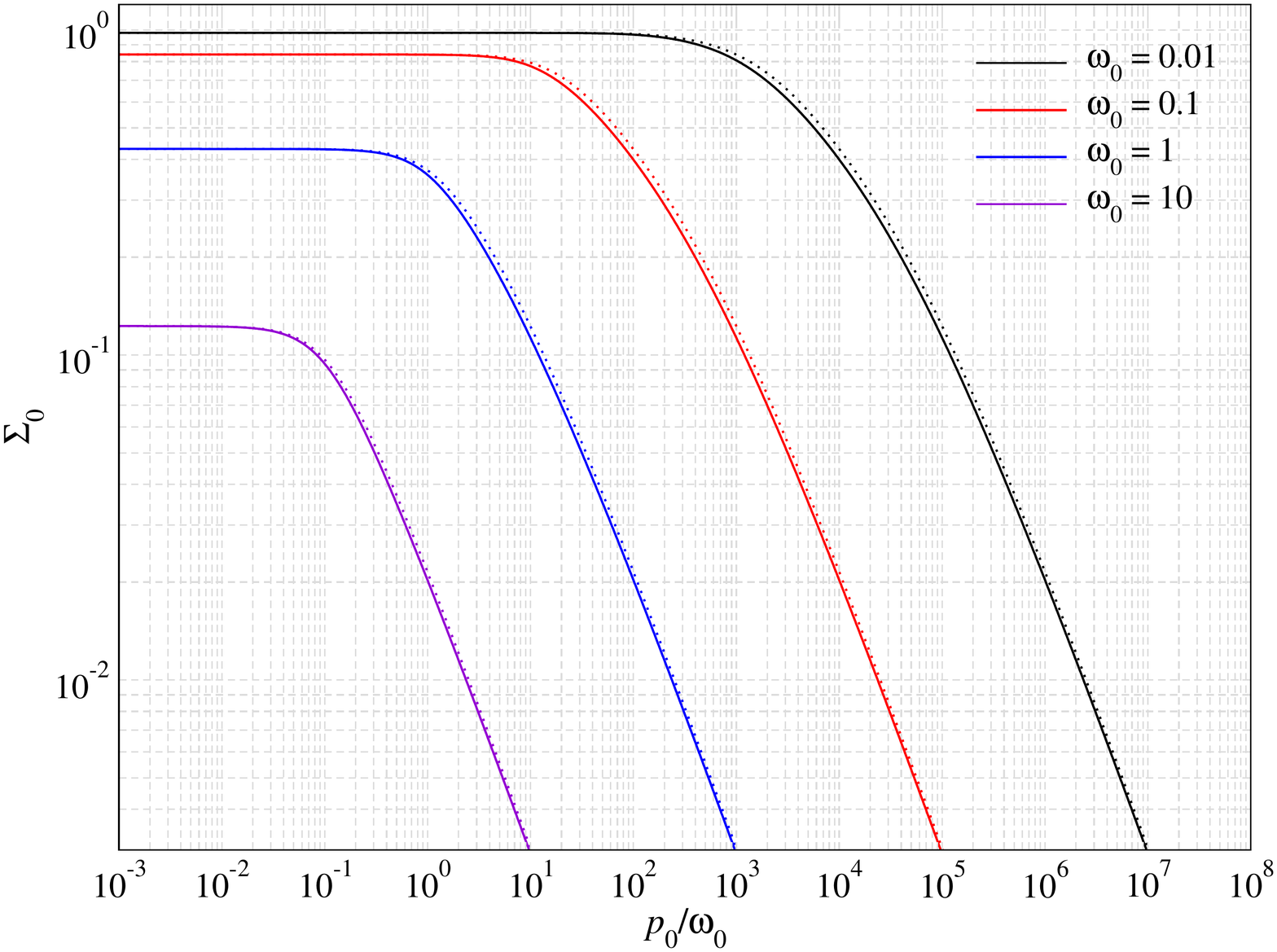}
\\[0.5mm]
\includegraphics[width=1.0 \linewidth ]{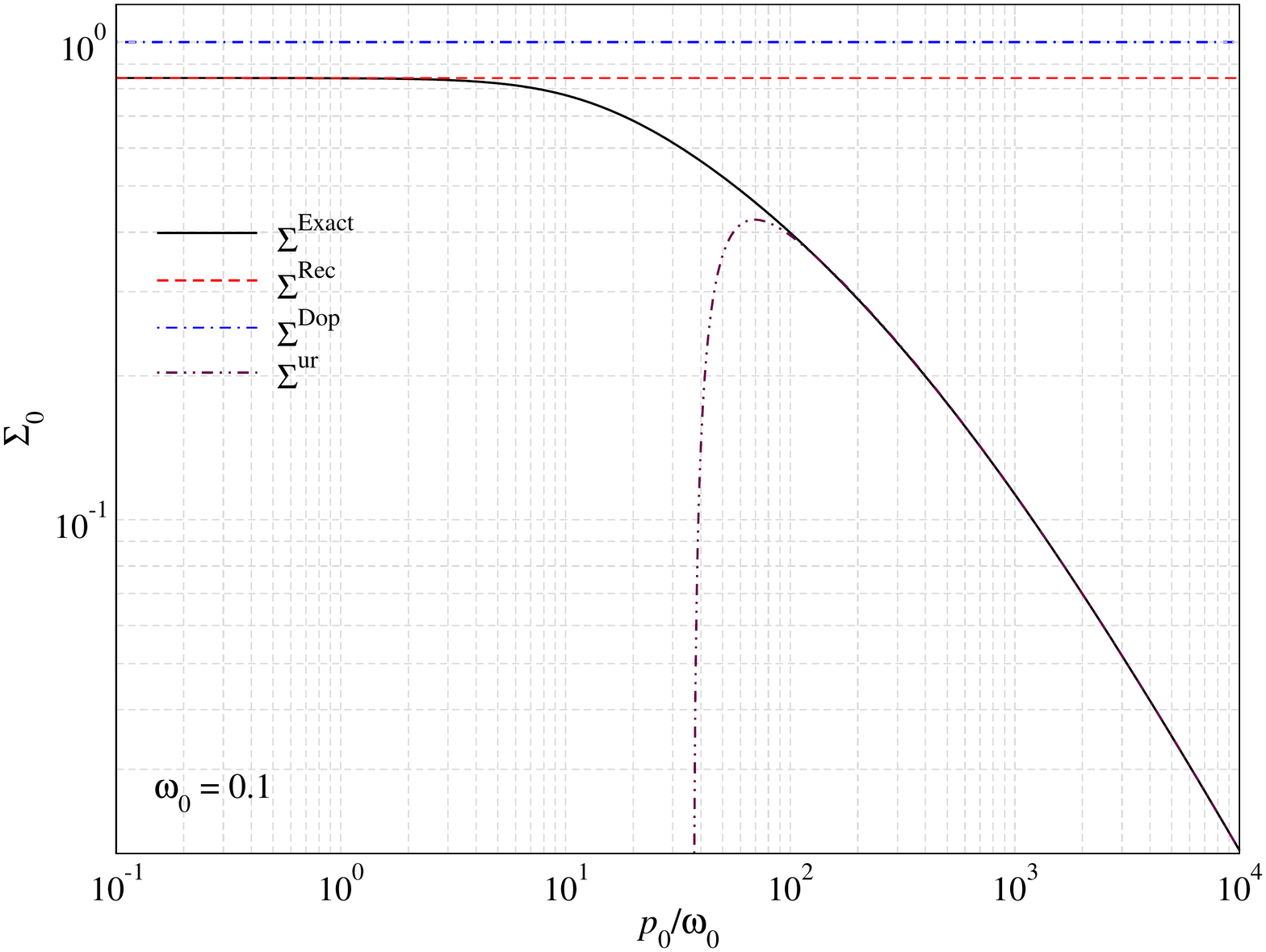}
\\[0.5mm]
\includegraphics[width=1.0 \linewidth ]{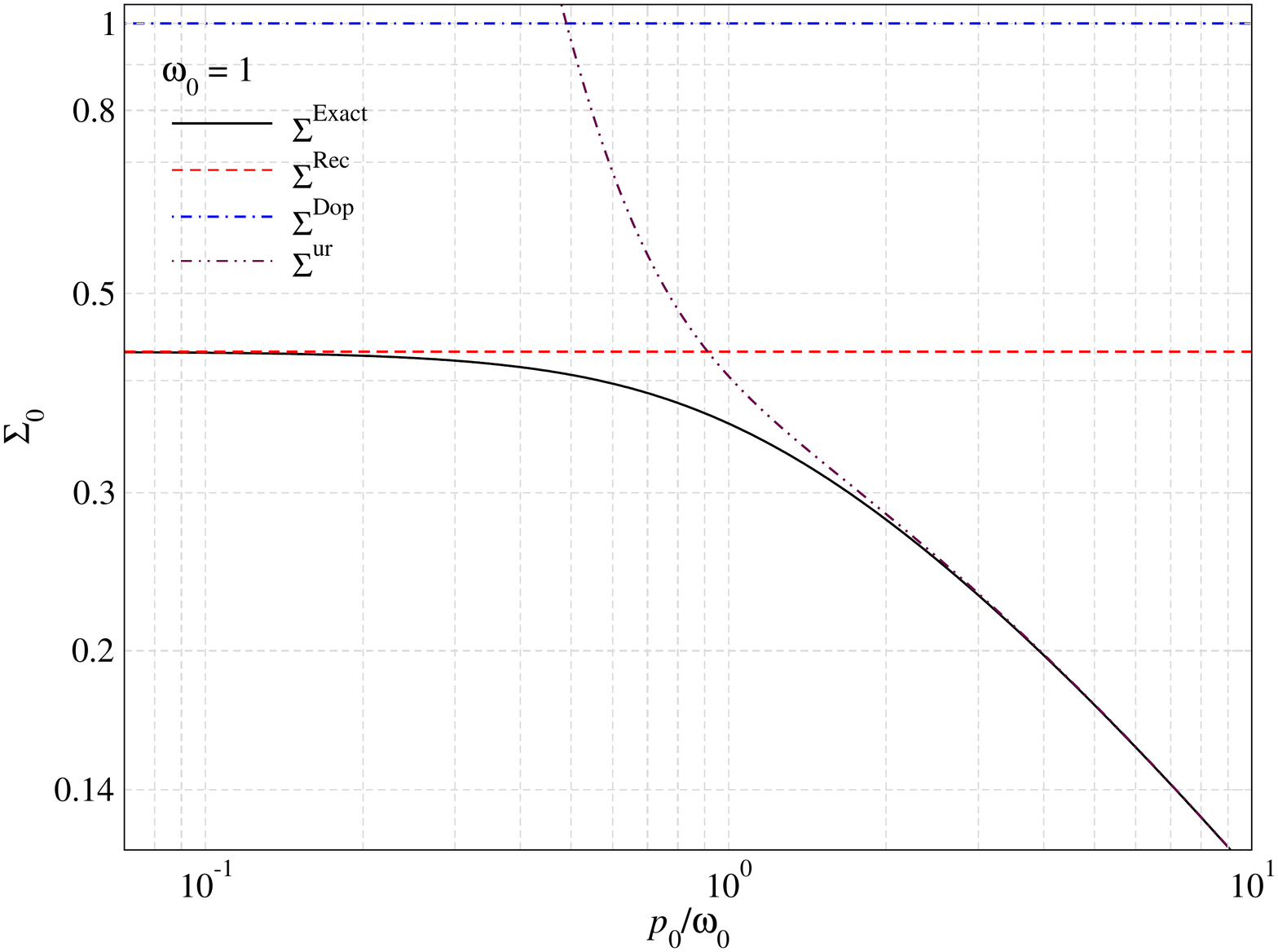}
\caption{Illustrations of the zeroth moment of the Compton scattering kernel. Solid lines represent the exact result, Eq.~\eqref{eq:mom0_exact}. In the upper panel, the moments are plotted as functions of $p_0$ for $\omega_0 =\{0.01, 0.1,1,10\}$.
One can clearly see the Klein-Nishina suppression of the cross section with $\omega_0$ at $p_0\ll1$, which further increases at larger $p_0$. In the upper panel, we also show the simple approximation, $\Sigma_0(\omega_0, p_0)\approx \Sigma_0^{\rm Rec}(\gamma_0\omega_0)$, finding excellent agreement.
The lower panels compare the exact analytical expression, with the moment from the recoil-dominated kernel, Eq.~\eqref{eq:moment_rec} (short-dashed red line), the Doppler-dominated kernel, Eq.~\eqref{eq:moment_dop} (short dot-dashed blue line) and our approximation for ultra-relativistic electrons, Eq.~\eqref{eq:mom0_ur} (long dot-dashed violet line) for $\omega_0=0.1$ and $\omega_0=1$, respectively.
}
\label{fig:moment0}
\end{figure}

Due to the asymptotic convergence of the Taylor series, Eq.~\eqref{eq:mom0_nr} has a rather limited range of applicability. However, for $p_0\gg1$, the leading order effect is a boost of the photon energy in the restframe of the electron. Assuming that the scattering (of the highly anisotropic) radiation field in the restframe is simply given by the Klein-Nishina cross section, one can expect $\Sigma_0(\omega_0, p_0)\approx \Sigma_0^{\rm Rec}(\gamma_0\omega_0)$, with $\Sigma_0^{\rm Rec}(\omega)$ from Eq.~\eqref{eq:moment_rec_a}, to work well. Indeed we find that this approximation captures the main behavior of the zeroth kernel moment at all energies (see Fig.~\ref{fig:moment0}).

For ultra-relativistic electrons ($\gamma_0 \gg 1, \, \gamma_0 \simeq p_0$), another approximation to Eq.~\eqref{eq:mom0_exact} can be deduced by studying the asymptotic behavior of the moment. This yields
\begin{align}
\label{eq:mom0_ur}
\Sigma_0^{\rm ur}(\omega_0, p_0)
&\approx 
\frac{6 \ln \chi-3}{4 \chi}
-\frac{42.239-(27-6\ln \chi)\ln \chi}{2 \chi^2}
\nonumber\\
&\qquad\qquad
+\frac{39+ 24\ln \chi}{2 \chi^3}
+\frac{(9-6\ln \chi)\omega_0^2 }{\chi^3}
\end{align}
with $\chi=4 p_0  \omega_0$. We confirmed this result by full numerical integration, finding excellent agreement in this regime (see Fig.~\ref{fig:moment0}). The leading order term $\Sigma_0^{\rm ur}(\omega_0, p_0)\approx 6 \ln \chi/[4 \chi]$ was also given in \citet{Coppi1990}.

In Fig.~\ref{fig:moment0} we illustrate the total cross-section as a function of $p_0$. In the upper panel we present exact results, while the other two highlight various approximations. The total cross section reduces with increasing photon energy due to Klein-Nishina corrections. These corrections are further amplified once $\gamma_0\omega_0$ becomes large, as mentioned above.
Turning to the performance of the various approximations (lower panels in Fig.~\ref{fig:moment0}), we see that the Doppler-dominated approximation ($\Sigma_0^{\rm Dop}(\omega_0, p_0)= 1$) can only work for $\omega_0,p_0\ll 1$, while the recoil-dominated approximation performs well as long as $p_0 \ll \omega_0$ and $p_0\ll1$. For approximation Eq.~\eqref{eq:mom0_ur} to work, one really needs $\omega_0 \gamma_0 \gg 1$. For instance, for $\omega_0=0.1$, this means $\gamma_0\simeq p_0\simeq 10$ or $p_0/\omega_0\simeq 10^2$, as also evident from Fig.~\ref{fig:moment0}.

\begin{figure}
\includegraphics[width=1.0 \linewidth ]{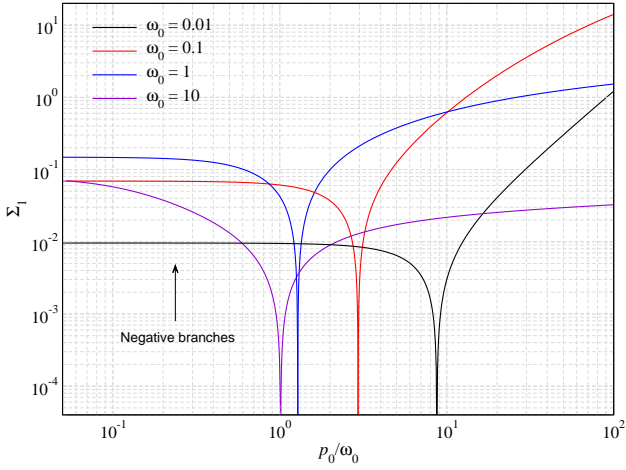}
\\[-0.5mm]
\includegraphics[width=1.0 \linewidth ]{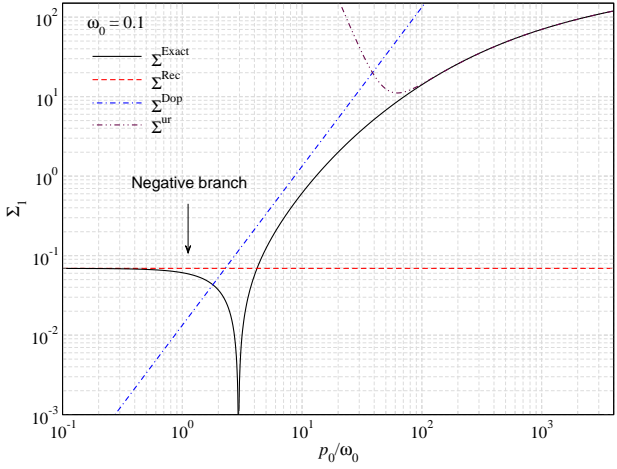}
\\[-0.5mm]
\includegraphics[width=1.0 \linewidth ]{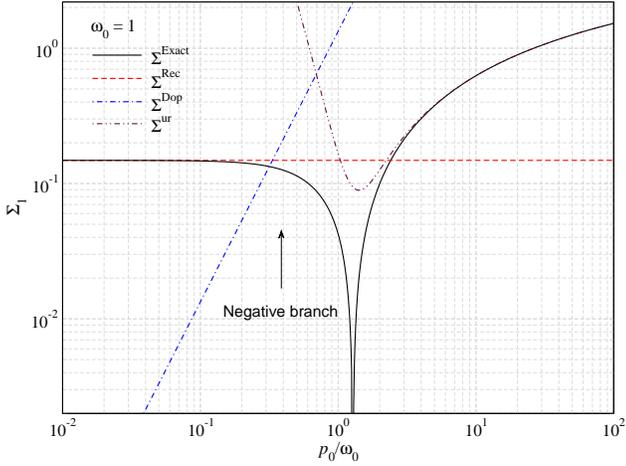}
\caption{Illustrations of the first moment of the Compton scattering kernel. In the upper panel, the exact moments (Eq.~\eqref{eq:mom1_exact}, solid lines) are plotted as functions of $p_0$, for $\omega_0 =\{0.01, 0.1,1,10\}$. 
In the other panels, we compare the exact analytical expression, with the approximations based on the recoil-dominated kernel (Eq.~\eqref{eq:moment_rec_b}, short-dashed red line), the Doppler-dominated kernel (Eq.~\eqref{eq:moment_dop}, short dot-dashed blue line) and our approximation for ultra-relativistic electrons (Eq.~\eqref{eq:mom1_ur}, long dot-dashed violet line) for $\omega_0=0.1$ and $\omega_0=1$. Even for $\omega_0=0.1$, the approximation based on the Doppler-dominated kernel breaks down, while the others work well in their respective regimes. 
}
\label{fig:moment1}
\end{figure}

\begin{figure}
\includegraphics[width=1.0 \linewidth ]{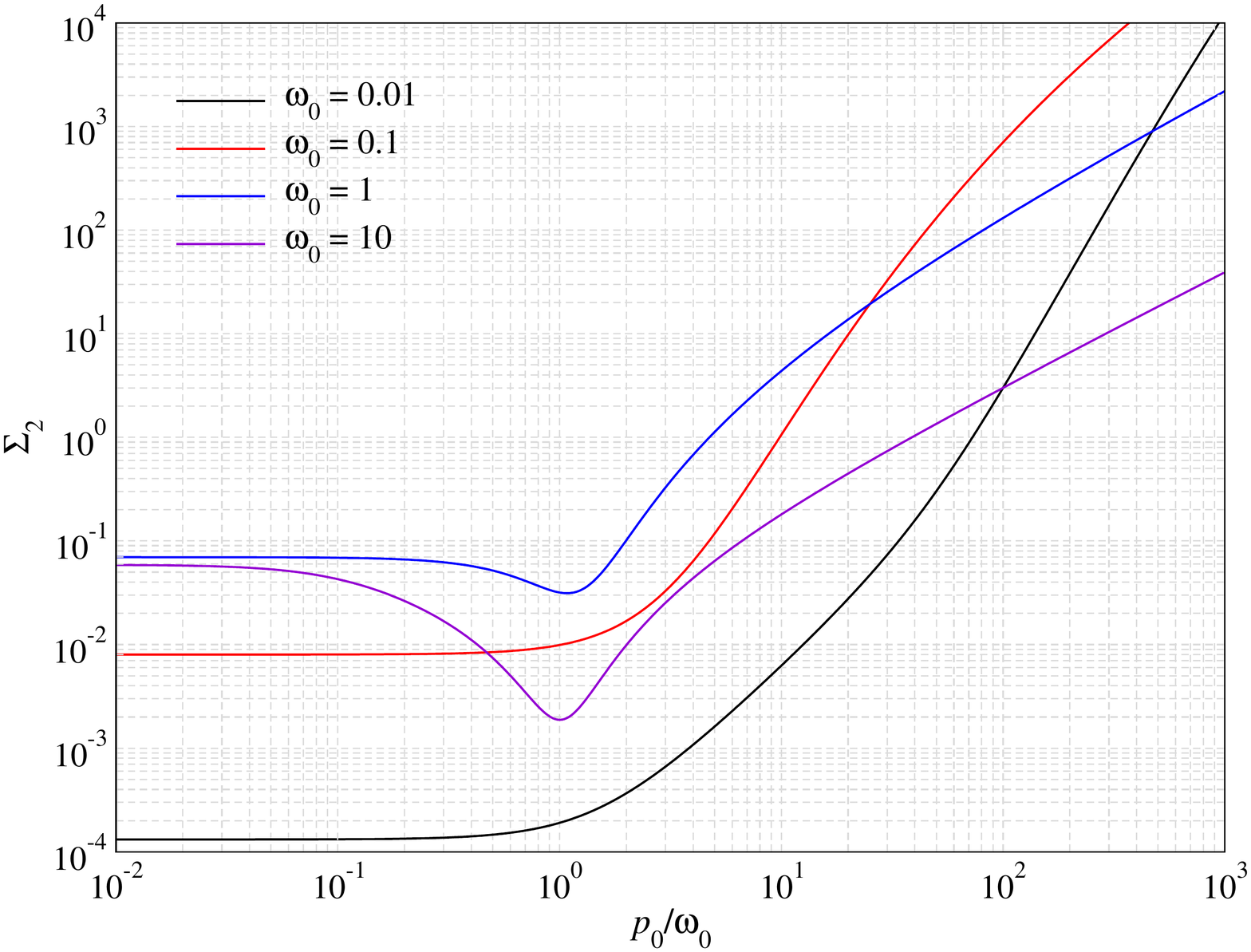}
\\[-0.5mm]
\includegraphics[width=1.0 \linewidth ]{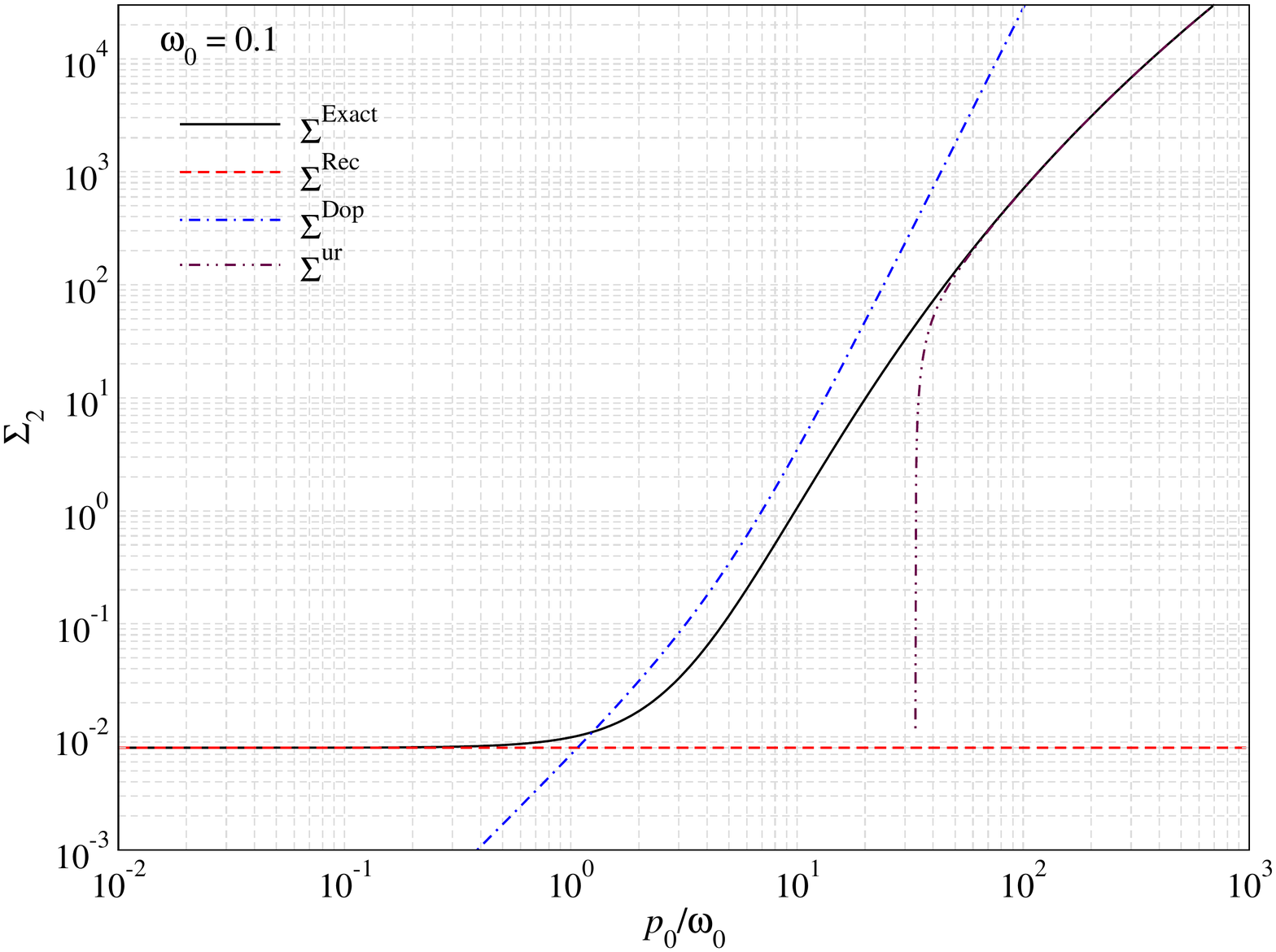}
\\[-0.5mm]
\includegraphics[width=1.0 \linewidth ]{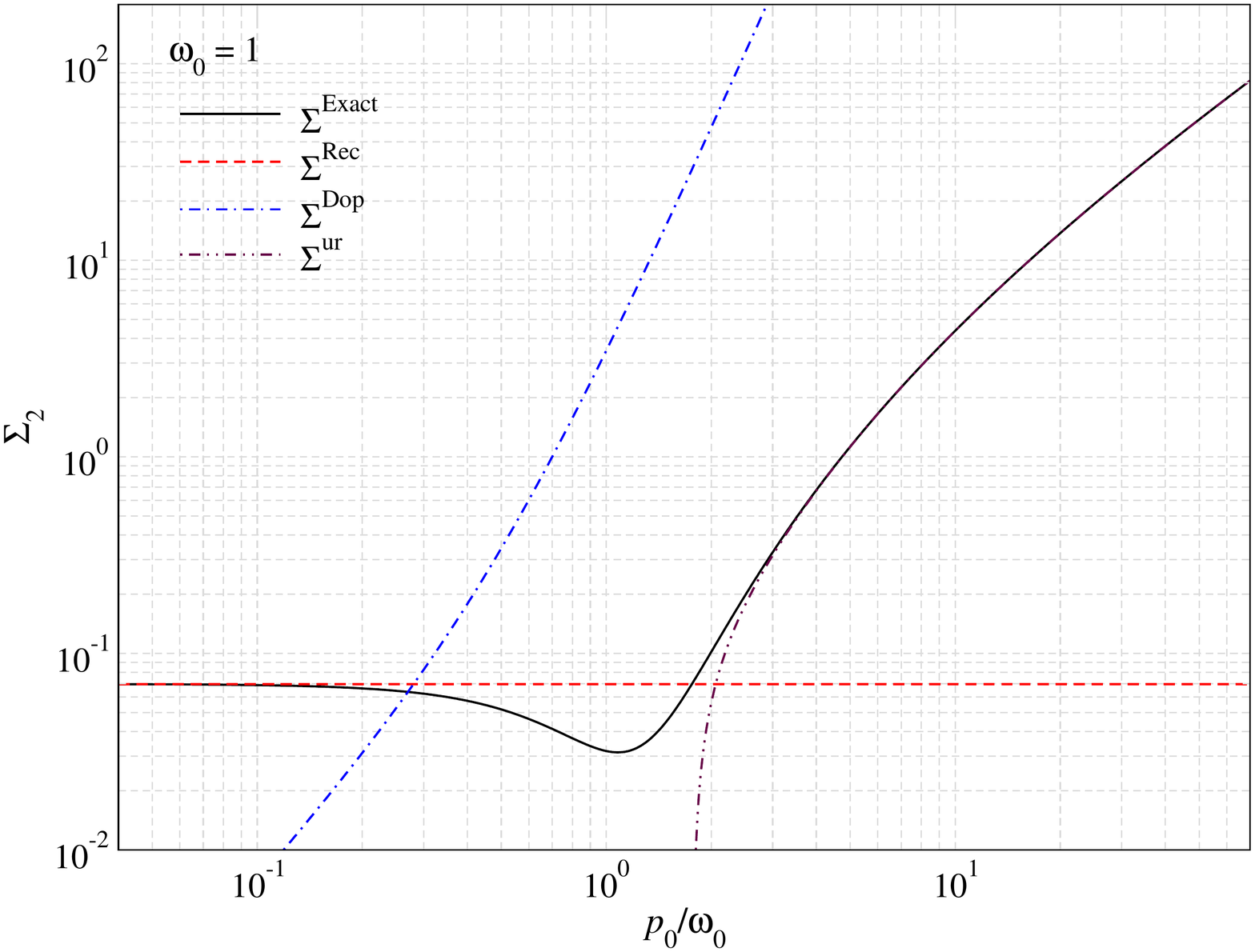}
\caption{Illustrations of the second moment of the Compton scattering kernel. In the upper panel, the exact moments (Eq.~\eqref{eq:mom2_exact}, solid lines) are plotted as functions of $p_0$, for $\omega_0 =\{0.01, 0.1,1,10\}$. 
In the other panels, we compare the exact analytical expression, with the approximations based on the recoil-dominated kernel (Eq.~\eqref{eq:moment_rec_c}, short-dashed red line), the Doppler-dominated kernel (Eq.~\eqref{eq:moment_dop}, short dot-dashed blue line) and our approximation for ultra-relativistic electrons (Eq.~\eqref{eq:mom2_ur}, long dot-dashed violet line) for $\omega_0=0.1$ and $\omega_0=1$. Even for $\omega_0=0.1$, the approximation based on the Doppler-dominated kernel fails, while the others work well in their respective regimes.
}
\label{fig:moment2}
\end{figure}

\vspace{-4mm}
\subsection{The First Moment}
The first moment can be identified with the net energy transferred to the scattered photon. Carrying out the required integrals yields
\begin{align}
\label{eq:mom1_exact}
\Sigma_1(\omega_0, p_0) &=  \frac{(2 \gamma_0 - \omega_0) }{2 \omega_0} \, \Sigma_0 
+\frac{3}{32 \gamma_0 \omega_0^4}
\Bigg\{(4 \gamma_0 + 8 \omega_0 - \omega_0^3) \ln \lambda
\nonumber\\[0mm]
&\quad
-\frac{1 + 4 p^2_0+5 \gamma_0 \omega_0 +\! \frac{35}{6} \omega_0^2 + \gamma_0\omega_0^3}{p_0} 
\ln\left(\frac{\alpha_+}{\alpha_-}\right)
\Bigg\}
\\
\nonumber
&
\qquad
-\frac{1}{64 \gamma_0 \omega_0^3} 
\Bigg[63+\frac{1}{\lambda^2}+8\lambda
-\left(70-\frac{6}{\lambda}+\frac{4}{\lambda^2} \right)\omega_0^2
\Bigg]
\end{align}
with all functions having their meanings as described in the last subsection. Similar to the zeroth moment, in the non-relativistic limit ($p_0, \omega_0 \ll 1$), the first moment  can be simplified as
\begin{align}
&\Sigma_1^{\rm nr}(\omega_0, p_0) \approx -\omega_0\,\Bigg(1 - \frac{21}{5}\omega_0 +\frac{147}{10} \omega_0^2- \frac{1616}{35}\omega_0^3 + \frac{940}{7}\omega_0^4\Bigg)\, +  \nonumber \\[1mm]
&\!
\Bigg(\frac{4}{3} - \frac{47}{6}\omega_0 + \frac{189}{5}\omega_0^2 - \frac{9551}{60}\omega_0^3\Bigg)\,p_0^2 - \frac{553}{120}\omega_0p_0^4 + \mathcal{O}(\omega_0^2p_0^4).
\label{eq:mom1_nr}
\end{align}
Comparing this with Eq.~\eqref{eq:moment_rec_low_b}, we can again confirm the leading order Klein-Nishina terms. Similarly, we find the pure Doppler term, $\propto p_0^2$ (see Eq.~\ref{eq:moment_dop}). The other cross terms are not present in the separate recoil- or Doppler-dominated regimes. Inserting $\left<p_0^2\right>\approx 3\The +\frac{15}{2}\The^2$ and $\left<p_0^4\right>\approx 15\The^2$ for the thermally-averaged values of the momenta, we can also confirm all terms given in Eq.~(25) of \citet{Sazonov2000}. 

For ultra-relativistic electrons ($\gamma_0 \gg 1, \, \gamma_0 \simeq p_0$), another approximation to Eq.~\eqref{eq:mom1_exact} can be deduced by studying the asymptotic behavior of the first moment. This yields
\begin{align}
\label{eq:mom1_ur}
\Sigma_1^{\rm ur}(\omega_0, p_0) 
&\!\approx\!
\frac{11 \!-\! 6 \ln \chi}{4 \chi}
\!+\!\frac{94.739-(56-6\ln \chi)\ln \chi}{4 \chi^2}
\!+\!\frac{18- 36\ln \chi}{\chi^3}
\nonumber
\\
\nonumber
&\!\!\!\!\!\!\!\!\!\!
-\frac{11 - 6 \ln \chi}{16 \omega_0^2}
-\frac{26.870-(18-3\ln \chi)\ln \chi}{4 \omega_0^2 \chi}
+\frac{12 + 9 \ln \chi}{2 \omega_0^2 \chi^2}
\\
&
+\frac{7}{4 \omega_0^2 \chi^3}
-\frac{(17-6\ln \chi)\omega_0^2 }{\chi^3}
\end{align}
with $\chi = 4 p_0\omega_0$, which works extremely well for large $\omega_0$ and $p_0$.

In Figure~\ref{fig:moment1} we illustrate the net energy transfer rate as a function of $p_0$. Focusing on the upper panel, at low electron momenta, photons are down-scattered due to recoil, while for high momentum, the opposite happens. No net energy transfer occurs close to $p^{\rm null}_0\approx \big[\frac{3}{4}\omega_0\big(1+\frac{4}{3}\omega_0\big)\big]^{1/2}$, which increases as $\omega_0$ increases, approaching $p^{\rm null}_0\approx \omega_0$, as visible in Fig.~\ref{fig:moment1}.
The other two panels in Fig.~\ref{fig:moment1} show the comparison of the exact expression for the first moment with various approximations. The recoil-dominated approximation works fairly well below the null, while the ultra-relativistic approximation captures the behavior for very high electron momenta. The Doppler-dominated approximation, on the contrary, never matches with the exact solution for the shown cases. It generally overestimates the energy exchange for large $p_0$ as already expected from the discussion of the kernel approximations (Fig.~\ref{fig:kernels_UR}).

\vspace{-4mm}
\subsection{The Second Moment}
The second moment represents the dispersion in the energy of the scattered photon with respect to the energy of the incident photon. The expression for general $\omega_0$ and $p_0$ is given by
\begin{align}
\label{eq:mom2_exact}
&\Sigma_2(\omega_0, p_0) =  \frac{2+\gamma_0(2 \gamma_0 - \omega_0)}{2 \omega_0^2} \, \Sigma_0 
+\frac{3}{16 \gamma_0 \omega_0^4}
\Bigg\{
\frac{2 - 7\omega^2_0+\omega_0^4}{\omega_0}\ln \lambda
\nonumber\\
&
-\frac{155-90\omega_0^2}{24 p_0} 
\ln\left(\frac{\alpha_+}{\alpha_-}\right)
\Bigg\}
+\frac{3}{32 \omega_0^5}
\Bigg\{
\left(4 \gamma_0 + 6\omega_0 -3\omega^3_0\right)\ln \lambda
\nonumber
\\[0mm]
&\quad
-\frac{12 \gamma^2_0 
+ 2\gamma_0\omega_0+25\omega^2_0+9\gamma_0\omega^3_0-6\omega_0^4}{3p_0} 
\ln\left(\frac{\alpha_+}{\alpha_-}\right)
\Bigg\}
\nonumber
\\
&
\qquad
+\frac{1}{512 \gamma_0 \omega_0^5} 
\Bigg[162-\frac{1}{\lambda^3}+\frac{5}{\lambda^2}-\frac{2}{\lambda}-141\lambda-23\lambda^2\Bigg]
\nonumber
\\
&
\qquad\qquad
+\frac{1}{64 \gamma_0 \omega_0^3} 
\Bigg[243+\frac{1}{\lambda^3}+\frac{12}{\lambda}+54\lambda\Bigg]
\nonumber
\\
&
\qquad\qquad\qquad
-\frac{1}{32 \gamma_0 \omega_0} 
\Bigg[114+\frac{1}{\lambda^3}+\frac{4}{\lambda^2}-\frac{3}{\lambda}\Bigg]
\end{align}
In the non-relativistic limit ($p_0, \omega_0 \ll 1$), the second moment can be simplified as
\begin{align}
\label{eq:mom2_nr}
&\Sigma_2^{\rm nr}(\omega_0, p_0) 
\approx \omega_0^2\,\Bigg(\frac{7}{5} - \frac{44}{5}\omega_0 
+\frac{1364\omega_0^2}{35}-\frac{1020\omega_0^3}{7}\Bigg) 
\\[1mm]
\nonumber
&\qquad
+ p_0^2\,\Bigg(\frac{2}{3} - \frac{42}{5}\omega_0 + \frac{161}{3}\omega_0^2- \frac{1886}{7}\omega_0^3\Bigg) 
+p_0^4\,\Bigg(\frac{14}{5}-\frac{763}{25}\omega_0\Bigg) 
\end{align}
Comparing this with Eq.~\eqref{eq:moment_rec_c}, we can again confirm the leading order Klein-Nishina terms, with no leading order correction in terms of $p_0$ alone. The lowest order terms can be identified with the expression for the second moment given in Eq.~(25) of \citet{Sazonov2000}, after carrying out the thermal averages. 

For ultra-relativistic electrons ($\gamma_0 \gg 1, \, \gamma_0 \simeq p_0$), another approximation to Eq.~\eqref{eq:mom2_exact} can be deduced by studying the asymptotic behavior of the moment. This yields
\begin{align}
\label{eq:mom2_ur}
\!\!\!\!\!\!& \Sigma_2^{\rm ur}(\omega_0, p_0) 
\approx 
-\frac{29 - 12 \ln \chi}{8 \chi}
-\frac{45 - 33 \ln \chi}{4 \chi^2}
-\frac{529- 300\ln \chi}{8\chi^3}
\nonumber
\\
\nonumber
&\!\!
-\!\frac{64.989-(43\!-\!6\ln \chi)\ln \chi}{32 \omega_0^4}
\!+\!\frac{109 +96 \ln \chi}{64 \omega_0^4 \chi}
\!+\!\frac{45}{64 \omega_0^4 \chi^2}
\!-\!\frac{35}{192 \omega_0^4 \chi^3}
\\
\nonumber
&
-\frac{(29 - 12 \ln \chi)\chi}{128 \omega_0^4}
+\frac{65 - 24 \ln \chi}{32 \omega_0^2}
+\frac{214.74-(119-6\ln \chi)\ln \chi}{16 \omega_0^2 \chi}
\\
&\;
-\!\frac{192.69+(21+36\ln \chi)\ln \chi}{8 \omega_0^2 \chi^2}
\!+\!\frac{23}{2 \omega_0^2 \chi^3}
+\frac{(41-12\ln \chi)\omega_0^2 }{2\chi^3},
\end{align}
with $\chi = 4 p_0\omega_0$. Like for the other moments, this approximation works extremely well when $p_0 \gg 1/[4\omega_0]$.

In Figure~\ref{fig:moment2} we illustrate  the energy dispersion of the scattered photon with respect to the incident photon energy for varying $p_0$. Focusing on the upper panel, we observe that the second moment shows a complex non-monotonic behavior. Increasing $p_0$ leads to a monotonic increase of $\Sigma_2$ for $\omega_0=0.01$ and $\omega_0=0.1$, however, the other two cases show a slight decrease around $p_0\simeq \omega_0$. This is the regime when $\omega_{\rm c}$ approaches $\omega_0$ and even from the lower panel of Fig.~\ref{fig:kernels} one can read off that around this combination the width of the kernel is reduced. Numerically, we found that for $\omega_0\lesssim 0.211$ there is no local minimum in $\Sigma_2$ when varying $p_0$.

The other two panels in Fig.~\ref{fig:moment2} show the comparison of the exact expression for the second moment with various approximations. In the recoil-dominated regime, Eq.~\eqref{eq:moment_rec_c} works very well, accurately capturing the increase in the dispersion with $\omega_0$. Similarly, in the ultra-relativistic limit, our approximation Eq.~\eqref{eq:mom2_ur} performs very well. The Doppler-dominated kernel is only useful for rather low values of $\omega_0$, but does not match the second moment in the illustrated examples.

\begin{figure}
\includegraphics[width=1.0 \linewidth ]{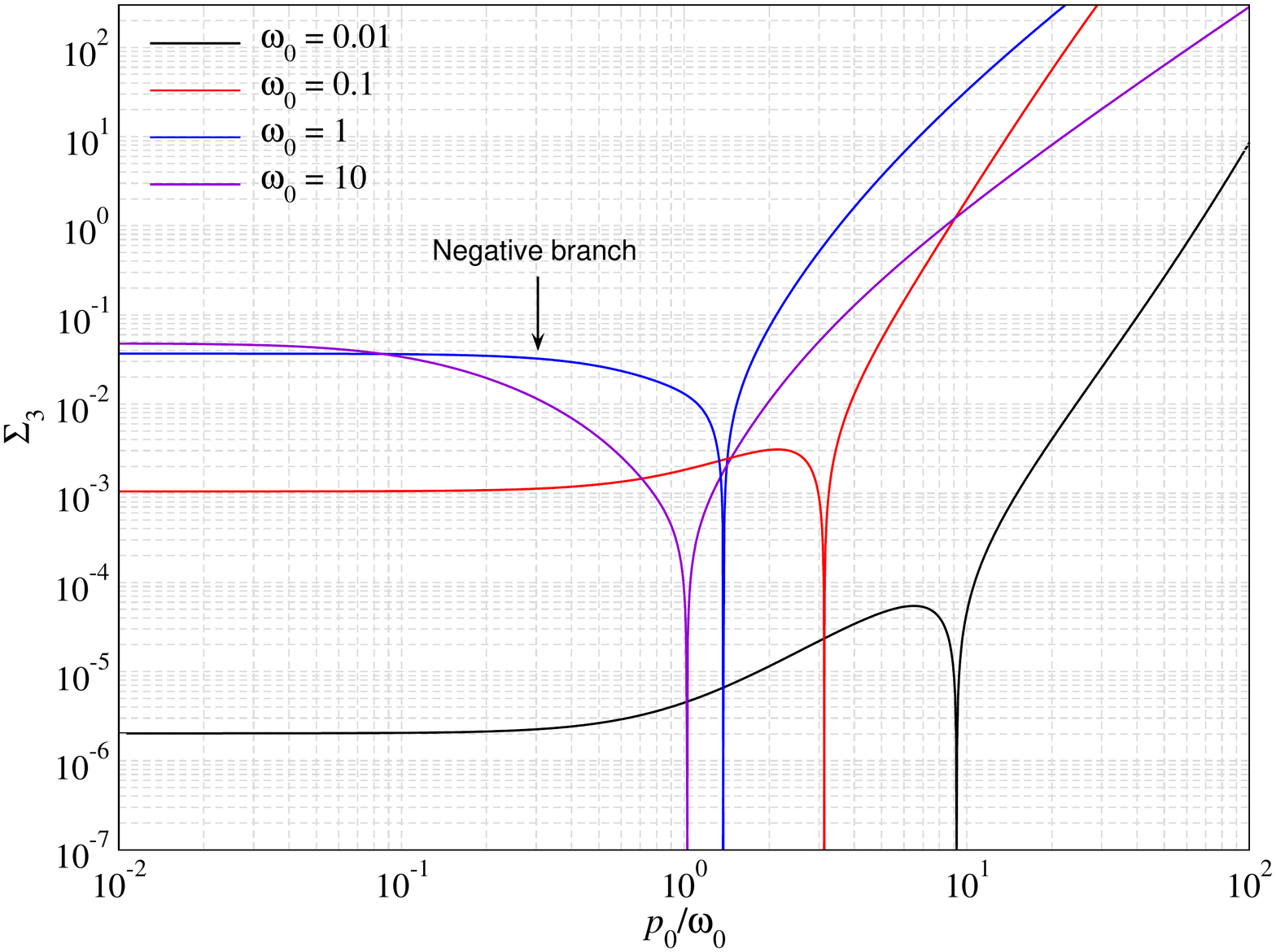}
\\[-0.5mm]
\includegraphics[width=1.0 \linewidth ]{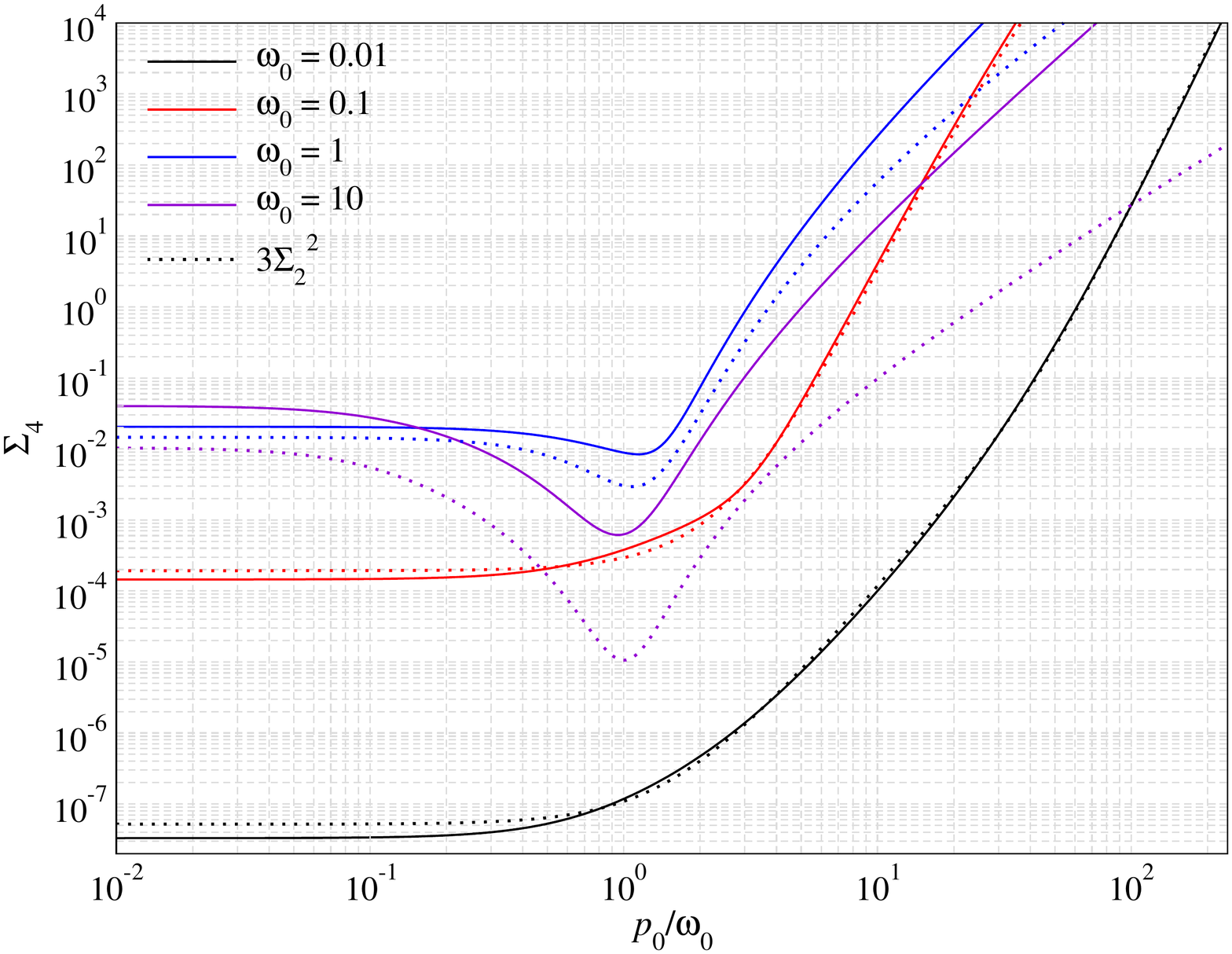}
\caption{The third (upper panel) and fourth (lower panel) moments of the Compton kernel for $\omega_0=\{0.01,0.1,1,10\}$. These were computed numerically from Eq.~\eqref{eq:moment} for $m=3,\,4$ and represent the skewness and kurtosis of the kernel, respectively. Similar to the first moment, the third moment has a negative branch at small $p_0$. In the lower panel we also show $3\Sigma_2^2$ (dotted lines) to illustrate the non-Gaussian nature of the kernel, which is more pronounced for larger values of $\omega_0$. }
\label{fig:moments_higher_num}
\end{figure}

\vspace{-3mm}
\subsection{Higher order moments}
\label{sec:higher_moments}
Higher order moments of the kernel can be important for Fokker-Planck treatments of the kinetic equation \citep{Sazonov1998, Itoh98}. Here we illustrate the behaviour of the third and fourth moments, representing the skewness and kurtosis of the kernel, respectively. We restrict ourselves to a numerical study only, as the explicit expressions for these moments become lengthy and numerical integration with {\tt CSpack} is very efficient.

In the upper panel of Fig.~\ref{fig:moments_higher_num}, we illustrate the third moments of the kernel for $\omega_0 = \{0.01, 0.1, 1, 10\}$. The behaviour of the third moment is quite similar to the first moment. At low $p_0$, it becomes negative, and the zero-crossing moves towards higher values of $p_0$ with increasing $\omega_0$, following $p_0^{\rm null} \approx [\frac{21}{25}\omega_0(1+\frac{25}{21}\omega_0)]^{1/2}$.
The fourth moment is illustrated in the lower panel of Fig.~\ref{fig:moments_higher_num} (solid lines) for the same set of $\omega_0$. The overall scaling with $p_0$ is similar to the second moment.
For a Gaussian distribution function, the fourth central moment is 3 times the square of the second central moment. For comparison, we also plotted the latter quantity with dotted line, showing that at low values of $\omega_0$ one indeed has $\Sigma_4 \approx 3 \Sigma^2_2$. However, at large values of $\omega_0$ the departures tend to increase, highlighting the non-Gaussian nature of the kernel.

\vspace{-4mm}
\subsection{Performance of the approximate expressions}
\label{sec:app_perform_mom}
In Figs.~\ref{fig:moment0}, \ref{fig:moment1}, and \ref{fig:moment2}, we directly compared the analytic expressions for the moments in the recoil-dominated [$\Sigma_m^{\rm Rec}$, Eq.~\eqref{eq:moment_rec}],  Doppler-dominated [$\Sigma_m^{\rm Dop}$, Eq.~\eqref{eq:moment_dop}] and ultra-relativistic [$\Sigma_m^{\rm ur}$, Eq.~\eqref{eq:mom0_ur}, \eqref{eq:mom1_ur} and \eqref{eq:mom2_ur}] limits, with the exact analytical expressions [$\Sigma_m$, Eq.~\eqref{eq:mom0_exact}, \eqref{eq:mom1_exact} and \eqref{eq:mom2_exact}]. Here, we more quantitatively assess the accuracy of these approximations. For this purpose, we study the error 
\begin{align}
\epsilon =\left|\frac{\Sigma_{\rm app}}{\Sigma_{\rm exact}}-1\right|
\end{align}
in the $\omega_0-p_0$ plane, determining when the approximation departs by $\epsilon = \{0.1, 1, 10\}\%$ from the exact expression.\footnote{In practice we perform a simple search for the electron momentum at which the approximation first exceeds a given error. For the recoil-dominated regime we start at very low momenta while for the ultra-relativistic approximation we start at very high electron momenta. For the Doppler-dominated case, a search in $\omega_0$ is performed.}
\begin{figure}
\includegraphics[width=1 \linewidth ]{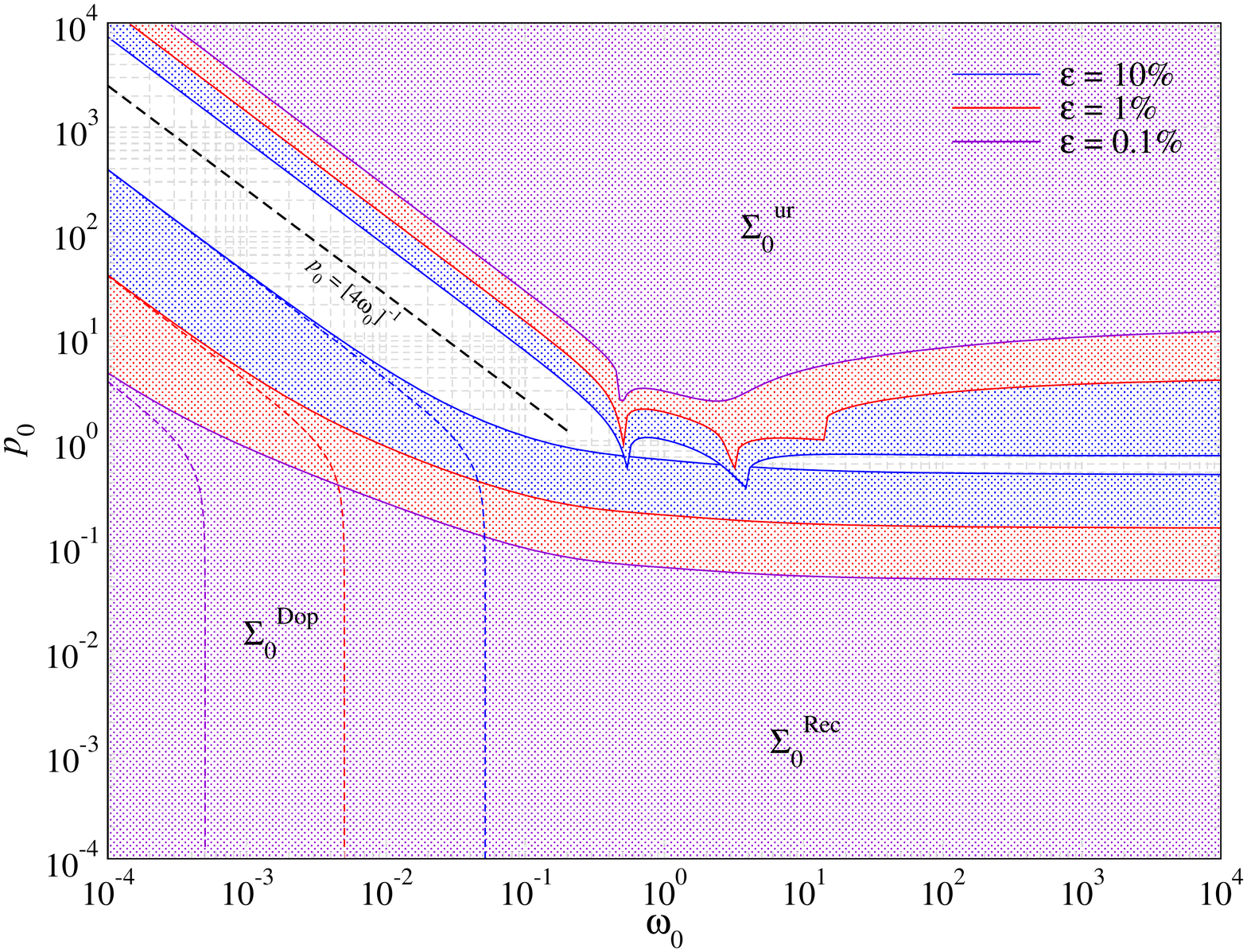}
\\[0.5mm]
\includegraphics[width=1 \linewidth ]{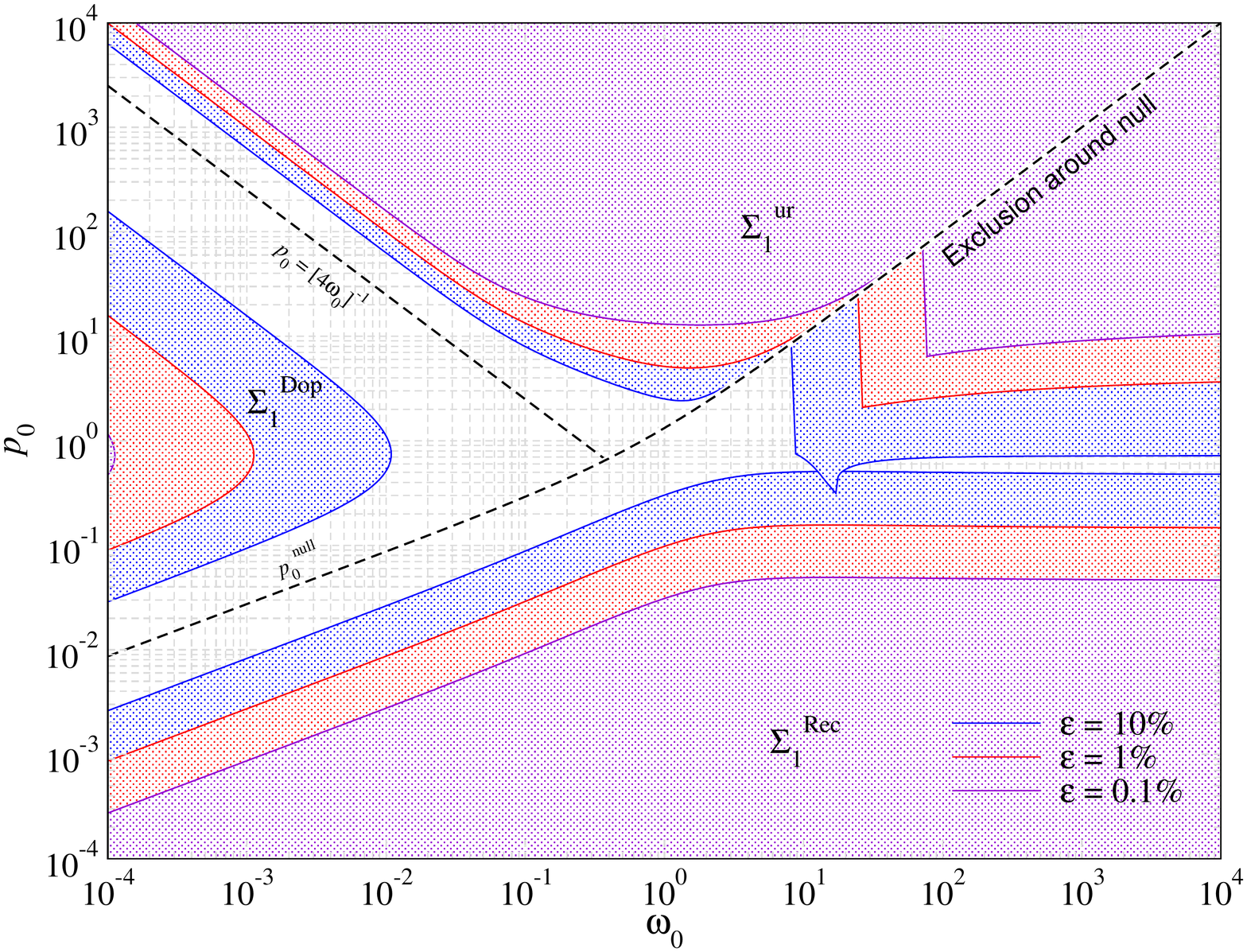}
\\[0.5mm]
\includegraphics[width=1 \linewidth ]{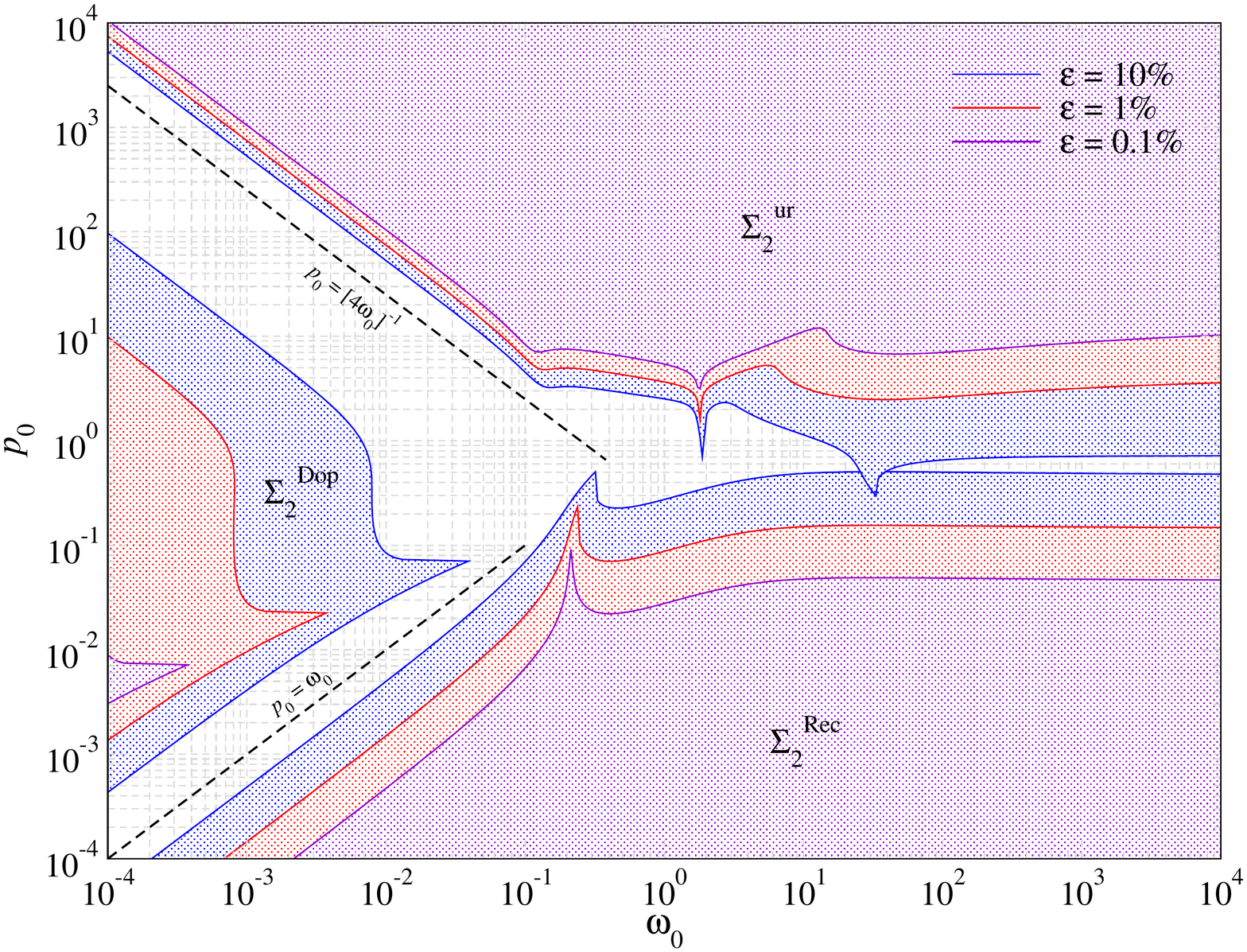}
\caption{Range of applicability for the recoil-dominated [$\Sigma_m^{\rm Rec}$, Eq.~\eqref{eq:moment_rec}],  Doppler-dominated [$\Sigma_m^{\rm Dop}$, Eq.~\eqref{eq:moment_dop}] and ultra-relativistic [$\Sigma_m^{\rm ur}$, Eq.~\eqref{eq:mom0_ur}, \eqref{eq:mom1_ur} and \eqref{eq:mom2_ur}] approximations of the first three kernel moments. Inside the shaded regions the agreement with the exact expressions [$\Sigma_m$, Eq.~\eqref{eq:mom0_exact}, \eqref{eq:mom1_exact} and \eqref{eq:mom2_exact}] is better than $\epsilon = 0.1\%, 1\%$ and $10\%$ respectively. In the upper panel, the dashed lines represent the respective boundaries for $\Sigma_0^{\rm Dop}=1$. We also show several critical lines, with $p^{\rm null}_0\approx \big[\frac{3}{4}\omega_0\big(1+\frac{4}{3}\omega_0\big)\big]^{1/2}$, as annotated. By combining all approximations, a fairly accurate representation of the first three kernel moments over a wide range of parameters can be achieved. For the first moment, a narrow exclusion region around the null, $p^{\rm null}_0\simeq \omega_0\gg 1$ is present in the ultra-relativistic regime.}
\label{fig:cont_l}
\end{figure}

In Figure~\ref{fig:cont_l}, we summarize the results. In each panel, the 
lines indicate the borders, within which the approximations can be applied to the respective accuracy. A combination of the approximations provides a fairly accurate representation of the first three kernel moments. 

\vspace{-4mm}
\section{Thermally-averaged moments}
\label{sec:th_moments}
In this section, we discuss the thermally-averaged moments for various temperatures, providing numerous illustrations and comparisons with analytic approximations. We mostly restrict ourselves to temperatures $k\Te < 511$~keV to avoid complications with Fermi-blocking. Since the thermal average involves an integral over the rMB distribution, we also explore several new approximations for the moments to speed up the computations and enable future applications. Some additional discussion can be found in \cite{Prasad1988}, \citet{1988JQSRT..40..577S} and NP94 with various levels of simplifications, limitations and illustrations.

\begin{figure}
\includegraphics[width=1.0 \linewidth ]{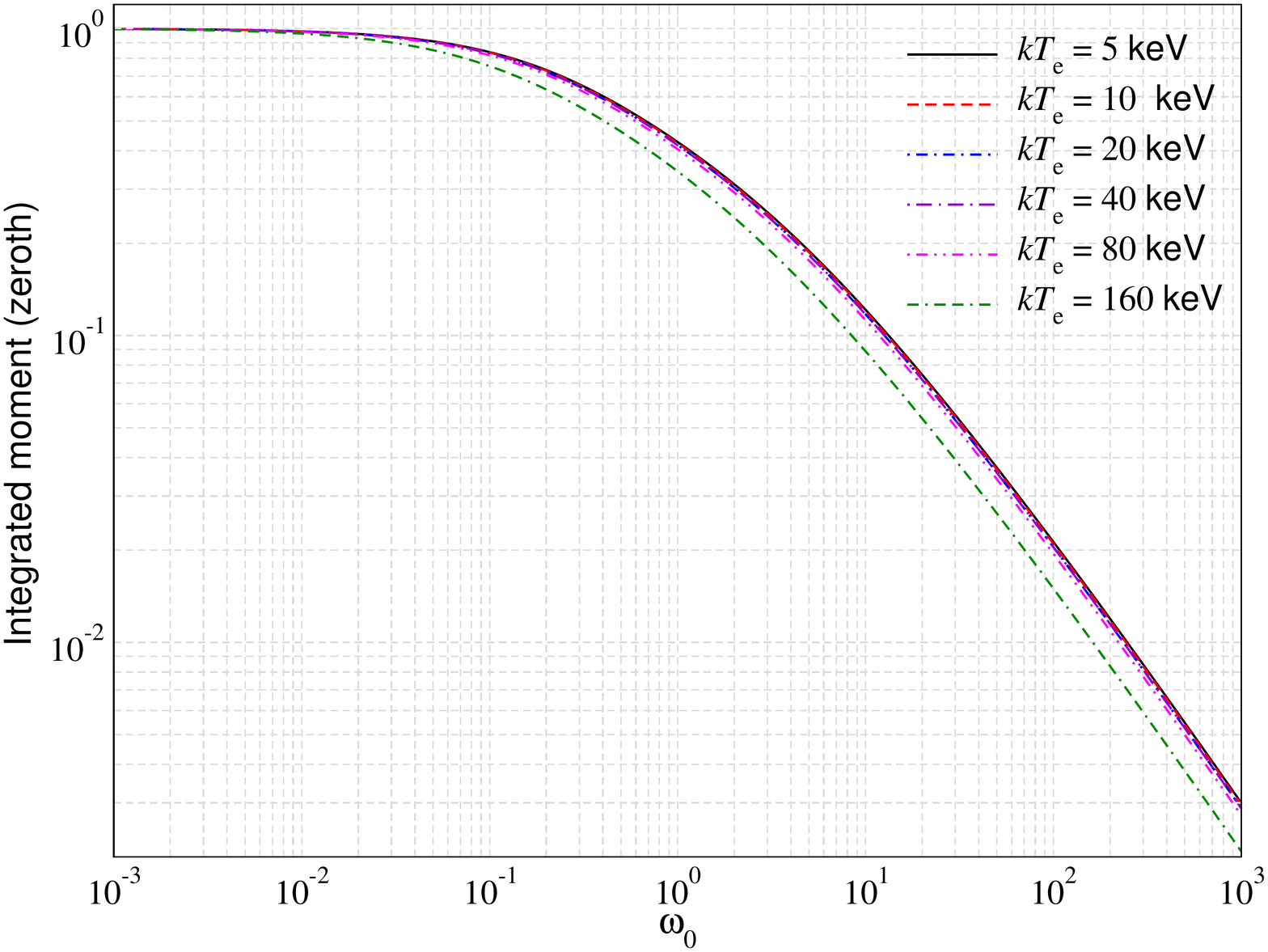}
\\[0.5mm]
\includegraphics[width=1.0 \linewidth ]{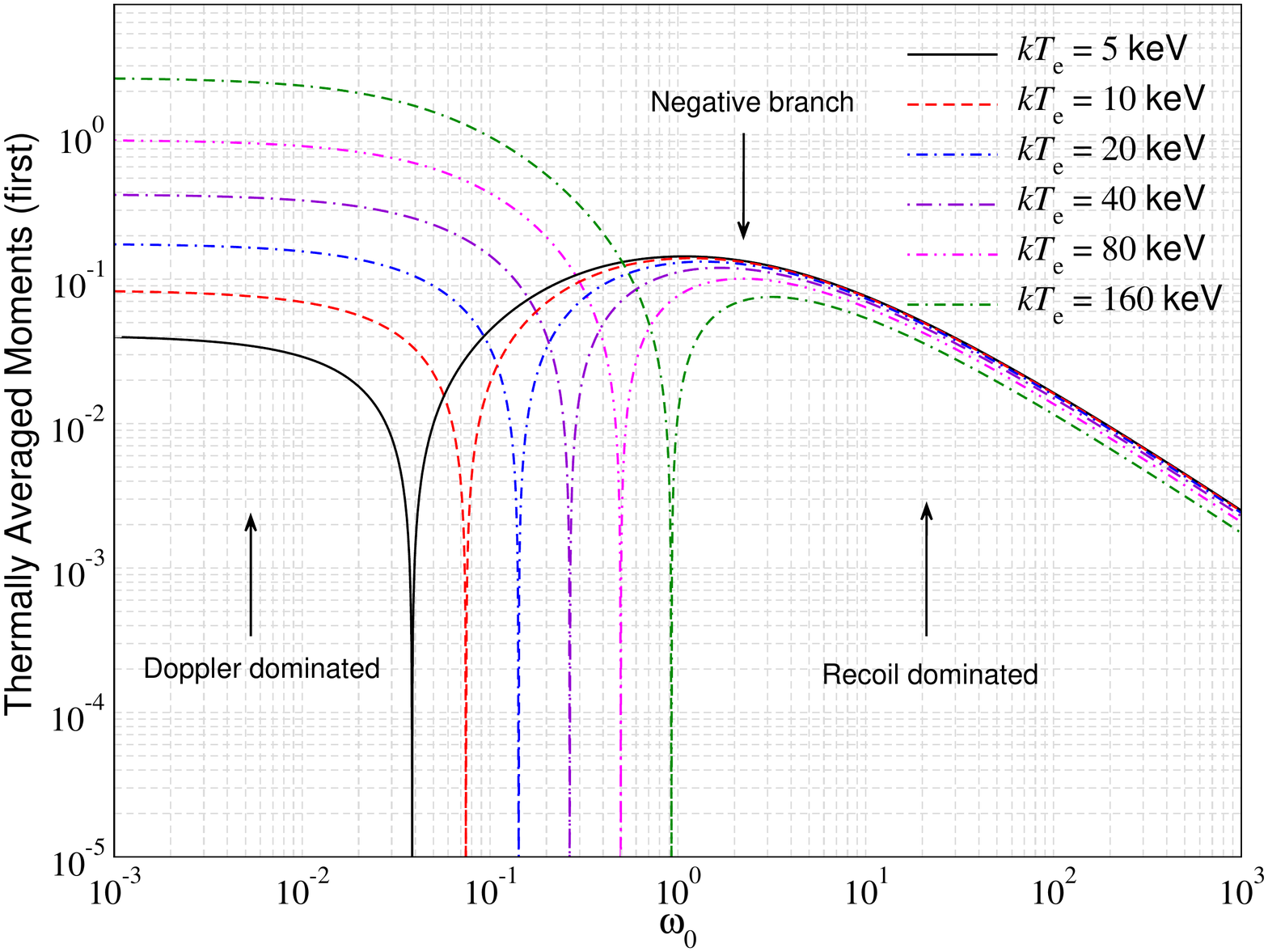}
\\[0.5mm]
\includegraphics[width=1.0 \linewidth ]{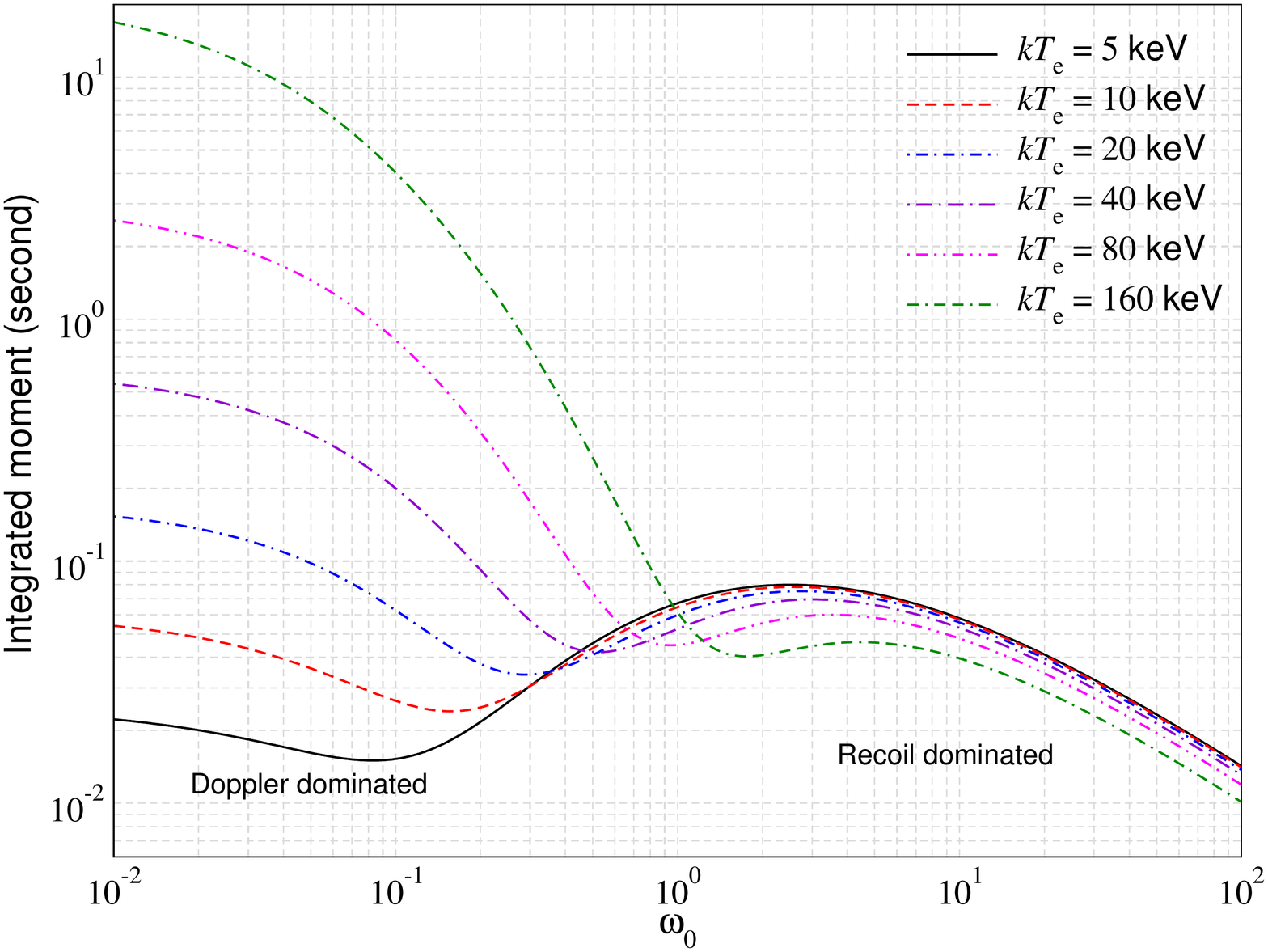}
\caption{Thermally-averaged moments up to $m=2$ for temperatures $kT_{\rm e} = \{5, 10, 20, 40, 80, 160\}$~keV. These were obtained by numerically averaging the exact expressions over a rMB distribution. For the shown examples, the thermal averages has a small effect on the zeroth moment. Temperature effects are much more significant for the first and second moments in particular in the Doppler-dominated regime.}
\label{fig:moments_therm}
\end{figure}

In Fig.~\ref{fig:moments_therm}, we illustrate the moments up to $m=2$ as a function of $\omega_0$ for various temperatures.  The zeroth moment is practically independent of temperature for the shown examples. The main effect becomes visible at high values of $\omega_0$ as also expected from the fact that $\Sigma_0\approx 1$ in the Doppler-dominated regime. Similarly, both the first and second moments show a rather mild dependence on the temperature in the recoil-dominated regimes (large $\omega_0$). This changes strongly around the transition to the Doppler-dominated regime, where the first and second moment show a strong dependence on the electron temperature (see lower panels of Fig.~\ref{fig:moments_therm}).

Turning to some key features, for the first moment, the position of the null is determined by $\omega_{\rm null}\approx 4\The/(1+76\,\The)^{0.1}$ (which was found numerically and works well up to $\The \simeq 2$). At $\omega_0>\omega_{\rm null}$, recoil dominates and the photons on average lose energy, while they gain energy from the electrons below the null. Interestingly, even the lowest order result, $\omega_{\rm null}\approx 4\The\ll 1$, provides a very good estimate for the position of the null, departing only by a factor of $\simeq 1.5$ from the correct value for $\The=1$.
One can also observe a dip in the second moment around the transition from the Doppler- to the recoil-dominated regime. Numerically, we find the position of this dip to be located around $\omega_{\rm m}\approx 0.27 \exp([\The/0.01]^{0.29}-1)$, which can be helpful for estimates. This critical value also indicates the regions where the approximate results tend to deviate the most from the exact numerical results (see Fig.~\ref{fig:cont_te_l}).

\begin{figure}
\includegraphics[width=1.0 \linewidth ]{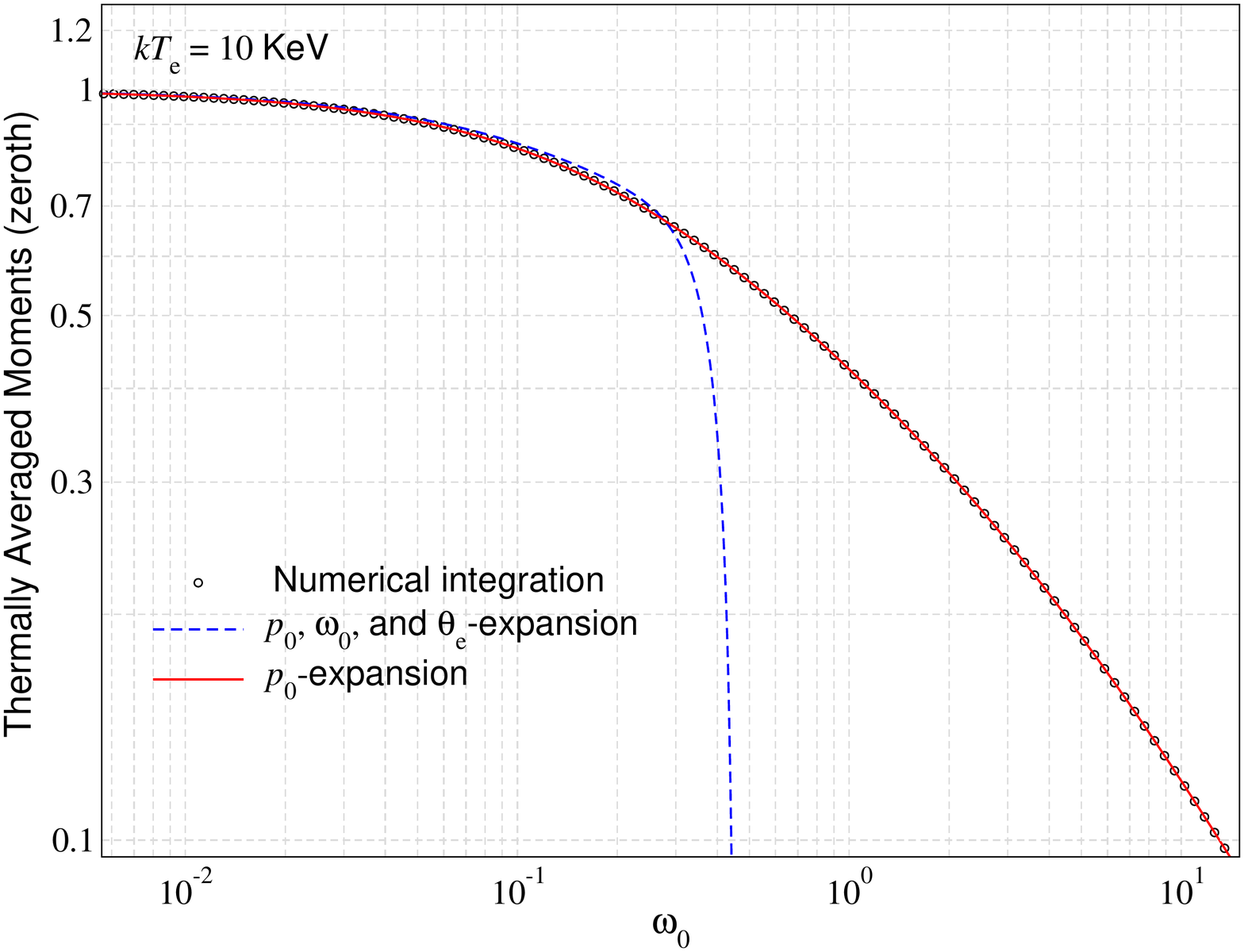}
\\[-0.5mm]
\includegraphics[width=1.0 \linewidth ]{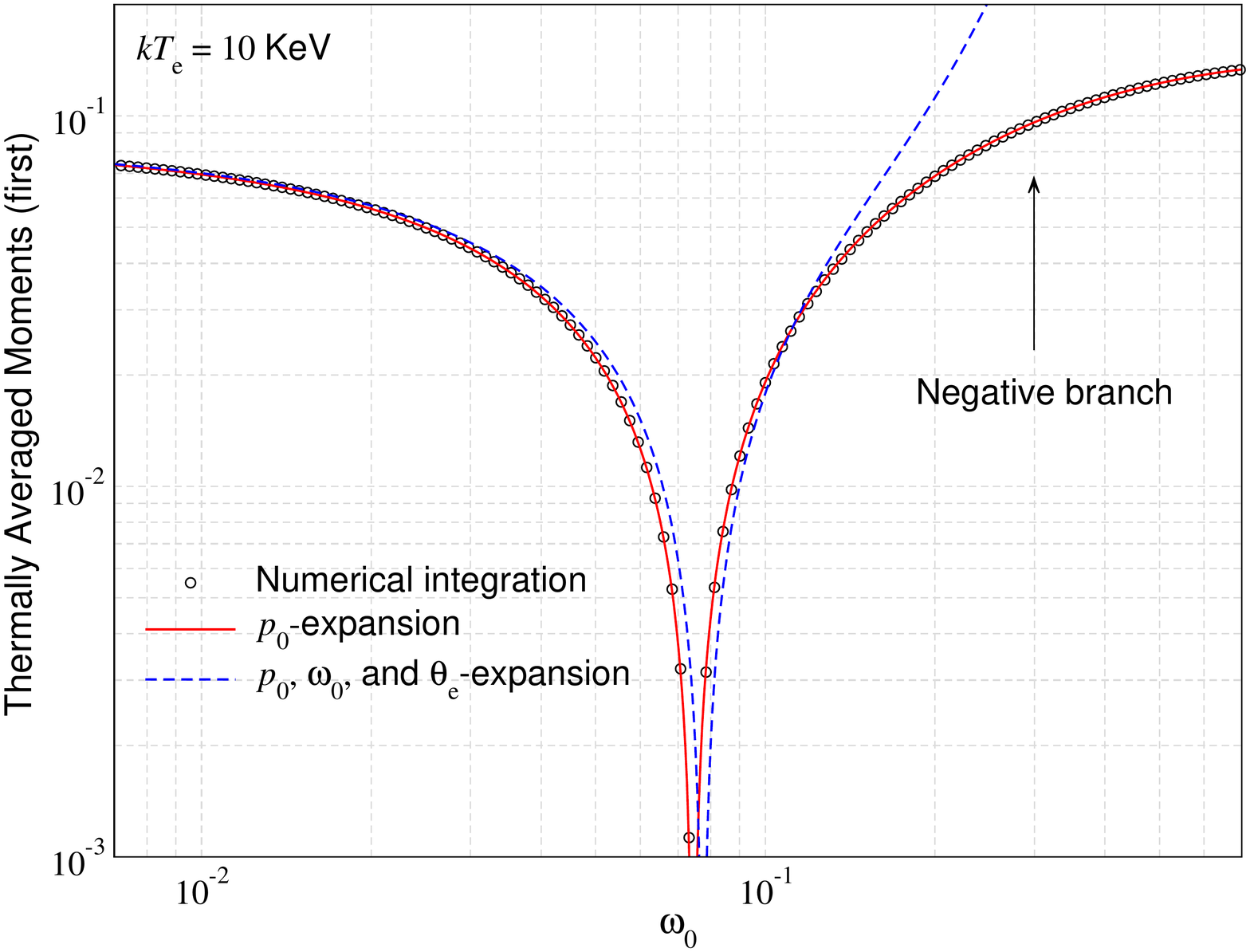}
\\[-0.5mm]
\includegraphics[width=1.0 \linewidth ]{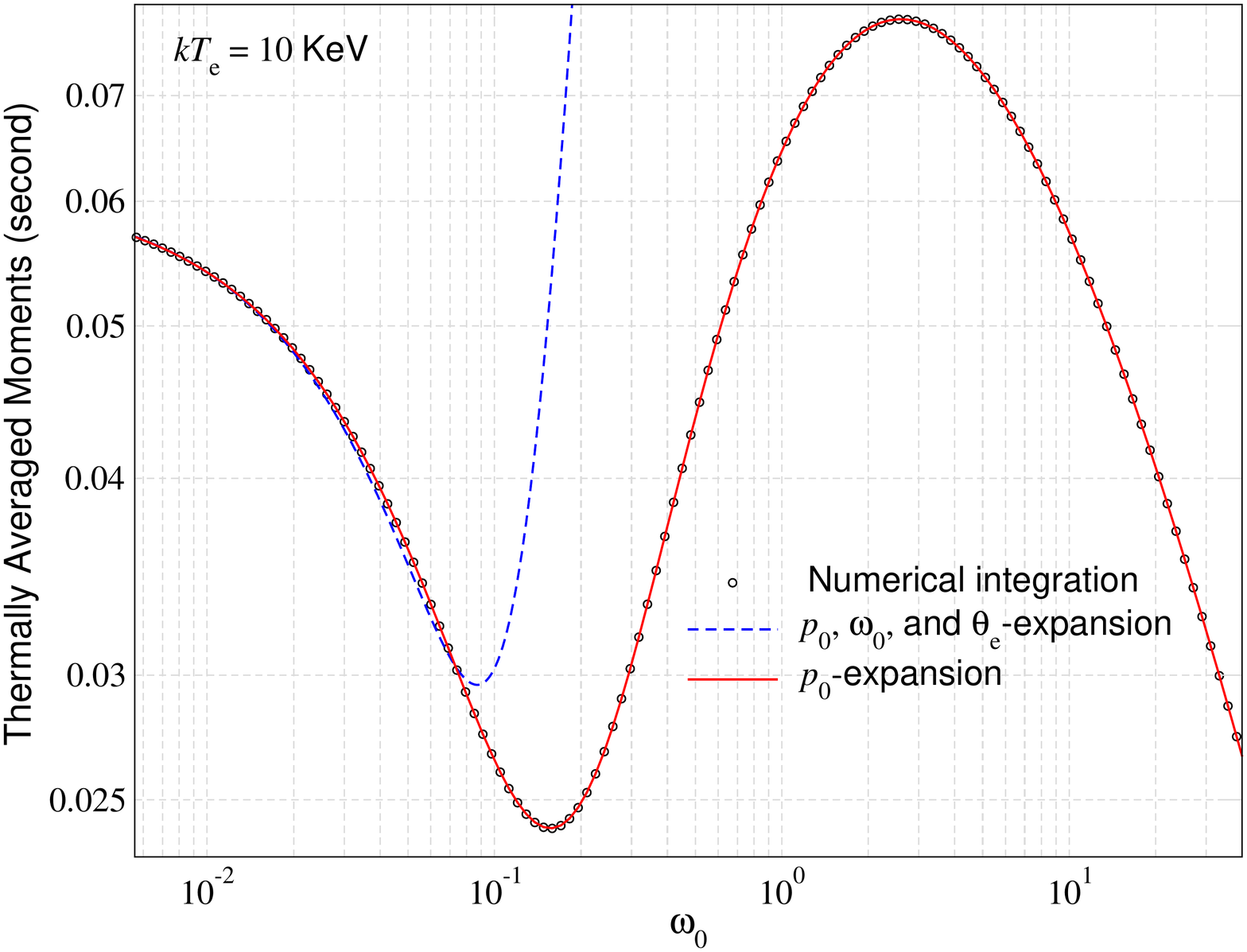}
\caption{Comparison of approximations for the thermally-averaged moments up to $m=2$ with the exact result (circles) for $k\Te=10$~keV. The blue dashed lines are for the full Taylor-series expressions, Eq.~\eqref{eq:moment_therm_low}, while the red solid lines show the approximation obtained by only expanding the exact expression in terms of $p_0$ and then replacing these with the exact values for the thermal averages. Here we included terms up to $\big<p_0^2\big>$, $\big<p_0^4\big>$ and $\big<p_0^8\big>$ for the zeroth, first and second moment expressions, respectively. The explicit Taylor-series converges poorly at $\omega_0\gtrsim 0.05$ in the shown examples, while the second approach works extremely well.}
\label{fig:moments_therm_Te10_appr}
\end{figure}

\vspace{-3mm}
\subsection{Analytic approximations assuming $\omega_0\ll1$ and $\The\ll 1$}
The thermal average of moments involves an integral over the rMB distribution. It is thus useful to explore simpler analytic approximations that avoid this integration. One simple approach is to assume $\omega_0\ll1$ and $\The\ll 1$. One can thus perform a Taylor-series expansion of the moment expressions in $\omega_0, p_0\ll 1$ and then replace the thermally-averaged values for $p^k_0$ by their Taylor-series in $\The$. This approach was taken by \citet{Sazonov2000}, which provided the first few terms in the series. From our expressions, we find
\begin{subequations}
\label{eq:moment_therm_low}
\begin{align}
\label{eq:moment_therm_low_a}
\left<\Sigma_0\right>
&\approx 
1 - 2\omega_0 + \frac{26}{5} \omega_0^2 - \frac{133}{10} \omega_0^3 + \frac{1144}{35}\omega_0^4-\frac{544}{7} \omega_0^5
\nonumber\\
&
-\omega_0\left(5- \frac{156}{5}\omega_0+\frac{2793}{20}\omega_0^2-\frac{18304}{35}\omega_0^3\right)  \The
\nonumber \\
&\quad
-\omega_0\left(\frac{15}{4}- 78\omega_0+\frac{53067}{80}\omega_0^2\right)  \The^2
\nonumber\\
&\qquad
+\omega_0\left(\frac{15}{4}+ \frac{117}{2}\omega_0\right)  \The^3 
-\frac{135}{64}\omega_0  \The^4
\\[1.5mm]
\label{eq:moment_therm_low_b}
\left<\Sigma_1\right>
&\approx 
-\omega_0\,\Bigg(1 - \frac{21}{5}\omega_0 +\frac{147}{10} \omega_0^2- \frac{1616}{35}\omega_0^3 + \frac{940}{7}\omega_0^4\Bigg)
\nonumber\\
&
+\left(4- \frac{47}{2}\omega_0+\frac{567}{5}\omega_0^2-\frac{9551}{20}\omega_0^3+\frac{63456}{35}\omega_0^4\right)  \The
\nonumber\\
&\quad
+\left(10- \frac{1023}{8}\omega_0+\frac{9891}{10}\omega_0^2-\frac{472349}{80}\omega_0^3\right)  \The^2
\\
\nonumber
&\qquad
+\left(\frac{15}{2}- \frac{2505}{8}\omega_0+\frac{177849}{40}\omega_0^2\right)  \The^3 
-\left(\frac{15}{2}+\frac{30375}{128}\omega_0\right)  \The^4
\\[1.5mm]
\label{eq:moment_therm_low_c}
\left<\Sigma_2\right>
&\approx 
\omega_0^2\,\Bigg(\frac{7}{5} - \frac{44}{5}\omega_0 
+\frac{1364\omega_0^2}{35}-\frac{1020\omega_0^3}{7}\Bigg)
\nonumber\\
&
+\left(2- \frac{126}{5}\omega_0+161\omega_0^2-\frac{5658}{7}\omega_0^3+\frac{123024}{35}\omega_0^4\right)  \The
\nonumber\\
&\quad
+\left(47- \frac{2604}{5}\omega_0+\frac{38057}{10}\omega_0^2-\frac{44769}{2}\omega_0^3\right)  \The^2
\\
\nonumber
&\qquad
+\left(\frac{1023}{4}- \frac{21294}{5}\omega_0+\frac{1701803}{40}\omega_0^2\right)  \The^3 
\\
\nonumber
&\qquad\quad
+\left(\frac{2505}{4}-\frac{187173}{10}\omega_0\right)  \The^4.
\end{align}
\end{subequations}
The lowest order terms agree with those from \citet{Sazonov2000}, however, here we obtained terms up to $\mathcal{O}(\omega^2_0\The^4)$, which extend the range of applicability slightly and provide more rigorous checks of analytic calculations. 

In Fig.~\ref{fig:moments_therm_Te10_appr}, we illustrate the performance of the approximations in Eq.~\eqref{eq:moment_therm_low} for different examples. At small values of $\omega_0$, these expressions work very well, but depart from the exact result at higher energies. Even for rather low temperatures (here $k\Te=10$~keV) these approximations converge quite slowly. However, they can be very helpful at low temperatures and for small $\omega_0$, where numerical issues can arise for the exact expressions. 

\subsection{Analytic approximations assuming $p_0\ll 1$}
\label{sec:p_expansion}
Another set of useful approximations can be obtained by only assuming that $p_0\ll 1$ but then applying the exact expressions for $\big<p_0^k\big>$ (Appendix~\ref{app:MMrMB}) with no additional assumption on $\omega_0$. As we will illustrate below, these approximations perform very well up to $k\Te \simeq 20-40\,$~keV (depending on the selected moment). At $k\Te \simeq 40\,$~keV, the average momentum reaches $\big<p_0\big>\simeq 0.5$, so that also the Taylor-series in $p_0$ starts to become non-perturbative and converges very slowly. Still the applicability is strongly extended and covers situations for a wide range of physical conditions.

With the exact expressions for the moments, it is straightforward to obtain the required expressions up to high powers in $p_0$. Only even powers in $p_0$ contribute and since the expressions quickly become lengthly, here we only give terms up to $\left<p_0^2\right>$:
\begin{subequations}
\label{eq:moment_therm_p}
\begin{align}
\label{eq:moment_therm_p_a}
\left<\Sigma_0\right>
&\approx \Sigma_0^{\rm Rec}
+\Bigg\{
\frac{6 -12\xi  - 3 \xi^2 + 27\xi^3-67\xi^4+\xi^5}{16 \xi^4 (1- \xi)^2} 
\nonumber\\
&\qquad\qquad\qquad\qquad\quad
- \frac{\big(5 + 22\xi - 3\xi^2\big) \ln \xi}{8 (1- \xi)^3} 
\Bigg\}\left<p_0^2\right>,
\\[1.5mm]
\label{eq:moment_therm_p_b}
\left<\Sigma_1\right>
&\approx \Sigma_1^{\rm Rec}
+\Bigg\{
\frac{24 -48\xi  +22 \xi^2 + 27\xi^3+33\xi^4+281\xi^5-27\xi^6}{48 \xi^4 (1- \xi)^2} 
\nonumber\\
&\qquad\qquad\qquad\qquad\quad
+ \frac{\big(41 + 14\xi - 3\xi^2\big) \ln \xi}{8 (1- \xi)^3} 
\Bigg\}\left<p_0^2\right>,
\\[1.5mm]
\label{eq:moment_therm_p_c}
\left<\Sigma_2\right>
&\approx \Sigma_2^{\rm Rec}
+\Bigg\{
\frac{60 -84\xi  +39 \xi^2 +16\xi^3+40\xi^4-348\xi^5-75\xi^6}{96 \xi^6 (1- \xi)} 
\nonumber\\
&
- \frac{\big(77 + 6\xi - 3\xi^2\big) \ln \xi}{8 (1- \xi)^3} 
-\frac{4-8\xi}{(1- \xi)^3}
+\frac{4\ln \xi}{(1- \xi)^4}
\Bigg\}\left<p_0^2\right>,
\end{align}
\end{subequations}
with $\xi=1+2\omega_0$. Here, we used the expressions for the recoil-dominated limit, $\Sigma_m^{\rm Rec}$ from Eq.~\eqref{eq:moment_rec}, to which everything reduces for $p_0\rightarrow 0$.
Once the Taylor-series in $p_0$ is performed, we replace the powers of $p_0^{2k}$ with the exact expressions for their thermal averages as given in Appendix~\ref{app:MMrMB}. We found that this procedure is superior to using Taylor-series expressions in $\The$.

\begin{figure}
\includegraphics[width=0.96 \linewidth ]{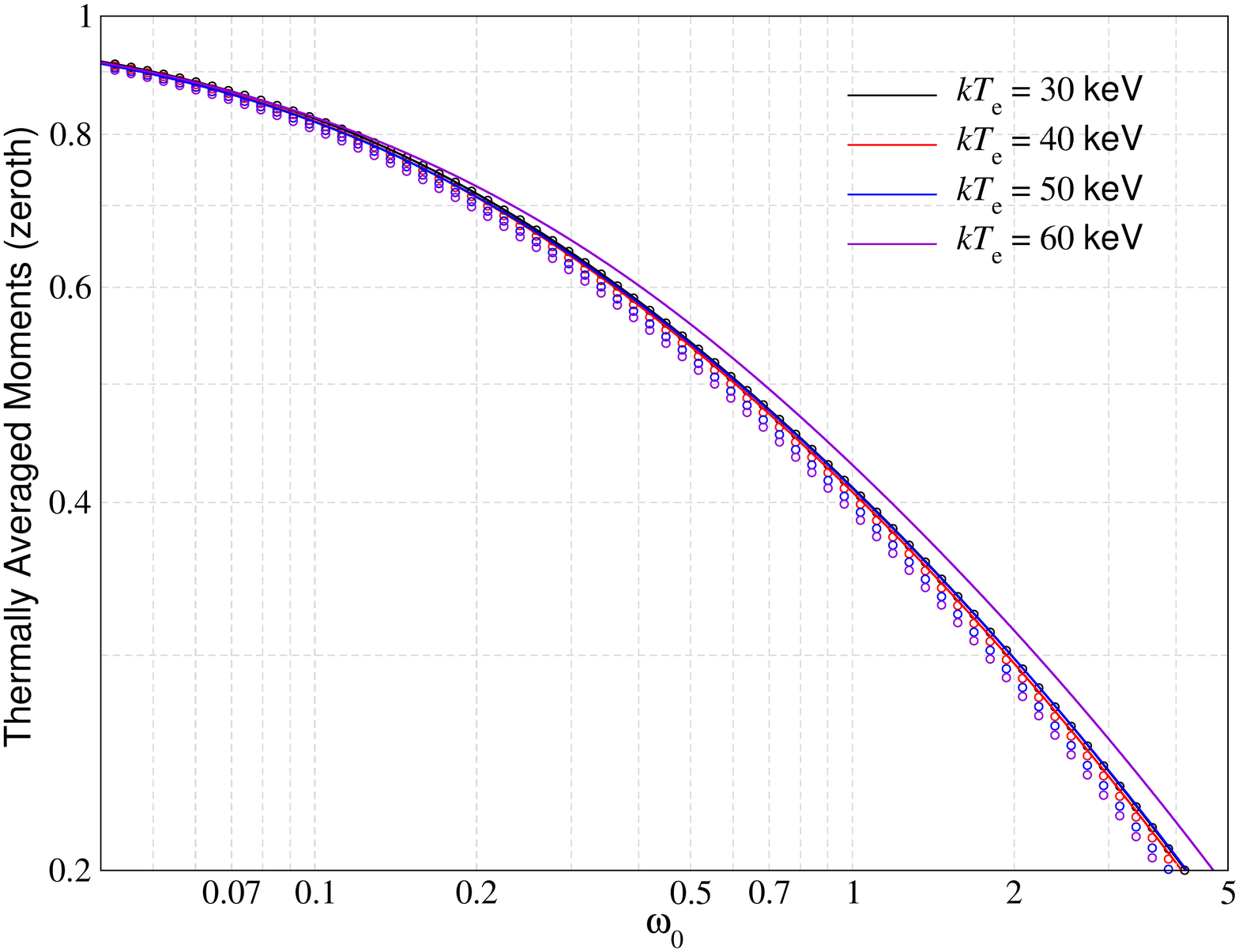}
\\[-2.5mm]
\includegraphics[width=0.96 \linewidth ]{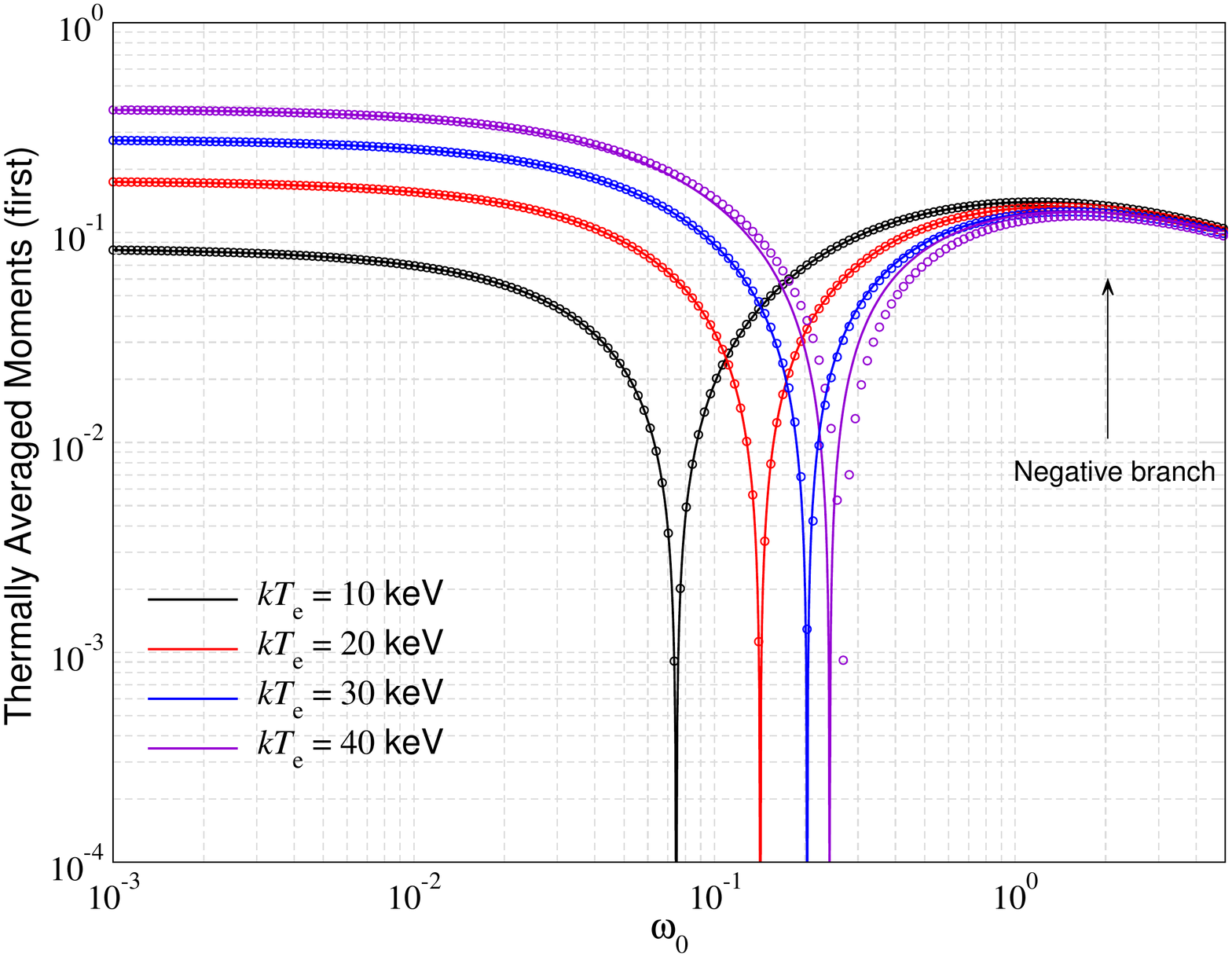}
\\[-2.5mm]
\includegraphics[width=0.96 \linewidth ]{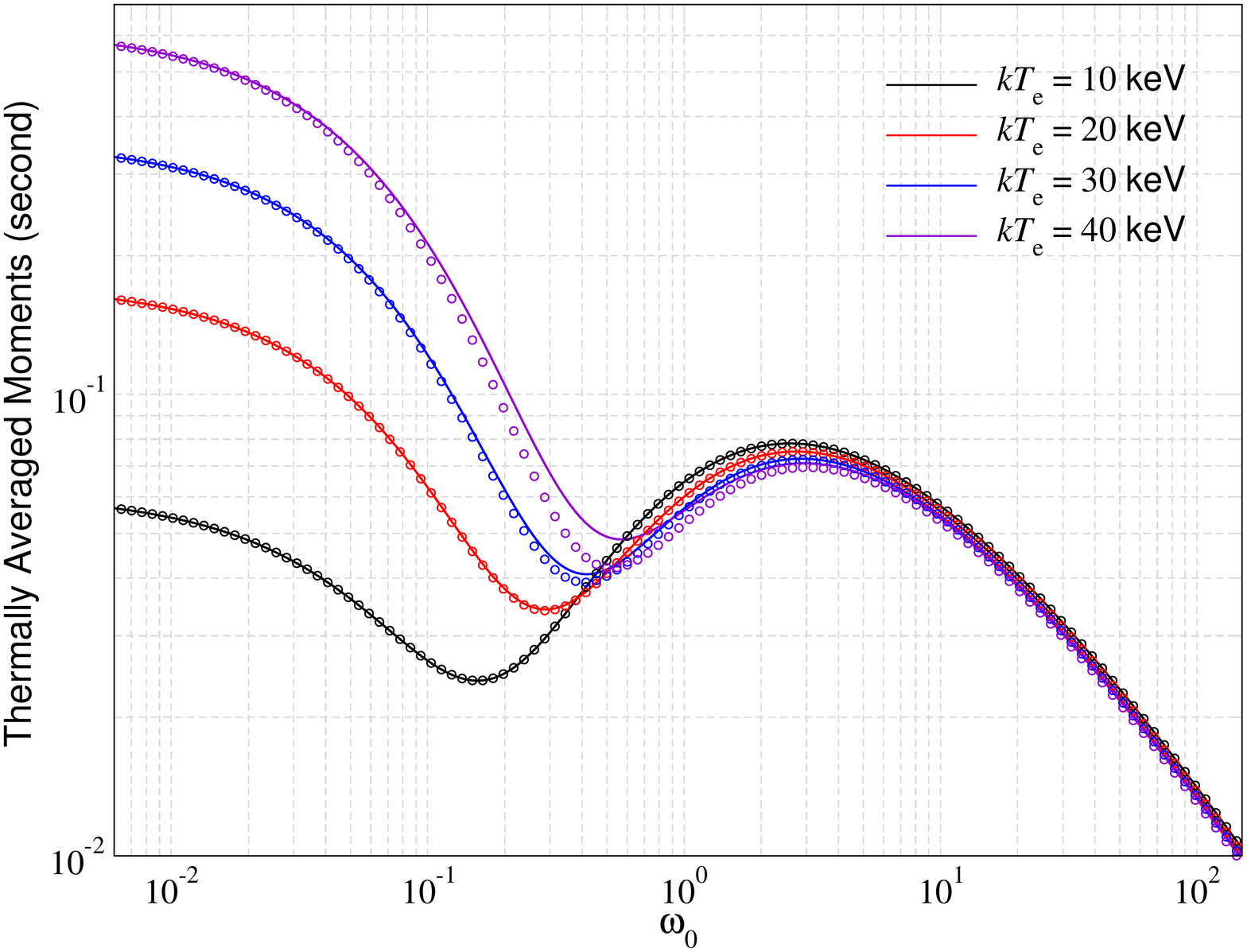}
\caption{Comparison of the approximations based on the $p_0$ series expansion for the thermally-averaged moments up to $m=2$ with the exact result for various temperatures. Terms up to $\big<p_0^8\big>$ were included for all cases. Overall very good agreement with the full numerical result is found. For the first moment, the approximations tend to start disagreeing most around the null, while for the second moment it is around the local minimum.}
\label{fig:moments_therm_Te_appr}
\vspace{-2mm}
\end{figure}

In Fig.~\ref{fig:moments_therm_Te10_appr} and \ref{fig:moments_therm_Te_appr}, we illustrate the performance of the approximations based on the $p_0$-series expansion for the thermally-averaged moments for various temperatures. In Fig.~\ref{fig:moments_therm_Te_appr}, we included terms up to $\big<p_0^8\big>$ in all cases using {\tt CSpack}. Adding higher orders did not seem to improve the convergence radius, and can even lead to worse results. The convergence of the results is best for the zeroth moment, where even for $k\Te\simeq 40$~keV accurate results are obtained. For the first moment, we find the approximation to break down around $k\Te\simeq 30$~keV, while for the second moment the match decreases at $k\Te\simeq 20$~keV (see Fig.~\ref{fig:moments_therm_Te_appr}).

We performed a detailed search regarding the minimal $\omega_0$ at which the $p_0$ series approximation breaks down. Including terms up to $\big<p_0^8\big>$, for the zeroth moment we find the approximation to work better than 10\%, 1\% and 0.1\% for $\The\lesssim 0.1, 0.07$ and $0.05$ at $\omega_0\lesssim 10^3$. For the first moment we find the respective critical temperatures to be $\The\lesssim 0.09, 0.05$ and $0.03$, although around the null the approximation is not as accurate. Finally, for the second moment we have the critical temperatures $\The\lesssim 0.07, 0.04$ and $0.026$. This demonstrates that the approach is extremely useful.

\begin{figure}
\includegraphics[width=1.0 \linewidth ]{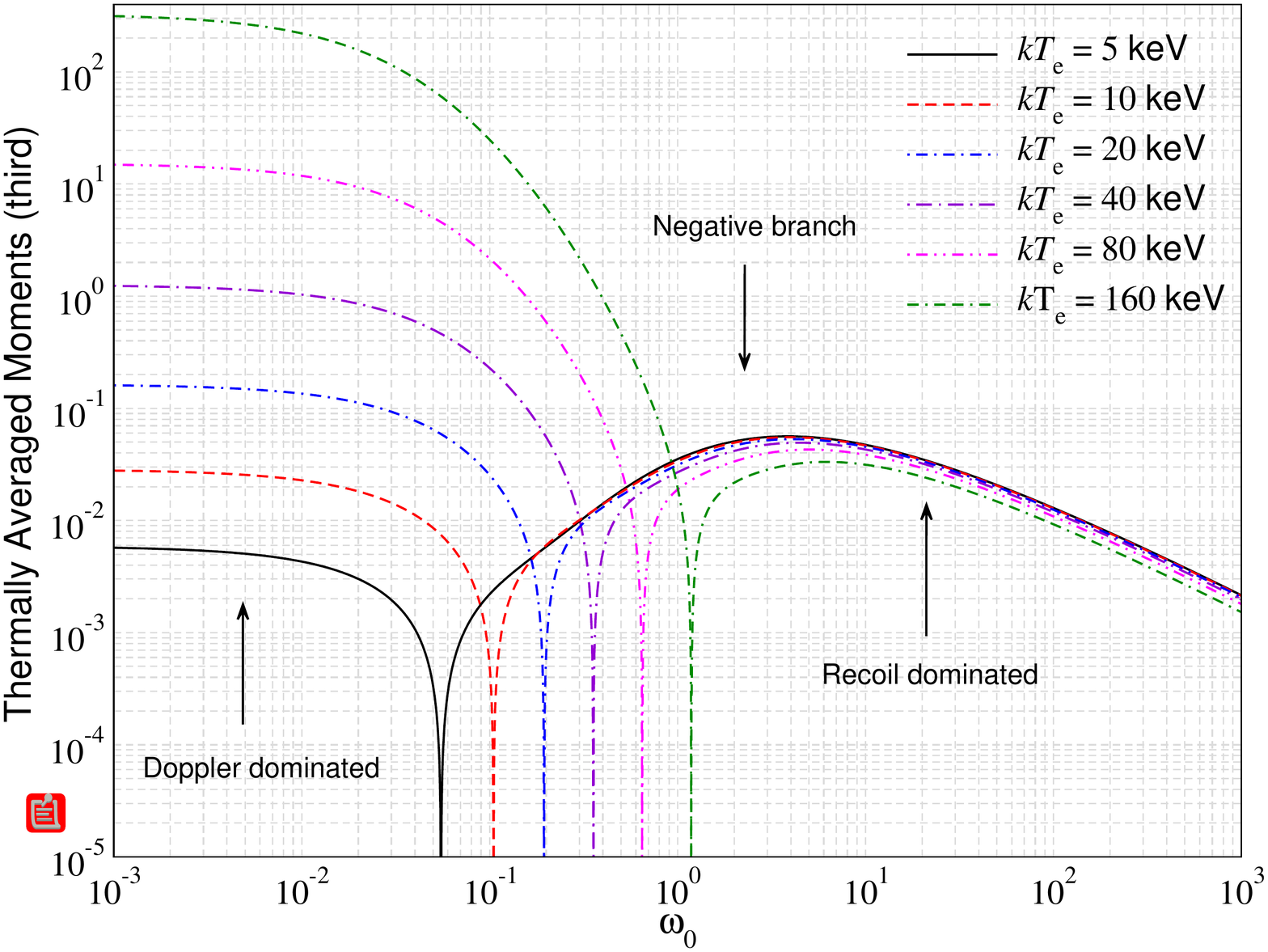}
\\[-2mm]
\includegraphics[width=1.0 \linewidth ]{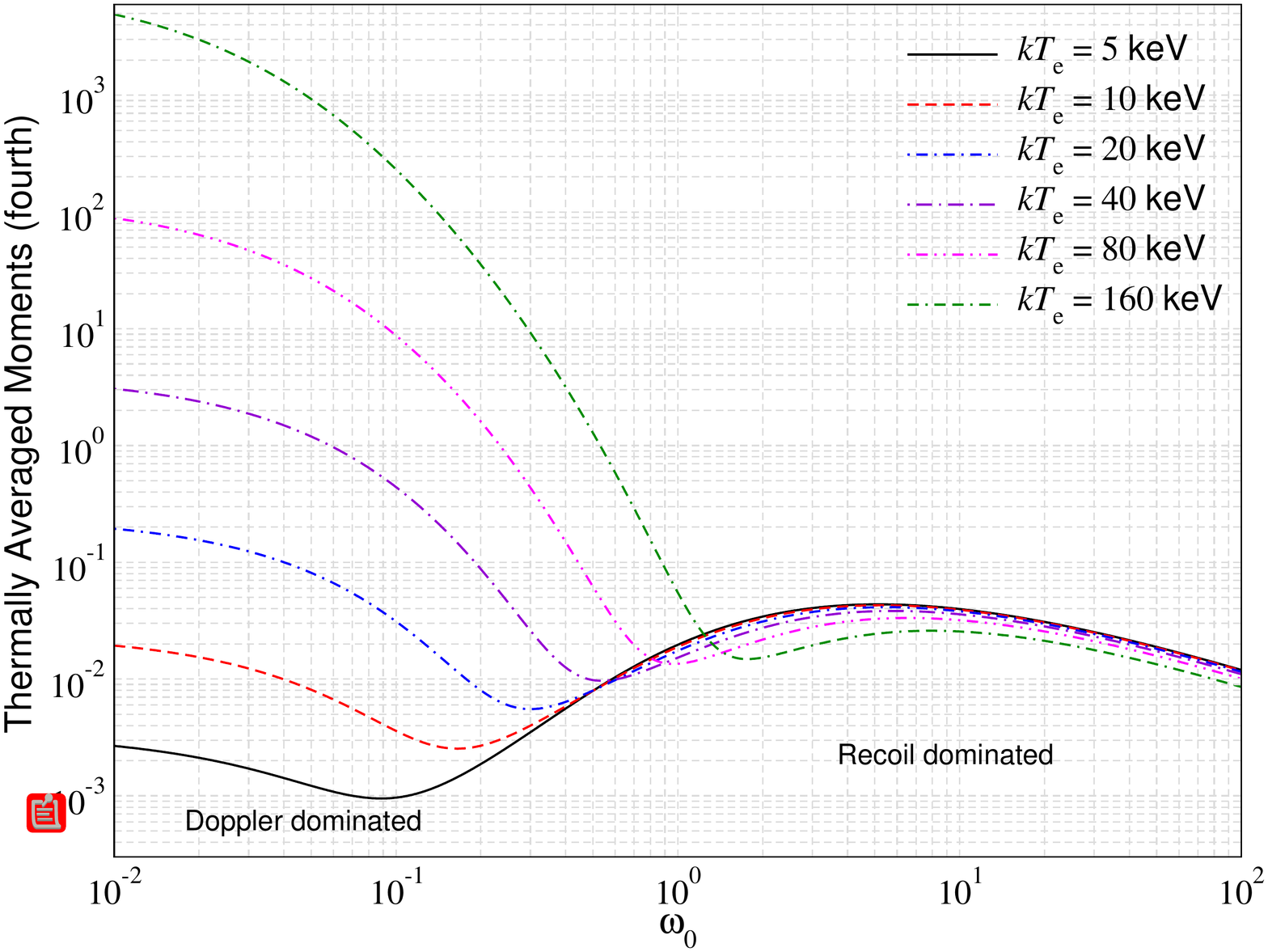}
\caption{Thermally-averaged moments for $m=3,\, 4$. These were obtained by numerically averaging Eq.~\eqref{eq:moment} over a rMB distribution function for the electrons. For the shown examples, temperature effects are more significant in the Doppler-dominated regime.
\label{fig:moments_higher_therm}}
\end{figure}

\vspace{-4mm}
\subsection{Analytic approximations introducing effective $p_0$}
We also explored approximations based on determining an effective value for $p_0$ that matched the thermally-averaged result very well when inserted into the expressions for $\Sigma_k(\omega_0,p_0)$.
One rough approximation can be obtained by simply replacing $p_0$ with the thermally-averaged value $\left<p_0\right>$. This roughly captures the global features of the thermally-averaged moments, but fails in detail. The lowest order corrections from thermal averaging are $\mathcal{O}(\The)$ so related to $\big<p^2_0\big>$. For small temperature, $\big<p_0\big>\simeq \sqrt{8\The/\pi}$ and hence $\big<p_0\big>^2\simeq  2.5\The$. This is close to $\big<p^2_0\big>\simeq  3\The$, such that it is not as surprising that this rather simplistic approach does not fail more strongly. 
A significant improvement can be achieved by setting $p_0\rightarrow\big<p^2_0\big>^{1/2}$, which works extremely well for the zeroth moment even up to $\The\simeq 1$ ($\equiv k\Te \simeq 511$~keV). For the first moment, this approach also works well until $\The\simeq 0.2$ ($\equiv k\Te \simeq 100$~keV), while for $\left<\Sigma_2\right>$ it underestimates its amplitude by $\simeq 10\%$ at low values for $\omega_0$ even for $\The\simeq 0.01$. 
To improve this, we explored additional modifications using $p_0\rightarrow \big<p^2_0\big>^{1/2}/(1+a\,\The)$. At $\omega_0\lesssim 10^3$, for better than $5\%$ precision we found $a=0.6$ to work well at $\The \lesssim 0.8$ for the zeroth moment. For the first and second moments, this approach was not successful and a frequency-dependent correction would be needed. However, we did not follow this idea any further.

\vspace{-5mm}
\subsection{Higher order thermally-averaged moments}
In Fig.~\ref{fig:moments_higher_therm}, we illustrate the thermally-averaged third (fourth) moments in the upper (lower) panel for a set of $kT_{\rm e}$. These can be useful for improved Fokker-Planck treatment of the collision term.
The third moment is negative in the recoil-dominated regime, as photons are more likely to be down-scattered. The zero-crossing point keeps moving to higher $\omega_0$ with increasing electron temperature and the position is approximately given by $\omega_0^{\rm null} \approx \frac{6\theta_{\rm e}}{\left(1+64\theta_{\rm e}\right)^{0.15}}$. This estimate works with $\simeq 10\%$ accuracy up to $\theta_{\rm e} \simeq 1$. The fourth moment shows a characteristic dip around $\omega_0^{\rm dip} \approx \frac{9.2\theta_{\rm e}}{\left(1+60\theta_{\rm e}\right)^{0.17}}$, where the transition from the Doppler- to recoil-dominated regime occurs.

\begin{figure}
\includegraphics[width=1 \linewidth ]{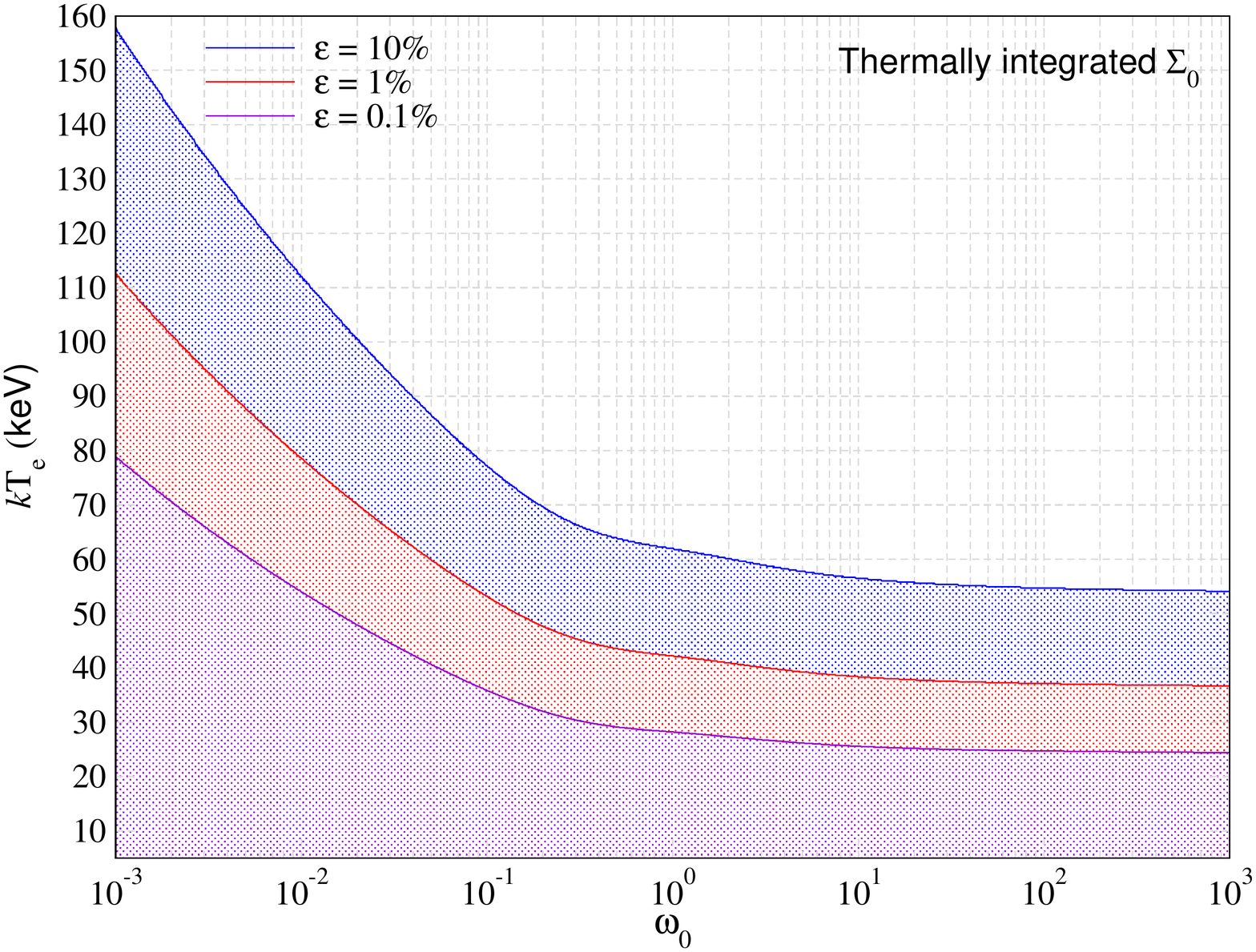}
\includegraphics[width=1 \linewidth ]{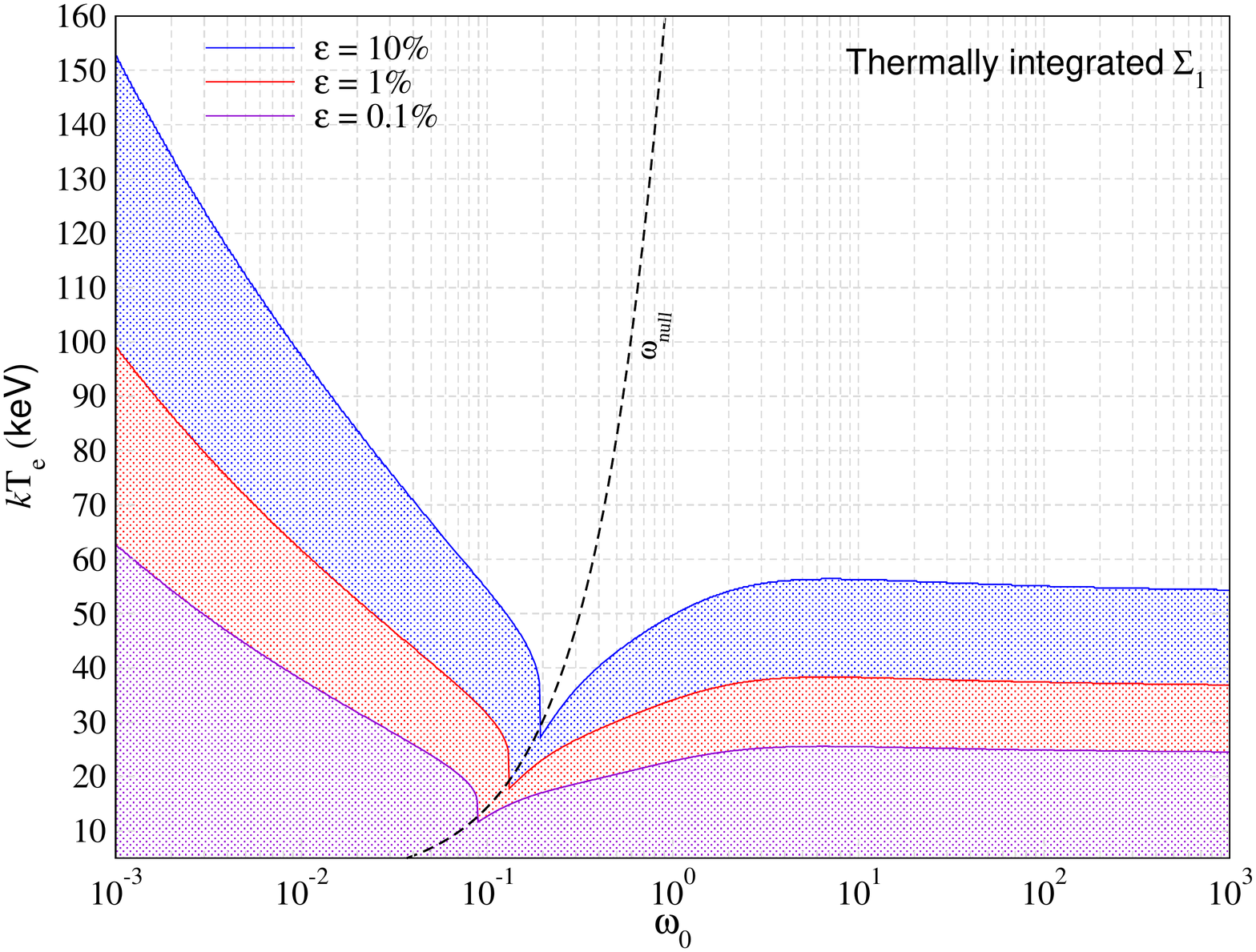}
\includegraphics[width=1 \linewidth ]{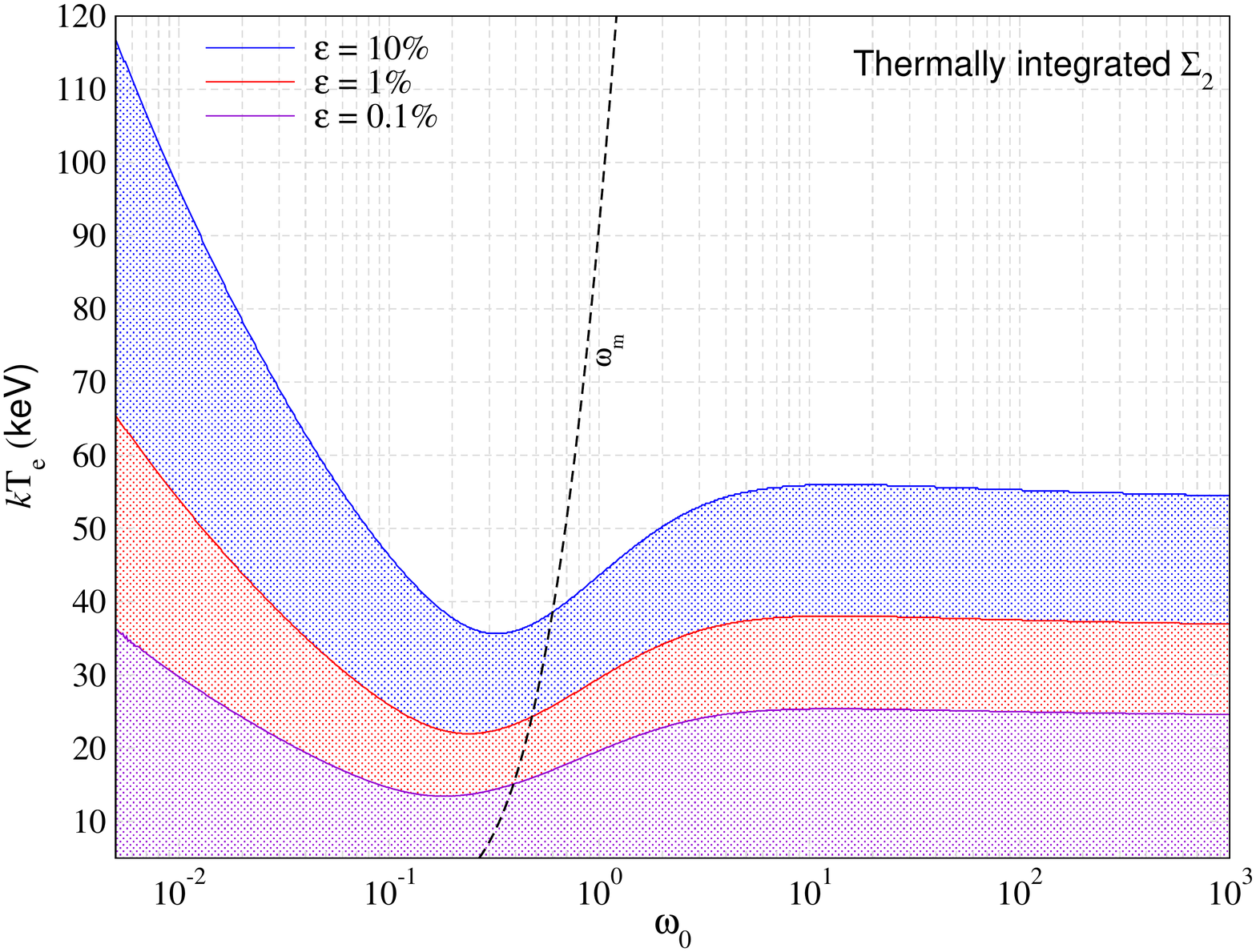}
\caption{Range of applicability for the thermally-averaged moment approximations using the $p_0$ series method described in Sect.~\ref{sec:p_expansion}. Terms up to $\langle p_0^8 \rangle$ were included in all cases using {\tt CSpack}. Inside the shaded regions the agreement with the exact result is better than $\epsilon = 0.1\%, 1\%$ and $10\%$ respectively. The simple quasi-analytic approximations cover a wide range of temperatures at all considered values of $\omega_0$. We also show the positions of the null of the first moment, $\omega_{\rm null}\approx 4\The/(1+76\,\The)^{0.1}$, and the local minimum, $\omega_{\rm m}\approx 0.27 \exp([\The/0.01]^{0.29}-1)$, of the second.} 
\label{fig:cont_te_l}
\end{figure}

\vspace{-3mm}
\subsection{Applicability of the approximations}
\label{sec:app_perform_therm}
In Fig.~\ref{fig:cont_te_l}, we illustrate the allowed regions in the $\omega_0-kT_{\rm e}$ plane for the moment approximation scheme studied in Sect.~\ref{sec:p_expansion}. In all cases, terms up to $\langle p_0^8 \rangle$ were included using {\tt CSpack}. The quasi-analytic approximation scheme works best at low values of $\omega_0$, with diminishing precision from the zeroth to the second moment. For the first moment, the deviation between the two schemes is the largest near the zero-crossing, as also seen in the central panel of Fig.~\ref{fig:moments_therm_Te_appr}. However, the approximations resume to approach the full numerical result at higher energy. A similar behaviour is found for the second moment, having largest departures around $\omega_0\simeq 0.1-0.5$. Overall the quasi-analytic approximation scheme described in Sect.~\ref{sec:p_expansion} works extremely well over a wide range of temperatures.

\begin{table}
\centering
\caption{List of expressions for the Compton scattering kernel and its first three moments discussed in this work.}
\begin{tabular}{|lcccccc} 
 \hline
 & Kernel & $\Sigma_0$ & $\Sigma_1$ & $\Sigma_2$ \\ [0.5ex] 
 \hline\hline
 Exact & 
\eqref{eq:kernel_all} & \eqref{eq:mom0_exact} & \eqref{eq:mom1_exact}  & \eqref{eq:mom2_exact}  
\\ 
 Non-relativistic
 & -- & \eqref{eq:mom0_nr} & \eqref{eq:mom1_nr} & \eqref{eq:mom2_nr} 
 \\
 Ultra-relativistic
 & -- & \eqref{eq:mom0_ur} & \eqref{eq:mom1_ur} & \eqref{eq:mom2_ur} 
 \\
 Recoil-dominated
 & \eqref{eq:P_rec} & \eqref{eq:moment_rec_a} & \eqref{eq:moment_rec_b} & \eqref{eq:moment_rec_c} 
 \\
 Doppler-dominated 
 & \eqref{eq:P_Doppler} & \eqref{eq:moment_dop} & \eqref{eq:moment_dop} & \eqref{eq:moment_dop} 
 \\ 
 Ultra-relativistic electrons 
 & \eqref{eq:P_Urel} & -- & -- & -- 
 \\ [1ex] 
 \hline
\end{tabular}
\label{table:eq_ref}
\end{table}

\vspace{-0mm}
\section{Discussion and conclusion}
\label{sec:conclusion}
We presented a strongly-simplified, numerically-stable expression for the Compton scattering kernel in an isotropic medium valid for all photon and electron energies. The properties of the kernel (e.g., its shape, symmetries and various moments) were studied and illustrated in detail, highlighting many features with an eye on their physical origin. We provided a comprehensive survey of existing approximations in the literature (see Table~\ref{table:eq_ref} for an overview), investigating their limitations in comparison to our exact expressions. We confirmed all our results using full numerical integration, finding exact agreement with our general expressions. 

By resolving the momentum conditions present in the kernel expressions given in B09 we showed that the kernel contains at most three independent energy zones (see Fig.~\ref{fig:domains}) that are separated by singular points. The expression for each zone can be obtained from a single algebraic expression by switching its arguments (see Eq.~\ref{eq:kernel_all}). For incident photon energy, $\omega_0<1/2$, the highest energy zone disappears for electron momenta $p_0\geq 2\omega_0(1-\omega_0)/(1-2\omega_0)$, while it always stays open for higher photon energies. This effect is visible in the kernel close to the maximal scattering energy at $\omega\simeq \gamma_0+\omega_0-1$ (see Fig.~\ref{fig:kernels} and Fig.~\ref{fig:kernels_comp_UR}).

In the recoil-dominated regime ($p_0\ll 1$ and $p_0\ll\omega_0$), the kernel is well represented by Eq.~\eqref{eq:P_rec} (see Fig.~\ref{fig:kernels_recoil_dom}). On the other hand, the widely-used Doppler-dominated kernel approximation ($p_0\gg \omega_0$), Eq.~\eqref{eq:P_Doppler} does not capture the behavior of the kernel at scattered photon energies $\omega>\omega_0$ once $p_0\gg 1/[4\omega_0]$ (see Fig.~\ref{fig:kernels_UR}). In this regime, recoil and Klein-Nishina corrections become important, as the photon in the boosted frame becomes highly energetic, $\omega_0'\simeq \gamma_0\omega_0\gtrsim 1$. This effect manifests itself as a pile-up of photons in the high-energy tail of the kernel (see Fig.~\ref{fig:kernels_UR}), which is reproduced by the kernel approximation for ultra-relativistic electrons, Eq.~\eqref{eq:P_Urel}. However, even this approximation fails to capture the exact shape of the kernel and the aforementioned high-energy cusp when the initial photon energy exceeds $\omega_0>1/2$ (see Fig.~\ref{fig:kernels_comp_UR}). 
While the various approximations can be useful for estimates and numerical tests, the general kernel expression should be preferred in physical applications. 

We also discussed the properties of thermally-averaged kernels for electrons following a relativistic Maxwell-Boltzmann (rMB) distribution function at various temperatures, $\The=k\Te/\me c^2$ (see Fig.~\ref{fig:kernels_therm}). We restricted ourselves to $\The < 1$ as for higher temperature Fermi-blocking of electrons and pair-production come into play (see discussion in Sect.~\ref{sec:extreme_Te}).  We provided explicit expressions for the minimal electron momentum required in a scattering event, Eq.~\eqref{eq:p0min_limits}. This eases the computation when numerically integrating the kernel over electron momenta. 

We explicitly demonstrate how electrons with different momenta contribute to the thermally-averaged scattering kernel in Fig.~\ref{fig:kernels_therm_cons}. We find that electrons with low momenta contribute mostly to the central peak of the thermally-averaged kernel, while high energy electrons can scatter the photon more strongly, producing the distant kernel wings. The shape of the kernel is strongly smoothed so that only one visible cusp at $\omega=\omega_0$ remains. Since it is difficult to provide general approximations for the thermally-averaged kernel \citep[e.g., see][]{Sazonov2000}, for numerical applications in radiative transfer calculations, it is thus best to independently tabulate the kernel at $\omega<\omega_0$ and $\omega\geq \omega_0$ if highly accurate results are required. This should allow the development of highly-efficient exact treatments of scattering problems.

We furthermore provided general analytic expressions for the first three moments of the scattering kernel, along with simpler approximations for extreme scenarios (see Table~\ref{table:eq_ref} for an overview). All of the exact expressions were derived independently and confirmed numerically. A brief discussion of the third and fourth moments is given in Fig.~\ref{fig:moments_higher_num}.
The approximations based on the recoil-dominated limit work extremely well when $p_0\ll {\rm min}(1,\omega_0)$, and even $p_0\ll {\rm max}(1, [4\omega_0]^{-1})$ for $\Sigma_0$. Similarly, our approximations for ultra-relativistic electrons match the exact result at $p_0\gg {\rm max}(1, [4\omega_0]^{-1})$ (see Fig.~\ref{fig:moment0}-\ref{fig:cont_l}). 
Just like for the kernel, we find the approximations assuming Doppler-domination to have limited applicability, requiring $\omega_0\ll p_0\ll 1/[4\omega_0]$. In particular, the net energy exchange and width of the kernel are overestimated if these approximations are applied above this domain (see Fig.~\ref{fig:moment1} and \ref{fig:moment2}).
The accuracy of the various approximations is discussed more quantitatively in Fig.~\ref{fig:cont_l}. For $\Sigma_0$, this clearly shows that the approximations are often valid somewhat beyond the expected regions, while remaining more constrained for the higher order moments.

Finally, we also discussed the first three thermally-averaged moments and various simpler approximations in Sect.~\ref{sec:th_moments}. For temperature $\The~<~1$, the zeroth moment does not depend significantly on temperature, while the other two moments show strong variations (see Fig.~\ref{fig:moments_therm}) in particular in the Doppler-dominated regime ($\omega_0\lesssim p_0$). We presented and compared three methods for approximating the thermally-averaged moments. Simple approximations assuming $\omega_0,\The\ll 1$ have a limited applicability, as these expressions (e.g., Eq.~\ref{eq:moment_therm_low}) converge very slowly (see Fig.~\ref{fig:moments_therm_Te10_appr}). A much better approach, valid up to $k\Te \simeq 20-40$~keV (depending on the considered moment) and general incident photon energy, is obtained by only assuming $p_0\ll 1$ but then keeping the remaining expressions general (see Sect.~\ref{sec:p_expansion} and Fig.~\ref{fig:moments_therm_Te_appr}). This approach should have useful applications in extended Fokker-Planck approximations of the Compton scattering problem \citep[e.g.,][]{Itoh98, Sazonov1998, Belmont2008}. It could furthermore help to improve the treatment of relativistic temperature corrections to the evolution of primordial CMB spectral distortions \citep{Chluba2005, Chluba2014}. In Fig.~\ref{fig:cont_te_l} we explicitly illustrate the range of applicability for this approach at different levels of precision. We also included a brief discussion of the third and fourth thermally-averaged moments (Fig.~\ref{fig:moments_higher_therm}), keeping in mind their importance for extended Fokker-Planck treatment of the Compton collision term.

To conclude, our analysis of the Compton scattering process should be very useful for computations of radiative transfer problems in astrophysical plasma.  High-energy electrons and photons can, for instance, be found in accretion-flows \citep[e.g.,][]{Shakura1973, Abramowicz1988, Narayan2003, McKinney2017} and electromagnetic particle cascades present during multiple phases in the evolution of the Universe \citep{Zdziarski1988, Shull1985, Slatyer2009, Valdes2010, Slatyer2015, Liu2019}.
Highly relativistic non-thermal electron populations are furthermore encountered in jets of active galactic nuclei, supernovae and $\gamma$-ray bursts \citep[e.g.,][]{Giannios2006, Mimica2009, Giannios2010}. They are also relevant to the non-thermal SZ effect \citep{Ensslin2000, Colafrancesco2003}.
Many of the common approximations can be completely avoided with the general kernel expressions given here, which fully describe the transition between the different scattering regimes at practically no extra cost. A new software package, {\tt CSpack}, for computing all the outputs regarding Compton scattering is under construction and will be made available within one month.
We hope to apply our results to many of the aforementioned problems in the future.

\vspace{-4mm}
{\small
\section*{Acknowledgments}
The authors would like to thank the referee for their comprehensive review and Andrea Ravenni for useful discussions.
This project has received funding from the European Research Council (ERC) under the European Union’s Horizon 2020 research and innovation program (grant agreement No 725456; CMBSPEC).
JC was also supported by the Royal Society as a Royal Society University Research Fellow at the University of Manchester, UK.
EL was supported by the Royal Society on grant No RGF/EA/180053.
}

{\small
\bibliographystyle{mn2e}
\bibliography{Lit}
}

\begin{appendix}

\small
\section{The Compton scattering kernel in terms of the scattering matrix element}
\label{sec:app:kerdef}
In this section, we will discuss the derivation of the scattering kernel starting from the kinetic equation of the photon occupation number, Eq.~\eqref{eq:kin_eq1}, which we repeat here for convenience:
\begin{align}
\label{app:kin_eq1}
\frac{1}{c}\frac{\text{d}n(\omega_0)}{\text{d}t}  
&= \frac{1}{2 E_{\gamma_0}}\!\int \!\frac{\text{d}^3p_0}{(2\pi)^32E_0}  \frac{\text{d}^3p}{(2\pi)^32E} \frac{\text{d}^3k}{(2\pi)^32E_{\gamma}}  
\nonumber
\\[1mm] 
\nonumber
&\quad\qquad\times
(2\pi)^4\delta^{(4)}(p+k-p_0-k_0) \,\lvert \mathcal{M} \rvert ^2 
\\[1mm] 
&\quad\qquad\qquad\times\Big[f n(1+n_0) - f_0 n_0(1+n)\Big].
\end{align}
Following \cite{Jauch1976}, the squared matrix element of the Compton process, $|\mathcal{M}|^2$, is given by
\begin{align}
|\mathcal{M}|^2 &= e^4 X = 2 e^4 \bar{X}, \quad
\bar{X} &= \frac{\kappa'}{\kappa} + \frac{\kappa}{\kappa'} + 2\left(\frac{1}{\kappa} - \frac{1}{\kappa'}\right) + \left(\frac{1}{\kappa} - \frac{1}{\kappa'}\right)^2.
\end{align}
Here $\bar{X}$ is averaged over photon polarization states and $e$ is the electron charge. 
The important four-vector invariants are
\bsub
\begin{align}
\kappa &=-\frac{p_0\cdot k_0}{m_{\rm e}^2 c^2} =-\frac{p\cdot k}{m_{\rm e}^2 c^2}=-\gamma_0 \omega_0 (1-\beta_0 \mu_0)
\\
\kappa'&=-\frac{p_0\cdot k}{m_{\rm e}^2 c^2} =-\frac{p\cdot k_0}{m_{\rm e}^2 c^2}=-\gamma_0 \omega (1-\beta_0 \mu)
\\
 \kappa'-\kappa &= \frac{k_0 \cdot k}{m_{\rm e}^2 c^2}=\omega_0\omega(1-\mu_{\rm sc}),
\end{align}
\esub
where the $\mu_i$ denote the corresponding direction cosines. The direction cosine between the incident electron and scattered photon can be eliminated using $\mu = \mu_0\mu_{\rm{sc}} + \cos[\phi_0-\phi_{\rm{sc}}](1-\mu_0^2)^{1/2} (1-\mu_{\rm{sc}}^2)^{1/2}$.
With these definitions, $\bar{X}$ can now be expressed as
\begin{align}
\bar{X} &= \frac{\kappa'}{\kappa} + \frac{\kappa}{\kappa'} - \frac{2(1-\mu_{\rm{sc}})}{\gamma_0^2(1-\beta_0\mu_0)(1-\beta_0\mu)} + \frac{(1-\mu_{\rm{sc}})^2}{\gamma_0^4(1-\beta_0\mu_0)^2(1-\beta_0\mu)^2}
\nonumber 
\\[0mm] 
\label{app:energy_omega}
\frac{\kappa'}{\kappa} &= \frac{\omega}{\omega_0} \frac{1-\beta_0\mu}{1-\beta_0\mu_0},
\qquad 
\frac{\omega}{\omega_0} = \frac{1-\beta_0\mu_0}{1-\beta_0\mu} \frac{1}{1+\frac{\omega_0}{\gamma_0 }\frac{1-\mu_{\rm{sc}}}{1-\beta_0\mu}}.
\end{align}
In Eq.~\eqref{app:kin_eq1}, we first carry out the integration over the scattered electron momenta, $\id^3 p$, making use of the Dirac $\delta$-function. This eliminates the three-vector, $\vek{p}$, resulting in $\vek{p}=\vek{p}_0+\vek{k}_0-\vek{k}$ and $\gamma=\gamma_0+\omega_0-\omega$ everywhere. We are then left with
\begin{align}
\frac{1}{c}\frac{\text{d}n(\omega_0)}{\text{d}t}  
&= \frac{e^4}{2^3\,(2\pi)^2\,(\me c^2)^2}\!\int \!\frac{\text{d}^3p_0}{(2\pi)^3} \,\text{d}^3k 
\;\delta(\gamma+\omega-\gamma_0-\omega_0)
\nonumber
\\[0mm] 
\nonumber
&
\qquad \times \frac{\bar{X}}{\gamma_0\,\gamma \, \omega_0\, \omega} 
\times \Big[f n(1+n_0) - f_0 n_0(1+n)\Big],
\end{align}
where we collected factors and also transformed to dimensionless variables. We next replace $e^4\rightarrow (4\pi)^2\,\me^2c^4\,r_0^2$, where $r_0$ is the classical electron radius. Since the Thomson cross section is given by $\sigT=8\pi r_0^2 / 3$, we have 
\begin{align}
\frac{1}{c}\frac{\text{d}n(\omega_0)}{\text{d}t}  
&= \frac{r_0^2}{2} \!\int \!\frac{\text{d}^3p_0}{(2\pi)^3} \,\text{d}^3k 
\;\delta(\gamma+\omega-\gamma_0-\omega_0)
\nonumber
\\[0mm] 
\nonumber
&
\qquad \times \frac{\bar{X}}{\gamma_0\,\gamma \, \omega_0\, \omega} 
\times \Big[f n(1+n_0) - f_0 n_0(1+n)\Big]
\\
\nonumber
&= \sigT \Ne \!\int \!\frac{\text{d}^3p_0}{(2\pi)^3\,\Ne} \int \text{d}^3k 
\;\delta(\gamma+\omega-\gamma_0-\omega_0)
\nonumber
\\[0mm] 
\nonumber
&
\qquad \times \frac{3\bar{X}}{16\pi \,\gamma_0\,\gamma \, \omega_0\, \omega} 
\times \Big[f n(1+n_0) - f_0 n_0(1+n)\Big].
\end{align}
Here, we introduced the electrons number density, $\Ne = \int \frac{\text{d}^3p_0}{(2\pi)^3} f(\vek{p}_0)$. 

The customary way forward is to carry out the integral over $\id\omega$ using $\id (\gamma + \omega)/\id \omega=\frac{\gamma_0\omega_0}{\gamma\omega}(1-\beta_0\mu_0)$ \citep[see][]{Jauch1976}. Aligning the $z$-axis with the direction of the incident photon, this then yields the kinetic equation in the form
\begin{align}
\label{app:Boltz_dsigma}
\frac{\text{d}n(\omega_0)}{\text{d}\tau}  
&= \int \!p_0^2 \text{d}p_0  
\int \frac{\text{d}\mu_0\,\text{d}\phi_0\,\text{d}\mu_{\rm sc}\text{d}\phi_{\rm sc}}{4\pi}\,(1-\beta_0\mu_0)\,\frac{\text{d}\sigma}{\text{d}\Omega}
\nonumber
\\
&
\qquad\qquad \times \Big[\tilde{f} n(1+n_0) - \tilde{f}_0 n_0(1+n)\Big],
\end{align}
where $\tau=\int c\Ne \sigT \id t$ is the Thomson optical depth. We furthermore introduced the differential Compton scattering cross section (in units of $\sigT$)
\begin{equation}
\label{eq:App_kerdef2}
\frac{\text{d}\sigma}{\text{d}\Omega} = \frac{3}{16\pi}\,\Bigg[\frac{\omega}{\omega_0}\Bigg]^2\frac{\bar{X}}{\gamma_0^2(1-\beta_0\mu_0)^2},
\end{equation}
and also renormalized the electron distribution function, such that $\int \frac{\text{d}^3p_0}{(2\pi)^3\,\Ne} f(\vek{p}_0)=\int \frac{p_0^2 \text{d}p_0}{2\pi^2\,\Ne} f(\gamma_0)=1$, meaning $\tilde{f}=f/(2\pi^2 \Ne)$. 

In the following we shall simply replace\footnote{We cordially thank Andrea Ravenni for clarifying this fact with us.} $\tilde{f}\rightarrow f$, bearing this in mind. In the limit, $p_0,\omega_0\ll 1$, one has $\bar{X}\approx (1+\mu_{\rm sc}^2)$, such that $\int \frac{\text{d}\mu_0\,\text{d}\phi_0\,\text{d}\mu_{\rm sc}\text{d}\phi_{\rm sc}}{4\pi}\,\frac{\text{d}\sigma}{\text{d}\Omega}\approx 1$, as expected. This also highlights the importance of the factor of $2$ for $|\mathcal{M}|^2 = 2 e^4 \bar{X}$.

\subsection{Kernel formulation of the scattering problem}
To obtain the formulation of the kinetic equation using the scattering kernel, instead of carrying out the integral over $\id \omega$ we select the integral over $\id \phi_{\rm sc}$. Regrouping terms, we then have
\begin{align}
\label{app:kernel_def_0}
\frac{\text{d}n(\omega_0)}{\text{d}\tau}  
&= \!\int \! p_0^2 \text{d}p_0 \!\int \!\!\id \omega \,P(\omega_0\rightarrow \omega, p_0)
\,\Big[f n(1+n_0) - f_0 n_0(1+n)\Big]
\nonumber
\\[1mm]
P(\omega_0\rightarrow \omega, p_0)
&=\int \frac{\text{d}\mu_0\,\text{d}\phi_0\,\text{d}\mu_{\rm sc}\text{d}\phi_{\rm sc}}{4\pi}
\,\frac{3\bar{X}\omega \,\delta(\gamma+\omega-\gamma_0-\omega_0)}{16\pi \,\gamma_0\,\gamma \, \omega_0}
\end{align}
This expression can be further simplified by eliminating the Dirac $\delta$-function. For this we need $\id (\gamma+\omega-\gamma_0-\omega_0)/\id \phi_{\rm sc}=\id \gamma/\id \phi_{\rm sc}=-\frac{p_0 \omega}{\gamma}\,\!\id \mu/\id \phi_{\rm sc}$ at fixed $\omega, \omega_0$ and $p_0$. With Eq.~\eqref{app:energy_omega} and 
\begin{align}
\frac{\!\id \mu}{\id \phi_{\rm sc}}
&=\sin[\phi_0-\phi_{\rm{sc}}](1-\mu_0^2)^{1/2} (1-\mu_{\rm{sc}}^2)^{1/2}
\nonumber
\\
\nonumber
\cos[\phi_0-\phi_{\rm{sc}}]
&=\frac{\gamma_0(\omega-\omega_0) + p_0 \mu_0 (\omega_0-\omega \,\mu_{\rm sc}) + \omega_0\omega(1-\mu_{\rm sc})}
{p_0 \omega(1-\mu_0^2)^{1/2} (1-\mu_{\rm{sc}}^2)^{1/2}}
\\
\label{app:mu_elim}
\mu_{\rm t}&\equiv \frac{\gamma_0(\omega-\omega_0) + p_0 \omega_0 \mu_0 + \omega_0\omega(1-\mu_{\rm sc})}{p_0 \omega}
\end{align}
we then readily find
\begin{align}
\frac{\!\text{d}\gamma}{\text{d}\phi_{\rm sc}} 
\label{eq:App_dphscpom}
&=-\frac{p_0 \omega}{\gamma}\,
\Big[(1-\mu_0^2 ) (1-\mu_{\rm sc}^2 )
-\big( \mu_t - \mu_0\mu_{\rm sc}\big)^2
\Big]^{1/2}.
\end{align}
Due to the symmetries, the argument of the $\delta$-function has two solutions when varying $\phi_{\rm sc}$, such that $\int \delta(\gamma+\omega-\gamma_0-\omega_0) \id \phi_{\rm sc} \rightarrow 2|\id \phi_{\rm sc}/\!\id \gamma|$ and $\mu\rightarrow \mu_t$ everywhere. 
Thus, the final expression for the kernel reads
\begin{align}
P(\omega_0\rightarrow \omega, p_0)
&=\int \frac{\text{d}\mu_0\,\text{d}\phi_0\,\text{d}\mu_{\rm sc}}{4\pi}\,2\left|\frac{\text{d}\phi_{\rm sc}}{\id\gamma}\right|
\,\frac{3\bar{X}\omega}{16\pi \,\gamma_0\,\gamma\,\omega_0}
\label{eq:simplified_kernel}
\\[1mm]
\nonumber
&\!\!\!\!=\frac{3}{16\pi\,\gamma_0 p_0\omega_0}\!\int\!
\frac{\text{d}\mu_0\,\text{d}\mu_{\rm sc}\,\bar{X}}{
\!\sqrt{(1-\mu_0^2 ) (1-\mu_{\rm sc}^2 )
-\big(\mu_{\rm t} - \mu_0\mu_{\rm sc}\big)^2}}.
\end{align}
with $\mu_{\rm t}$ as in Eq.~\eqref{app:mu_elim}. Since after the elimination of $\cos[\phi_0-\phi_{\rm{sc}}]$ the remaining expression no longer depends on $\phi_0$, in the second line we directly carried out the integral over $\id \phi_0$. 

\vspace{-4mm}
\subsubsection{Numerical evaluation of the kernel}
The integrals over $\id \mu_0$ and $\id \mu_{\rm sc}$ have to be performed to obtain the desired kernel. The limits of the integration can be found by requiring that $\frac{\text{d}\phi_{\rm sc}}{\text{d}\gamma}$ must be real. In other words, the quantity inside the square root of Eq.~\eqref{eq:App_dphscpom} must be greater than or equal to zero. Solving for this, the following limits for $\mu_{\rm sc}$ can be derived:
\begin{subequations}
\begin{align}
\mu_{\rm sc}^{\rm min} 
&= \text{max}\left[-1, \frac{\mathcal{L}_{\mu_{\rm sc1}} - \mathcal{L}_{\mu_{\rm sc2}}}{\omega \lambda_0} \right], 
\\[0.5mm]
\mu_{\rm sc}^{\rm max} 
&= \text{min}\left[1, \frac{\mathcal{L}_{\mu_{\rm sc1}} + \mathcal{L}_{\mu_{\rm sc2}}}{\omega \lambda_0} \right],
\\[0.5mm]
\mathcal{L}_{\mu_{\rm sc1}} 
&= \lambda_1 (\omega_0 + p_0 \mu_0 ),
\\[0.5mm]
\mathcal{L}_{\mu_{\rm sc2}} 
&= p_0 \sqrt{ (1-\mu_0^2 )(\lambda_0\omega^2-\lambda_1^2)}.
\\[0.5mm]
\lambda_0&=p_0^2 + 2 p_0\mu_0 \omega_0 + \omega_0^2
\\[0.5mm]
\lambda_1&=\gamma_0 (\omega - \omega_0) + p_0 \omega_0 \mu_0+\omega_0 \omega
\label{eq:App_musclim}
\end{align}
\end{subequations}
The limits for the integration over $\mu_0$ can be found in a similar way requiring that the limits for $\mu_{\rm sc}$ have to be real. No special condition arises from $\lambda_0$, which does not vanishes inside the range $\mu_0\in [-1,1]$; however, for $p_0=\omega_0$ it vanishes on the boundary at $\mu_0=-1$, indicating that this case is special.
Requiring that $\mathcal{L}_{\mu_{\rm sc2}}$ remains real, one finds 
\bsub
\begin{align}
\mu^{\rm min} &= \text{max}\Bigg[-1, \frac{\gamma_0 \omega_0 - (\gamma+p) \omega}{p_0\omega_0}\Bigg], 
\\[1mm]
\mu^{\rm max} &= \text{min}\Bigg[1, \frac{\gamma_0 \omega_0 - (\gamma-p) \omega}{p_0\omega_0}\Bigg], 
\label{eq:App_mulim}
\end{align}
\esub
We numerically calculated the kernel using the prescription mentioned above and compared the result with the expression given in Sect.~\ref{sec:ker_domains}, finding excellent agreement.

\vspace{-3mm}
\subsubsection{Important properties of the kernel}
\label{app:symmetries}
How does the kernel for the forward direction, $P(\omega_0\rightarrow \omega, p_0)$, relate to the one of the backward direction, $P(\omega\rightarrow \omega_0, p)$? Since $\gamma=\gamma_0+\omega_0-\omega$ in the interaction and since the number of photons has to be conserved, we can thus write 
\begin{align}
\nonumber
P(\omega_0\rightarrow \omega, p_0)\omega_0^2 \id \omega_0 p_0^2 \id p_0 \id \omega 
\equiv P(\omega\rightarrow \omega_0, p)\omega^2 \id \omega p^2 \id p \id \omega_0
\end{align}
which then implies
\begin{align}
P(\omega_0\rightarrow \omega, p_0)
&= \frac{\omega^2 }{\omega_0^2} \frac{p^2\id p}{p_0^2\id p_0}
P(\omega\rightarrow \omega_0, p)
\nonumber
\\
\label{app:kernel_prop}
&= \frac{\omega^2}{\omega_0^2}\frac{\gamma p}{\gamma_0 p_0}\,P(\omega \rightarrow \omega_0, p) 
\end{align}
with $p=\sqrt{\gamma^2-1}=\sqrt{(\gamma_0+\omega_0-\omega)^2-1}$ as usual. We confirmed this important relation using the full kernel. It can also be read-off from the explicit form of the kernel, Eq.~\eqref{eq:P_ww0<1} and \eqref{eq:P_ww0>1}.

\section{Summary and further reduction of expressions from B09}
\label{app:Belmont}
\citet{2009A&A...506..589B} worked out improved expressions for the general Compton scattering kernel. We shall follow their notation closely here, but will redefine a few functions for convenience. We also removed all switches [e.g., present in Eq.~\eqref{eq:xaplus}, \eqref{eq:xbminus} and \eqref{eq:Delta_def}] in the final expression to ease the physical interpretation, numerical stability and other computations. According to Eq.~(27) and (28) of B09, the Compton scattering kernel for the scattering of a photon with energy $\omega_0=h\nu_0/\me c^2$ by an electron with momentum $p_0=(\gamma^2_0-1)^{1/2}$ can be written as
\begin{subequations}
\begin{align}
\mathcal{P}(\omega_0 \rightarrow \omega, p_0)
&=\frac{3\sigT \Delta}{8\gamma_0 p_0 \omega_0^2} G(\omega_0, \omega, p_0)
\\
\label{eq:G_redef_B09}
G(\omega_0, \omega, p_0) &=
2+(1+\omega\omega_0)\left[z_+ +z_- - \frac{2}{\omega\omega_0}\right]
\\ \nonumber
&\quad
+2\left[\!\sqrt{z} \,\mathcal{S}\!\left(\lambda z \Delta^2 \right)\right]^+_-
+ (1+\omega\omega_0) \!\left[\frac{\sqrt{z}}{\lambda}\,\mathcal{F}\!\left(\lambda z \Delta^2 \right) \right]^+_- .
\end{align}
Here $z_\pm=x^a_\pm x^b_\pm$ with the auxiliary functions
\begin{align}
\label{eq:xaplus}
x^a_+ &= {\rm min}\left[\frac{\gamma+p}{\omega}, \frac{\gamma_0+p_0}{\omega_0}\right],
\quad x^a_-=\frac{1}{\omega\omega_0 x^a_+}
\\
\label{eq:xbminus}
x^b_- &= {\rm min}\left[\frac{\gamma+p}{\omega_0}, \frac{\gamma_0+p_0}{\omega}\right],
\quad x^b_+=\frac{1}{\omega\omega_0 x^b_-}
\\
\label{eq:lamplus}
\lambda_{+}&=(\gamma_0+\omega_0)^2-1=(\gamma+\omega)^2-1,
\\
\label{eq:lamminus}
\lambda_{-}&=(\gamma_0-\omega)^2-1=(\gamma-\omega_0)^2-1,
\\[1mm]
\mathcal{S}(x)
&=\frac{\sinh^{-1}\!\!\sqrt{x}}{\sqrt{x}}
\equiv\frac{\sin^{-1}\!\!\sqrt{-x}}{\sqrt{-x}}
\\
\label{eq:def_F}
\mathcal{F}(x) &= \mathcal{S}(x) - \sqrt{1+x}.
\end{align}
Most importantly, the variable $\Delta$ is given by
\begin{align}
\label{eq:Delta_def}
\Delta(\omega_0, \omega, p_0, p)
&= {\rm min}\left[{\rm min}[p, p_0], \Lambda(\omega_-, \omega_+, p_0, p) \right]
\\[1mm]
\nonumber
\Lambda(\omega_0, \omega, p_0, p)
&=\frac{\gamma+\gamma_0+p+p_0}{2(p+p_0)}\left(\omega-\frac{\omega_0}{(\gamma+p)(\gamma_0+p_0)}\right),
\end{align}
\end{subequations}
with $\omega_-={\rm min}[\omega,\omega_0]$ and $\omega_+={\rm max}[\omega,\omega_0]$. In comparison to B09 a few changed were introduced. Firstly, the terms of the function $S(x)=\frac{1}{x}[\mathcal{S}(x)-1/\sqrt{1+x}]$ given by Eq.~(25) of B09 were rearranged to isolate $\mathcal{S}(x)$ for the parts without overall pole $\propto 1/\lambda_-$. For those terms with leading pole $\propto 1/\lambda_-$ we replaced $S(x)=\frac{1}{x}[\mathcal{F}(x)+\sqrt{1+x}-1/\sqrt{1+x}]$. This allowed eliminating some of the extra terms in Eq.~(28) of B09, leading to the shorter form of $G$ according to Eq.~\eqref{eq:G_redef_B09}. 
We also used the identity $ x_+ x_-=(\omega\omega_0)^{-1}$ to write $z_+=x^a_+/(\omega\omega_0 x^b_-)$ and $z_-=x^b_-/(\omega\omega_0 x^a_+)$ and hence $z_+ + z_- =[x^a_+ / x^b_- + x^b_-/ x^a_+ ]/(\omega\omega_0)$. Both changes ease the following discussion significantly.

\subsection{Further simplification of the kernel expression}
To further simplify the kernel expression of B09 (i.e, remove all the conditions), we only need to focus on the variables $x_a^+$ and $x_b^-$, as given by Eq.~\eqref{eq:xaplus} and ~\eqref{eq:xbminus}. In the various energy zones (see Sect.~\ref{sec:energy_zones}), these can take the following values
\begin{subequations}
\begin{align}
\label{xb}
x^a_+ &=
\begin{cases}
\frac{\gamma_0+p_0}{\omega_0} &\text{for}\quad \omega_{\rm min}\leq \omega<\omega_0
\\
\frac{\gamma+p}{\omega} &\text{for}\quad \omega_0 \leq \omega \leq \omega_{\rm max}
\end{cases},
\\[1mm]
x^b_- &=
\begin{cases}
\frac{\gamma+p}{\omega_0} &\text{for}\quad \omega_{\rm min}\leq \omega<\omega_{\rm I}
\\
\frac{\gamma_0+p_0}{\omega} &\text{for}\quad \omega_{\rm I}\leq \omega < \omega_{\rm II}
\\
\frac{\gamma+p}{\omega_0} &\text{for}\quad \omega_{\rm II}\leq \omega \leq \omega_{\rm max}
\end{cases},
\end{align}
\end{subequations}
where the zone boundaries, $\omega_{\rm I}$ and $\omega_{\rm II}$, are $\omega_{\rm I}={\rm min}(\omega_{\rm c}, \omega_0)$ and $\omega_{\rm II}={\rm max}(\omega_{\rm c}, \omega_0)$. For $p_0=\omega_0$ one has $\omega_{\rm I}=\omega_{\rm II}=\omega_0$, so that zone II is not present in this case. This shows again that one singular point is always found at $\omega=\omega_0$, while the other is caused solely by conditions in the upper integration boundary at $\omega = \omega_{\rm c}$.

Assuming $\omega_{\rm c}\leq\omega_0$ (or $p_0\leq\omega_0$) and introducing the energies
\begin{subequations}
\begin{align}
\bar{\omega}&=\sqrt{\frac{\omega\omega_0(\gamma+p)}{\gamma_0+p_0}},
\quad \bar{\omega}_0=\sqrt{\frac{\omega\omega_0(\gamma_0+p_0)}{\gamma+p}},
\end{align}
\end{subequations}
with identity $\bar{\omega}\bar{\omega}_0 \equiv \omega \omega_0$ one can then write
\begin{subequations}
\begin{align}
\label{zdef}
z_+ &=
\begin{cases}
\frac{1}{\bar{\omega}^2} &\text{for}\quad \omega_{\rm min}\leq \omega<\omega_{\rm c}
\\
\frac{1}{\omega_0^2} &\text{for}\quad \omega_{\rm c}\leq \omega < \omega_0
\\
\frac{1}{\omega^2} &\text{for}\quad \omega_0\leq \omega\leq \omega_{\rm max}
\end{cases},
\\[-0.5mm]
z_- &=
\begin{cases}
\frac{1}{\bar{\omega}^2_0} &\text{for}\quad \omega_{\rm min}\leq \omega<\omega_{\rm c}
\\
\frac{1}{\omega^2} &\text{for}\quad \omega_{\rm c}\leq \omega < \omega_0
\\
\frac{1}{\omega_0^2} &\text{for}\quad \omega_0\leq \omega\leq \omega_{\rm max}
\end{cases},
\\[-0.5mm]
\Delta &=
\begin{cases}
\Lambda(\omega_0, \omega, p_0, p) &\text{for}\quad \omega_{\rm min}\leq\omega<\omega_{\rm c}
\\
p_0 &\text{for}\quad \omega_{\rm c}\leq \omega < \omega_0
\\[-0.5mm]
p &\text{for}\quad \omega_0\leq \omega\leq \omega_{\rm max}
\end{cases}.
\end{align}
\end{subequations}
Inserting this into Eq.~\eqref{eq:G_redef_B09}, for given $\omega_0$ and $\omega$ we then find one convenient form for the auxiliary function, $G$, as given in Eq.~\eqref{eq:kernel_main}. 

Similarly, for $\omega_{\rm c}>\omega_0$ (or $p_0>\omega_0$), by explicitly writing the cases for $x_+^a$ and $x_-^b$ as well as $z_\pm$ and $\Delta$, one can again show that the same function $G$, as given in Eq.~\eqref{eq:kernel_main}, is applicable. These findings then lead to the final kernel expressions, Eq.~\eqref{eq:P_ww0<1} and \eqref{eq:P_ww0>1}.

\vspace{-5mm}
\subsection{Useful identities}
For analytic derivations, we note that $(\gamma+p)^{-1}=(\gamma-p)$. We also have the identity $z_+ z_-=(\omega \omega_0)^{-2}$. With this we can write
\begin{align}
\label{eq:Lambda_rewrite}
\Lambda(\omega_0, \omega, p_0, p)
&=\frac{\gamma_0(\omega-\omega_0)}{(p+p_0)}-\frac{(\omega-\omega_0)^2}{2(p+p_0)}+\frac{\omega+\omega_0}{2},
\end{align}
which is convenient for Taylor series expansions. Another useful identity is
\begin{align}
\label{eq:Lambda_rewrite_II}
\Lambda(\omega_0, \omega, p_0, p)
&=\frac{\omega_0-t^+ t^+_0\,\omega}{1 - \sqrt{t^+ t^+_0}}.
\end{align}
with $t^+_0=(\gamma_0+p_0)/(\gamma_0-p_0)$ and $t^+=(\gamma+p)/(\gamma-p)$. This can be obtained by using
\begin{align}
\label{eq:pp0_rewrite}
2p_0&=\sqrt{t^+_0}-\frac{1}{\sqrt{t^+_0}}\equiv \gamma_0+p_0-\frac{1}{\gamma_0+p_0},
\\
2p&=\sqrt{t^+}-\frac{1}{\sqrt{t^+}}\equiv \gamma+p-\frac{1}{\gamma+p}.
\end{align}
in the computation of $\Lambda$. We can also write $2\gamma_0=\sqrt{t^+_0}+1/\sqrt{t^+_0}$ and similarly for $\gamma$. We thus find
\begin{align}
\label{eq:lambda_pm_rewrite}
\lambda_\pm
&=\left[\frac{1+t^+}{4 \sqrt{t^+}}+\frac{1+t^+_0}{4 \sqrt{t^+_0}} \pm \frac{\omega+\omega_0}{2}\right]^2-1,
\end{align}
which is another useful relation for derivations.

Finally, by explicitly inserting $p$ and regrouping terms one can greatly reduce the expression for $\Lambda$, yielding
\begin{align}
\label{eq:Lambda_rewrite_III}
\Lambda(\omega_0, \omega, p_0, p)
&=\frac{p_0-p+\omega_0+\omega}{2}.
\end{align}
We will use $\Lambda$ in this form for our final kernel expression.

\vspace{-3mm}
\section{Momentum moments of the rMB distribution function}
\label{app:MMrMB}
The momentum moments for a rMB distribution are defined as
\begin{align}
\label{app:MMrMB_def}
\left<p^k\right>&=\int p^{k+2} f(\gamma) \id p = \int \gamma \sqrt{\gamma^2-1}^{k+1} f(\gamma) \id \gamma
\end{align}
For the first few moments we have
\bsub
\label{app:MMrMB_k}
\begin{align}
\left<p^0\right>&=1
\\
\left<p^1\right>&=\frac{2\The[1+3\The+3\The^2]}{K_2(1/\The)\,\expf{1/\The}}
\approx 2\sqrt{\frac{2\The}{\pi}}\left[1+\frac{9}{8}\The+\frac{9}{128}\The^2\right]
\\
\left<p^2\right>&=\frac{3\The K_3(1/\The)}{K_2(1/\The)}
\approx 3\The \left[1+\frac{5}{2}\The+\frac{15}{8}\The^2\right]
\\
\left<p^3\right>&=\frac{8\The^2[1+6\The+15\The^2+15\The^3]}{K_2(1/\The)\,\expf{1/\The}}
\approx 8\sqrt{\frac{2\The^3}{\pi}}\left[1+\frac{33}{8}\The\right]
\\
\left<p^4\right>&=\frac{15\The^2 K_4(1/\The)}{K_2(1/\The)}
\approx 15\The^2 \left[1+6\The+15\The^2\right]
\\
\left<p^5\right>&=\frac{48\The^3[1+10\The+45\The^2+105\The^3+105\The^4]}{K_2(1/\The)\,\expf{1/\The}}
\\
\left<p^6\right>&=\frac{105\The^3 K_6(1/\The)}{K_2(1/\The)}
\approx 105\The^3 \left[1+\frac{21}{2}\The+\frac{399}{8}\The^2\right]
\\
\left<p^{k}\right>&=\frac{2\,(2\The)^{k/2} K_{(k+4)/2}(1/\The)}{\sqrt{\pi}K_2(1/\The)}\,\Gamma\left(\frac{k+3}{2}\right),
\end{align}
\esub
where in the last line we introduced the $\Gamma$ function.

\end{appendix}

\end{document}